\documentclass[acmsmall]{acmart}
\setcopyright{none}
\AtBeginDocument{%
  \providecommand\BibTeX{{%
    \normalfont B\kern-0.5em{\scshape i\kern-0.25em b}\kern-0.8em\TeX}}}

\usepackage{subcaption}
\usepackage{adjustbox}
\usepackage{tikz}
\usetikzlibrary{quantikz}

\usepackage{wrapfig}
\usepackage{footmisc}

\usepackage{natbib}
\setcitestyle{square, comma, numbers,sort&compress, super}

\begin{document}

\title{Distributed Quantum Computing: a Survey}

\author{Marcello Caleffi}
\authornote{These authors contributed equally to this article.}
\orcid{0000-0001-5726-5489}
\affiliation{
    \department{Department of Electrical Engineering and Information Technologies (DIETI)}
    \institution{University of Naples Federico II}
    \city{Naples}
    \country{Italy}
}
\additionalaffiliation{
    \institution{\href{http://www.quantuminternet.it}{www.QuantumInternet.it} research group, FLY: Future communications Laboratory, University of Naples Federico II}
    \city{Naples}
    \state{Italy}
}
\additionalaffiliation{
    \institution{CNIT, National Inter-university Consortium for Telecommunications}
    \city{Naples}
    \state{Italy}
}
\email{marcello.caleffi@unina.it}
\author{Michele Amoretti}
\authornotemark[1]
\orcid{0000-0002-6046-1904}
\affiliation{
    \department{Department of Engineering and Architecture}
    \institution{University of Parma}
    \city{Parma}
    \country{Italy}
}
\additionalaffiliation{
    \institution{\href{https://www.qis.unipr.it/quantumsoftware.html}{Quantum Software Laboratory}, University of Parma}
    \city{Parma}
    \state{Italy}
}
\email{michele.amoretti@unipr.it}
\author{Davide Ferrari}
\authornotemark[1]
\authornotemark[4]
\orcid{0000-0002-4777-7234}
\affiliation{
    \department{Department of Engineering and Architecture}
    \institution{University of Parma}
    \city{Parma}
    \country{Italy}
}
\email{davide.ferrari1@unipr.it}
\author{Daniele Cuomo}
\authornotemark[2]
\orcid{0000-0002-9361-5797}
\affiliation{
    \department{Department of Physics}
    \institution{University of Naples Federico II}
    \city{Naples}
    \country{Italy}
}\email{daniele.cuomo@unina.it}
\author{Jessica Illiano}
\authornote{Jessica Illiano acknowledges support from TIM S.p.A. through the PhD scholarship.}
\authornotemark[2]
\orcid{0000-0002-7688-9593}
\affiliation{
    \department{Department of Electrical Engineering and Information Technologies (DIETI)}
    \institution{University of Naples Federico II}
    \city{Naples}
    \country{Italy}
}
\email{jessica.illiano@unina.it}
\author{Antonio Manzalini}
\orcid{0000-0003-1633-3099}
\affiliation{
\institution{TIM}
    \city{Turin}
    \country{Italy}
}
\author{Angela Sara Cacciapuoti}
\authornotemark[1]
\authornotemark[2]
\authornotemark[3]
\orcid{0000-0002-0477-2927}
\affiliation{
    \department{Department of Electrical Engineering and Information Technologies (DIETI)}
    \institution{University of Naples Federico II}
    \city{Naples}
    \country{Italy}
}

\renewcommand{\shortauthors}{Caleffi et al.}

\begin{abstract}
      Nowadays, quantum computing has reached the engineering phase, with fully-functional quantum processors integrating hundred of noisy qubits available. Yet -- to fully unveil the potential of quantum computing out of the labs and into business reality -- the challenge ahead is to substantially scale the qubit number, reaching orders of magnitude exceeding the thousands (if not millions) of noise-free qubits. To this aim, there exists a broad consensus among both academic and industry communities about considering the \textit{distributed computing} paradigm as the key solution for achieving such a scaling, by envision multiple moderate-to-small-scale quantum processors communicating and cooperating to execute computational tasks exceeding the computational resources available within a single processing device. 
    The aim of this survey is to provide the reader with an overview about the main challenges and open problems arising with distributed quantum computing, and with an easy access and guide towards the relevant literature and the prominent results from a computer/communications engineering perspective.
\end{abstract}



\maketitle

\section{Introduction}
\label{Sec:1}

The \textit{Quantum Internet}~\cite{CacCalTaf-20,Rohde2021} is envisioned as the final stage of the quantum revolution, opening new communication and computing capabilities. In synergy with the classical Internet, the Quantum Internet will connect quantum processors and devices to achieve capabilities that are provably impossible using classical communication. 

Within the last few years, a major effort is being undertaken by the research community and by the major ICT companies towards the Quantum Internet design and deployment.

Within the EU, The Quantum Flagship's projects are developing some of the most advanced physical quantum computing and quantum communication platforms in the world~\cite{EUQF}. Specifically, long-term ambition of the Quantum Internet Alliance project is to build a Quantum Internet that enables quantum communication applications between any two points on Earth~\cite{QIA}. In this decade, a quantum communication infrastructure (EuroQCI) will cover the whole EU, including its overseas territories~\cite{EuroQCI}. 

Overseas, the US government opened a number of publicly funded centers for quantum research in the last decade~\cite{Finigan2022}. One of them is Q-NEXT~\cite{QNEXT}, whose mission is to ``deliver quantum interconnects and establish a national foundry to provide pristine materials for new quantum devices''. Q-NEXT's vision is to help developing the technology that will enable applications in secure communication, distributed sensing, and scaling quantum computers. Another US public center is the Hybrid Quantum Architectures and Networks (HQAN)~\cite{HQAN}, whose goal is ``[tackling] the challenge of scaling quantum processors by pursuing an alternative paradigm: distributed quantum processing and networks composed of a hybrid architecture''. Furthermore, the Center for Quantum Networks (CQN)~\cite{CQN} is working directly on key challenges facing the construction of large-scale quantum networks. CQN's goals include developing foundational technology -- such as optical fibers, quantum repeaters, and switches -- for metropolitan-scale quantum networks. 

But Quantum Internet research is not limited to western countries. China is actively advancing research on quantum communications, with technology implementations such as their quantum secure communication networks (regional ``Trunks'') combined with their quantum satellite project (nicknamed \textit{Micius})~\cite{Yin2017}. It is estimated that there has been at least a 25 billion dollars Chinese government investment from the mid-1980s until 2022 into quantum technology~\cite{Graps2022}. One of Beijing's aims for its 14th five-year plan, which ends in 2025, is to establish an intercity quantum demonstration network based on secure relays. 

\begin{figure}[t]
    \centering
    \includegraphics[width=0.55\linewidth]{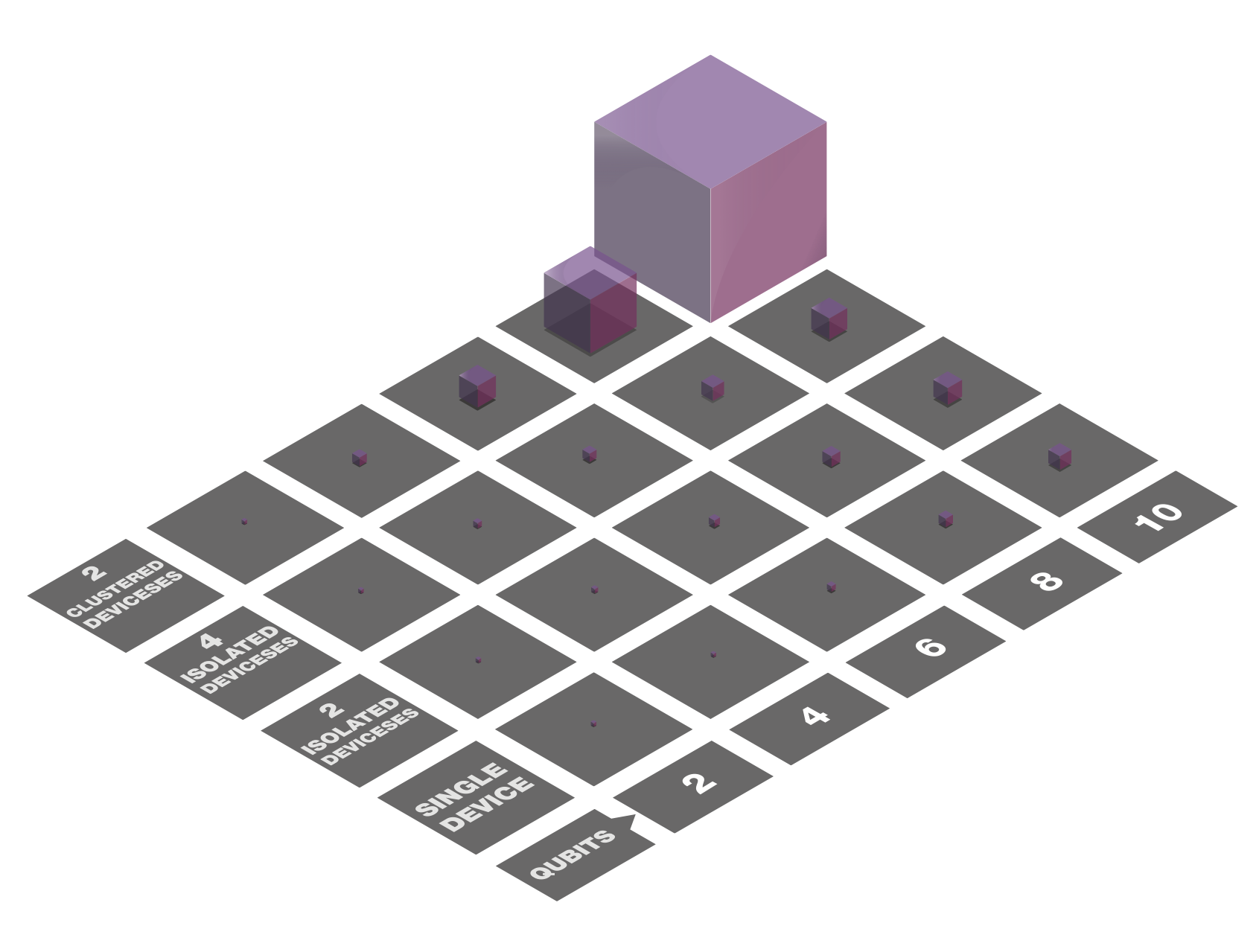}
    \caption{Quantum computing power of isolated vs interconnected processors. The volume of each cube graphically represents the ideal -- i.e., noise free -- quantum computing power as the number of qubits within each processor scales. Figure reproduced from~\cite{CuoCalCac-20}.}
    \label{Fig:01}
    \hrulefill
\end{figure}

Meanwhile, companies like IBM, Google, and Amazon are making significant investments in quantum computing and quantum networking. In May 2017, AT\&T announced that it was working with the California Institute of Technology to build out its quantum networking technology to offer more secure communications. British Telecommunications (BT), Toshiba Research, ADVA Optical Networking, and the UK National Physical Laboratory, are collaborating to research and implement quantum encryption. 
\begin{wrapfigure}[40]{r}{0.45\linewidth}
    \centering
        \usetikzlibrary{trees}
        \tikzstyle{every node}=[thick,anchor=west]
        \begin{tikzpicture}
            [level distance=2cm, grow via three points={one child at (0.3,-0.45) and two children at (0.3,-0.4) and (0.3,-0.95)}, edge from parent path={([xshift=0.0mm] \tikzparentnode.south west) |- (\tikzchildnode.west)}, growth parent anchor=south west, edge from parent/.style = {draw, -latex}]
            \node {}
            child { node {\small I. Introduction}}
            child { node {\small II. DQC: Distributed Quantum Computing}}
            child { node {\small III. Quantum Preliminaries}
                child[xshift=0.1cm] { node {\small III-A. Qubits in a Nutshell}}
                child[xshift=0.1cm] { node {\small III-B. Quantum Circuits}}
                child [xshift=0.1cm]{ node {\small III-C. Monolithic Execution}}
                child [xshift=0.1cm]{ node {\small III-D. Distributed Execution}}
            }
            child [missing] {}
            child [missing] {}
            child [missing] {}
            child [missing] {}
            child { node {\small  IV. Quantum Algorithms}
                child [xshift=0.1cm]{ node {\small IV-A. Partitioning of Quantum Algorithms}}
                child [xshift=0.1cm]{ node {\small IV-B. Execution Management}}
                child [xshift=0.1cm]{ node {\small IV-C. Open Issues and Research Directions}}
            }
            child [missing] {}
            child [missing] {}
            child [missing] {}
            child { node {\small V. Quantum Networking}
                child [xshift=0.1cm]{ node {\small V-A. Quantum Internet}}
                child [xshift=0.1cm]{ node {\small V-B. Quantum Teleportation}} 
                child[xshift=0.1cm] { node {\small V-C. Teledata vs Telegate}}
                child[xshift=0.1cm] { node {\small V-D. Physical vs Virtual Quantum Link}}
                child[xshift=0.1cm] { node {\small V-E. Open Issues and Research Directions}}
            }
            child [missing] {}
            child [missing] {}
            child [missing] {}
            child [missing] {}
            child [missing] {}
            child { node {\small VI. Quantum Compiling}
                child [xshift=0.1cm] { node {\small VI-A. Qubit Assignment}}
                child [xshift=0.1cm] { node {\small VI-B. Non-local Gate Handling}}
                child[xshift=0.1cm] { node {\small VI-C. Open Issues and Research Directions}}
            }
            child [missing] {}
            child [missing] {}
            child [missing] {}
            child { node {\small VII. Simulation Tools}
                child [xshift=0.1cm]{ node {\small VII-A. Hardware-Oriented}}
                child [xshift=0.1cm]{ node {\small VII-B. Protocol-oriented}}
                child [xshift=0.1cm]{ node {\small VII-C. Application-oriented}}
                child [xshift=0.1cm]{ node {\small VII-D. Open Issues and Research Directions}}
            }
            child [missing] {}
            child [missing] {}
            child [missing] {}
            child [missing] {}
            child { node {\small VIII. Conclusions and Future Perspectives}
                child [xshift=0.1cm] { node {\small VII-A. Industrial and Standardization Perspective}}
                child [xshift=0.1cm]{ node  {\small VII-B. The Journey Ahead}}
            }
            child [missing] {}
            child [missing] {}
            ;
        \end{tikzpicture}
    \caption{Structure of the Survey}
    \label{Fig:02}
    \hrulefill
\end{wrapfigure}
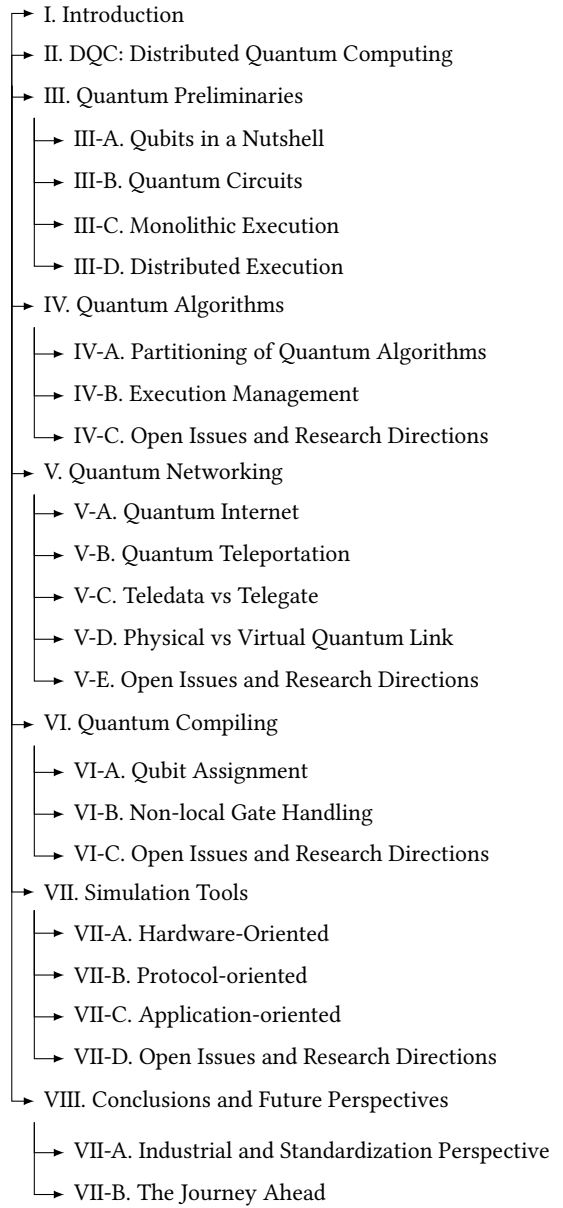

In June 2018, BT announced that it had built a ``quantum-secured'' internet network that spanned between Cambridge, UK and BT’s laboratory in Ipswich, a distance of around 50 miles~\cite{QCommHub}. Amazon has recently established the AWS Center for Quantum Networking~\cite{AWSQN}. Researchers at the center will work on technologies like quantum repeaters and transducers, to allow for the creation of global quantum networks.

Among the killer applications of the Quantum Internet~\cite{WanRahLi-22}, broad interest has been lately devoted to \textit{distributed quantum computing}, where individual quantum processors -- limited in the number of qubits -- work together to solve computational tasks exceeding the computational resources available within a single processing device~\cite{VanMeter2016,CalCacBia-18,CuoCalCac-20,FerCacAmo-21,Avron2021}. This interest came as no surprise since -- differently from classical distributed computing -- a linear increase in the number of interconnected quantum processors unlocks an exponential increase of the quantum computational power~\cite{CalCacBia-18,CuoCalCac-20}, as summarized in Figure~\ref{Fig:01}.

Unfortunately, the existing literature on distributed quantum computing is spread among different research communities -- ranging from the physics through the communications/computer engineering to the computer science community -- leading to a fundamental gap. The aim of this survey is precisely to bridge this gap, by introducing the reader to the astonishing and intriguing properties of distributed quantum computing.

Stemming from this, in the following we shed the light on the distinctive characteristics of distributed quantum computing, with the objective of allowing the reader:
\begin{itemize}
    \item [i)] to own the implications
    of the novel, astonishing and intriguing properties of quantum information for understanding the differences between distributed (classical) computing vs. distributed quantum computing;
    \item[ii)] to grasp the challenges as well as to appreciate the marvels arising with the paradigmatic shift from monolithic to distributed quantum computing
\end{itemize}

Indeed, due to the fast grow of this research field, such an understanding will serve the computer science community and the communications/computer engineering community alike to have an easy access and guide towards the relevant literature and to the prominent results, which will be of paramount importance for advancing the state-of-the-art.

To the best of authors' knowledge, a tutorial of this type is the first of its own. The paper is structured as depicted in Figure~\ref{Fig:02}. Specifically, in Section~\ref{Sec:2}, we introduce the rationale for distributed quantum computing, and we present the four different perspectives -- algorithms, networking, compiling and simulation -- discussed within the survey. Then, in Section~\ref{Sec:3}, we provide the reader with a concise overview about quantum computing, with reference to the quantum circuit model. In Section~\ref{Sec:4}, we focus on quantum algorithms, the extent to which they can be distributed as well as their execution management, once they are executed according to a distributed paradigm. In Section~\ref{Sec:5}, we detail the pivotal role played by quantum networking for enabling distributed quantum computing, by discussing in detail the unconventional features of quantum communications.
In Section~\ref{Sec:6}, we describe some relevant approaches to the problem of compiling quantum algorithms for distributed execution, i.e., splitting them conveniently to fit the available networking and computing hardware. In Section~\ref{Sec:7}, we provide an overview of the most advanced simulation tools for quantum networking, discussing their suitability for the design and analysis of distributed quantum computing architectures. Finally, we conclude our survey in Section~\ref{Sec:8} by first providing an industrial perspective on distributed quantum computing, and then by discussing the possible stages of distributing quantum computing development.

\section{DQC: Distributed Quantum Computing}
\label{Sec:2}

Nowadays, all major quantum computing technologies -- e.g., ion traps, superconductors, quantum dots, etc. -- exhibit hard technological limitations on the number of qubits that can be embedded in a single quantum chip~\cite{VanMeter2016}. Accordingly, the consensus of both academic and industry communities for realizing large-scale quantum processors goes toward a quantum computing paradigm-shift, which consists in relying on a quantum network infrastructure to cluster together modular and small quantum chips in order to scale the number of qubits~\cite{CalCacBia-18,WehElkHan-18,CacCalTaf-20,CalChaCuo-20}.

The aforementioned vision is expected to be realized in a very near future. For instance, IBM plans to introduce in 2025 \textit{Kookaburra} -- a 1386 qubit multi-chip processor with communication link support for quantum parallelization -- with three Kookaburra chips inter-connected into a 4158-qubit system~\cite{IBM2025}. With such modular systems, an ordinarily monolithic quantum computation can be ``split into pieces'' and executed on multiple inter-connected processors by following the distributed quantum computing (DQC) paradigm~\cite{CuoCalCac-20,Parekh2021}. Despite being expected further into the future, metropolitan-area and wide-area quantum networks are also under research and development~\cite{ZhoChaBie-21,PomHerBai-21,HerPomBeu-22}, which would enable DQC among geographically-distributed quantum device. 

\begin{figure}[t]
    \centering
    \includegraphics[width=0.67\linewidth]{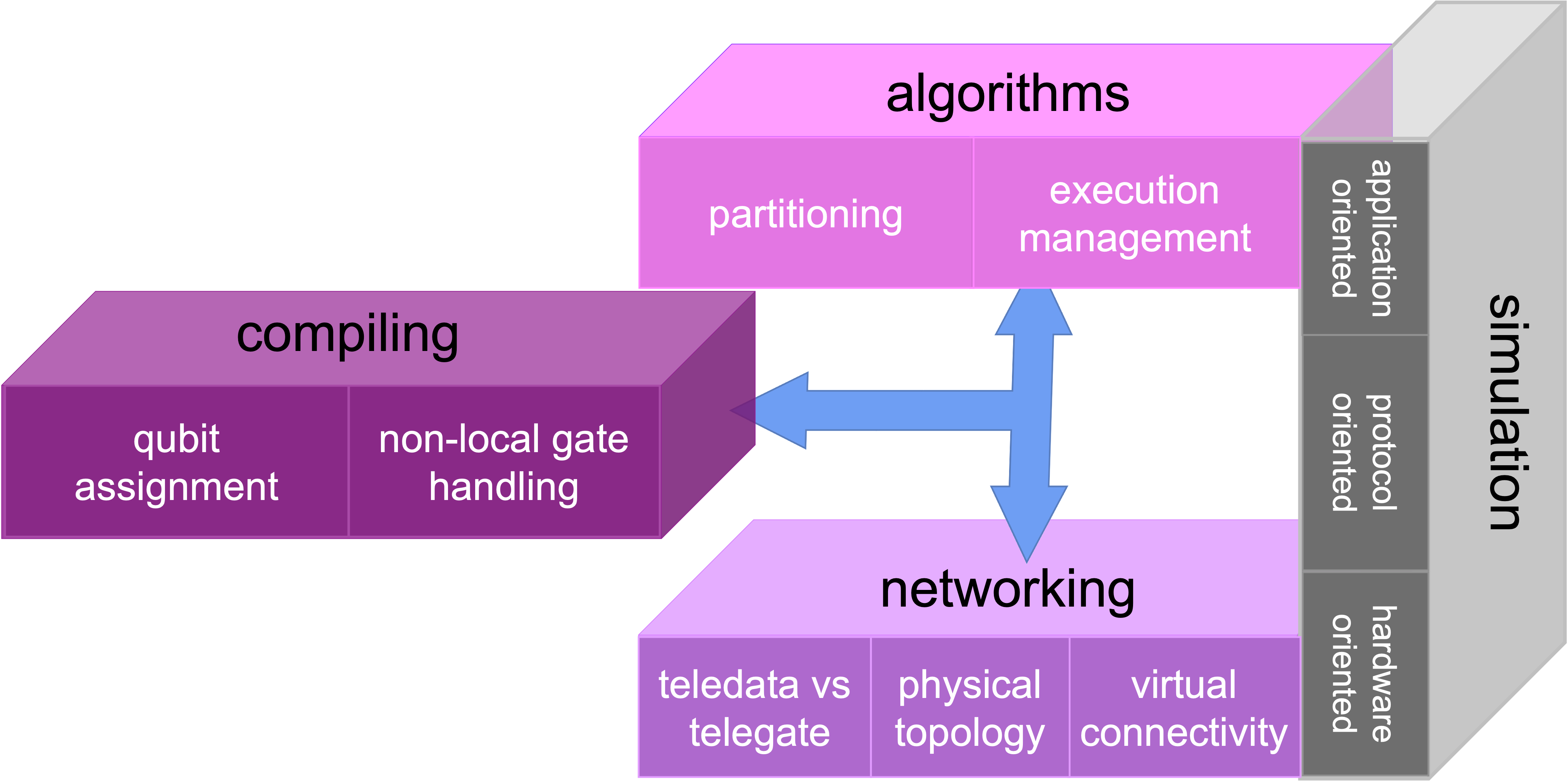}
    \caption{Distributed Quantum Computing. The four different perspectives overviewed within the survey, with \textit{algorithms}, \textit{compiling} and \textit{networking} represented as interdependent layers of a distributed quantum computing architecture, and \textit{simulation} represented as an inter-layer enabler, covering layers exhibiting quantumness, namely, \textit{algorithms} and \textit{networking}.}
    \label{Fig:03}
    \hrulefill
\end{figure}

This survey focuses on DQC, and it analyzes the state-of-the-art and challenges arising by looking at the DQC according to four main different perspectives, namely: \textit{algorithms}, \textit{networking}, \textit{compiling}, and \textit{simulation}. 

As illustrated in Figure~\ref{Fig:03}, for each of these four pillars, the most relevant aspects are discussed. Regarding algorithms, the focus is on the crucial and specific challenges arising when moving from monolithic to distributed quantum computing, namely, quantum algorithm partitioning and execution management. Being DQC an application of quantum networking -- namely, being some sort of (quantum) network a fundamental pre-requisite for any form of distributed (quantum) computing -- through the survey we will shed the light on the challenges arising with inter-networking different quantum processors, by introducing the reader to the fundamental differences between interconnecting remote \textit{classical} processors versus interconnecting remote \textit{quantum} processors. As regards to compiling, it deals with translating a hardware-agnostic description of the algorithm -- namely, the quantum circuit -- into a functionally equivalent description that takes into account the physical constraints of the underlying computing architecture~\cite{FerCacAmo-21}. Indeed, within the context of DQC, the compiling must account also for the network constraints, which impact on the strategy adopted for splitting the circuit into ``portions'' to be concurrently executed on the individual quantum processing units (QPUs)\footnote{Throughout the manuscript, the two terms quantum processor and quantum process unit are used as synonyms.}. As a matter of fact, a key goal is to minimize the number of remote operations, i.e., operations involving different QPUs. Last but not least, the design of DQC architectures can be highly facilitated by adequate simulation tools, as discussed and detailed in the manuscript.

\section{Quantum Preliminaries}
\label{Sec:3}

In this section, we provide a short overview about quantum computing, with reference to the quantum circuit model. We refer the reader to~\cite{CacCalVan-20} for a concise introduction to the peculiarities and the challenges arising with quantum information, whereas an overview about quantum computing and an in-depth treatise about quantum information and quantum computation are provided in~\cite{RiePol-00,RiePol-11} and~\cite{NieChu-11}, respectively.

\subsection{Qubits in a Nutshell}
\label{Sec:3.1}

\begin{figure}[t]
    \centering
    \resizebox{.55\textwidth}{!}{
        \begin{tikzpicture}[line cap=round, line join=round, >=Triangle,scale=1]
    		\draw(0,0) circle (2cm);
	    	\draw [rotate around={0.:(0.,0.)},dash pattern=on 3pt off 6pt] (0,0) ellipse (2cm and 0.9cm);
    		\draw [fill] (0,0) circle (1.5pt);
		    \draw [->] (0,0) -- (0,2.65);
    		\draw (0,3.1) node[anchor=north] {$\mathbf {\hat{z}}$};
		    \draw (-0.05,2.5) node[anchor=north west] {$\ket{0}$};
		    \draw [fill] (0,2) circle (1.5pt);
		    \draw (-0.05,-1.95) node[anchor=north west] {$\mathbf {\ket{1}}$};
		    \draw [fill] (0,-2) circle (1.5pt);
		    \draw [->] (0,0) -- (2.65,0);
		    \draw (2.65,0.3) node[anchor=north west] {$\mathbf {\hat{y}}$};
		    \draw [shift={(0,0)}, red, fill, fill opacity=0.1] (0,0) -- (56.7:0.4) arc (56.7:90.:0.4) -- cycle;
		    \draw (0,0)-- (0.70,1.07);
		    \draw [red] (-0.08,-0.3) node[anchor=north west] {$\phi$};
		    \draw [red] (0.01,0.9) node[anchor=north west] {$\theta$};
		    \draw [dotted] (0.7,1)-- (0.7,-0.46);
		    \draw [shift={(0,0)}, red, fill, fill opacity=0.1] (0,0) -- (-135.7:0.4) arc (-135.7:-33.2:0.4) -- cycle;
		    \draw [red] [dotted] (0,0)-- (0.7,-0.46);
		    \draw (0.65,1.7) node[anchor=north west] {$\ket{\varphi}$};
		    \draw [fill] (0.7,1.1) circle (1.5pt);
		    \draw [->] (0,0) -- (-1.053,-1.027);
		    \draw (-1.4,-0.95) node[anchor=north west] {$\mathbf {\hat{x}}$};
		    \draw [fill] (-0.84,-0.82) circle (1.5pt);
		    \draw (-0.95,-0.80) node[anchor=north west] {$\ket{+} = \frac{\ket{0} + \ket{1}}{\sqrt{2}}$}; 
		    \draw [fill] (2,0) circle (1.5pt);		
		    \draw (1.95,0) node[anchor=north west]     {$\ket{+_{\pi/2}} = \frac{\ket{0} + i \ket{1}}{\sqrt{2}}$};
	    \end{tikzpicture}
	}
    \caption{Bloch sphere: geometrical representation of a qubit in spherical coordinates. A pure state $\ket{\varphi} = \alpha \ket{0} + \beta \ket{1}$ is represented by a point on the sphere surface, with $\alpha = \cos{\frac{\theta}{2}}$ and $\beta = e^{i \phi} \sin{\frac{\theta}{2}}$.}
    \label{Fig:04}
    \hrulefill
\end{figure}
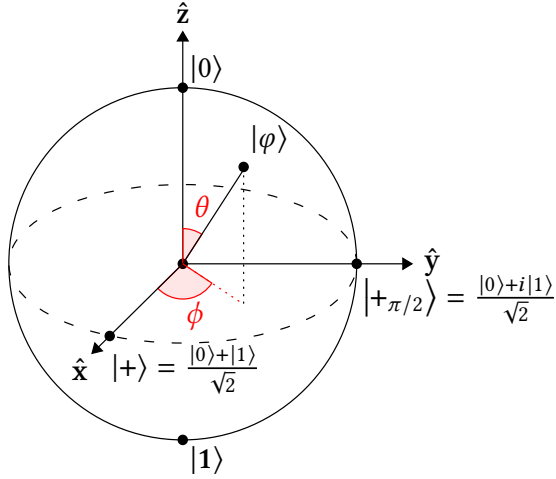

Information, either classical or quantum, can be encoded in the state of the simplest quantum mechanical system, namely, the quantum bit (qubit). Examples of two-level quantum systems are the spin of the electron, the polarization of a photon, or an atom with a ground state and an excited state.

Mathematically, the state of a qubit is defined as a vector in a two-dimensional complex Hilbert space. Hence qubit states can be treated as mathematical objects, thus enabling the construction of a general theory of quantum computing, which does not rely on the particulars of the underlying technology. By adopting the \textit{bra-ket} notation\footnote{The bra-ket notation, also known as Dirac’s notation~\cite{Dirac1939}, is a standard notation for describing quantum states. In a nutshell, a \textit{ket} $\ket{\cdot}$ represents a column vector, whereas a \textit{bra} $\bra{\cdot} \doteq \ket{\cdot}^\dagger$ represents the conjugate transpose of the corresponding ket.}, the state of an arbitrary qubit can be expressed as a linear combination -- namely, as a \textit{superposition} -- of two basis states $\ket{0}$ and $\ket{1}$:
\begin{equation}
    \label{Eq:1}
    \ket{\varphi} = \alpha \ket{0} + \beta \ket{1}.
\end{equation}
with $\alpha, \beta \in \mathbb{C}: |\alpha|^2 + |\beta|^2 = 1$ (normalization condition). The state of a single-qubit system is often represented geometrically in spherical coordinates by the \textit{Bloch sphere}, illustrated in Figure~\ref{Fig:04}. Specifically, a pure state is represented by a point on the sphere’s surface, with $\theta$ and $\phi$ denoting the spherical coordinates: 
\begin{equation}
    \label{Eq:1bis}
    \ket{\varphi} = \cos \frac{\theta}{2} \ket{0} + e^{i\phi} \sin \frac{\theta}{2} \ket{1}. 
\end{equation}
$\phi$ is known as relative phase of the quantum state and it is crucial in creating the interference patterns exploited for instance by quantum algorithms. 

The above can be generalized for a composite system of n-qubits, to which an Hilbert space of dimension $2^n$ can be associated, since the vector spaces associated with the constituent quantum systems are combined through the tensor product. The state of an n-qubit system can be in a superposition of all the $2^n$ basis states:
\begin{equation}
    \label{Eq:1ter}
    \ket{\phi}=\sum_{k=0}^{2^n-1}\alpha_k\ket{k},
\end{equation}
with $\alpha_k \in \mathbb{C}: \sum_{k=0}^{2^n-1}|\alpha_k|^2=1$. It must be noted, though, that the vast majority of n-qubit states cannot be written as the tensor product of n single-qubit states. These states are referred to as \textit{entangled states} and they represent the key ingredient in quantum computing~\cite{JozLin-03,MarSan-11}. As an example, a popular two-qubit entangled state, referred to as \textit{Bell state} or \textit{EPR pair}\footnote{With Bell states named in honor of Bell~\cite{Bel-64}, and EPR pairs named in honor of Einstein, Podolsky and Rosen~\cite{EinPodRos-35}.}, is given by:
    \begin{equation}
    \label{Eq:2}
    \ket{\Phi^+} = \frac{\ket{00} + \ket{11}}{\sqrt{2}}.
\end{equation}

\subsection{Quantum Circuits}
\label{Sec:3.2}

The \textit{quantum circuit}~\cite{NieChu-11} is the most popular model of quantum computation, where quantum operators are described as quantum gates. More into details, by sequentially interconnecting different quantum gates, a quantum circuit models the processing of quantum information corresponding to a specific \textit{quantum algorithm} is obtained~\cite{FerCacAmo-21}. Indeed, there exist several equivalent quantum circuits modeling the same computation with a different arrangement or different ordering of gates.

A very simple example of quantum circuit is provided in Figure~\ref{Fig:05}, where each horizontal line represents the time evolution of the state of a single (logical) qubit, with time flowing from left to right, dictating the order of execution of the different gates.

Quantum gates (and, overall, quantum circuits) are described by \textit{unitary matrices} relative to some basis, i.e., matrix $U$ such that $U^{\dagger}U = I$. It follows that quantum computation is \textit{reversible}: it is always possible to invert a quantum computation.

\begin{table*}[t]
    \scriptsize
    \centering
    \begin{tabular}{ | c | c | l |} 
		\toprule
        \textbf{Gate Name} & \textbf{Gate Matrix} & \textbf{Operation}\\ 
		\toprule
		Identity & $I = \begin{pmatrix} 1 & 0 \\ 0 & 1\end{pmatrix}$ & does not modify the quantum state\\
		\midrule
		Pauli-X & $\text{\texttt{X}} = \begin{pmatrix} 0 & 1 \\ 1 & 0\end{pmatrix}$ & bit-flip\\
		\midrule
		Pauli-Y & $\text{\texttt{Y}} = \begin{pmatrix} 0 & -i \\ i & 0\end{pmatrix}$ & combined bit- and phase-flip\\
		\midrule
		Pauli-Z & $\text{\texttt{Z}} = \begin{pmatrix} 1 & 0 \\ 0 & -1\end{pmatrix}$ & phase-flip\\
		\midrule
		Square-root-of-X & $\text{\texttt{SX}} = \frac{1}{\sqrt{2}} \begin{pmatrix} 1+i & 1-i \\ 1-i & 1+i\end{pmatrix}$ & $\text{\texttt{SX}}$ such that $\text{\texttt{X}} = \text{\texttt{SX}} \circ \text{\texttt{SX}}$ \\
		\midrule
		Hadamard & $\text{\texttt{H}} = \frac{1}{\sqrt{2}} \begin{pmatrix} 1 & 1 \\ 1 & -1\end{pmatrix}$ & maps an element of the computational basis  -- either $\ket{0}$ or $\ket{1}$ -- into an\\
		& & even superposition of the basis elements (and vice versa)\\
		\midrule
		S & $\text{\texttt{S}} = \begin{pmatrix} 1 & 0 \\ 0 & i\end{pmatrix}$ & $\pi/2$ phase shift\\
		\midrule
		T & $\text{\texttt{T}} = \begin{pmatrix} 1 & 0 \\ 0 & e^{i\pi/4}\end{pmatrix}$ & $\pi/4$ phase shift\\
		\midrule
		Phase shift & $\text{\texttt{P}}_\theta = \begin{pmatrix} 1 & 0 \\ 0 & e^{i \theta}\end{pmatrix}$ & $\theta$ phase shift\\
		\midrule
		x-Rotation & $\text{\texttt{R}}_x(\theta) = e^{-i\frac{\theta}{2}X} = \cos \frac{\theta}{2}I - i \sin \frac{\theta}{2}X$ & rotation by $\theta$ along $\hat{x}$-axis of the Bloch sphere \\
		\midrule
		y-Rotation & $\text{\texttt{R}}_y(\theta) = e^{-i\frac{\theta}{2}Y} = \cos \frac{\theta}{2}I - i \sin \frac{\theta}{2}Y$ & rotation by $\theta$ along $\hat{y}$-axis of the Bloch sphere \\
		\midrule
		z-Rotation & $\text{\texttt{R}}_z(\theta) = e^{-i\frac{\theta}{2}Z} = \cos \frac{\theta}{2}I - i \sin \frac{\theta}{2}Z$ & rotation by $\theta$ along $\hat{z}$-axis of the Bloch sphere \\
		\midrule
		Controlled-NOT & $\text{\texttt{CNOT}} = \begin{pmatrix} 1 & 0 & 0 & 0 \\ 0 & 1 & 0 & 0 \\ 0 & 0 & 0 & 1 \\ 0 & 0 & 1 & 0 \end{pmatrix}$ & bit-flips the second (target) qubit whenever the first (control) qubit is $\ket{1}$ \\
		\bottomrule
    \end{tabular}
    \caption{Common quantum gates.}
    \label{Tab:01}
    \hrulefill
\end{table*}

\begin{figure}[t]
	\centering
	\begin{minipage}[c]{.48\textwidth}
		\centering
		\begin{quantikz}[row sep={1.3cm,between origins},column sep=0.6cm]
            \lstick{$\ket{0}$}& \gate{\texttt{H}} & \ctrl{1} & \qw \rstick[2]{$\ket{\Phi^+} = \frac{\ket{00} + \ket{11}}{\sqrt{2}}$}\\
            \lstick{$\ket{0}$}& \qw & \targ{} & \qw
        \end{quantikz}
		\caption{Quantum circuit for generating a two-qubit entangled state -- namely, the Bell state in \eqref{Eq:2} -- starting from the input state $\ket{00}$. Time flows from left to right: the first qubit undergoes through a single-qubit Hadamard gate -- denoted with \texttt{H} -- followed by a two-qubit \texttt{CNOT} gate -- represented by $\bullet$ and $\oplus$ symbols interconnected by a vertical line -- both defined in Table~\ref{Tab:01}.}
        \label{Fig:05}
	\end{minipage}
	\hspace{.02\textwidth}
	\begin{minipage}[c]{.48\textwidth}
		\centering
        \begin{quantikz}[row sep={1.3cm,between origins},column sep=0.6cm]
            \lstick{$\ket{\psi}$} & \meter{$\ket{0},\ket{1}$} & \cw{0,1}
        \end{quantikz}
        \caption{Quantum circuit for measuring a qubit. Single wires denote quantum states, whereas double wires denote classical states, namely, bits. The measurement of a qubit -- whose output is a classical bit -- induces the state of the qubit to collapse into the measured state.}
    \label{Fig:06}
	\end{minipage}\\
	\hrulefill
\end{figure}

Some widely-used gates\footnote{
It is worth noting that most quantum gates are self-inverse (like Hadamard ad Pauli gates) or determining the inverse is straightforward (e.g., by taking the negated rotation angle for rotation gates).} are reported in Table~\ref{Tab:01} and, as a matter of fact, every unitary operator $U$ on a single qubit can be formulated as:
\begin{equation}
    \label{Eq:3}
    U = e^{i\theta_1}\text{\texttt{R}}_x(\theta_2)\text{\texttt{R}}_y(\theta_3)\text{\texttt{R}}_z(\theta_4), \quad \theta_i \in \mathbb{R}
\end{equation}
with $\text{\texttt{R}}_i$ denoting the $i$-axis rotation operator, defined in Table~\ref{Tab:01}. More precisely, the possibility of implementing two arbitrary rotation operators is sufficient, as their combined application can be exploited to obtain the third type of rotation in \eqref{Eq:3}.

Among two-qubit gates, highly relevant are the \textit{controlled} ones. The generic \texttt{Controlled-U} gate operates on two\footnote{\textit{Controlled} operations can be also defined for multi-qubit targets.} qubits, namely a \textit{control qubit} (controlling the operation) and a \textit{target qubit} (subjected to the operation). By denoting with $\ket{\varphi_c}$ and $\ket{\varphi_t}$ the control and target qubits respectively, the effect of the controlled U gate on the target qubit is the following:
\begin{equation}
    \label{Eq:4}
    \begin{cases}
        \text{\texttt{I}}\ket{\varphi_t} & \text{ if } \ket{\varphi_c} = \ket{0} \\
        \text{\texttt{U}}\ket{\varphi_t} & \text{ if } \ket{\varphi_c} = \ket{1}.
    \end{cases}
\end{equation}

The most famous example of controlled gate is represented by the \texttt{CNOT} gate, which is a \texttt{Controlled-U} gate where the \texttt{U} gate is a \texttt{Pauli-X} one. The \texttt{CNOT} gate can be used to create or destroy entanglement among the inputs. Specifically, to obtain an entangled state, we may start from the separable input $\ket{00}$ and, by applying \texttt{H} to the first qubit, obtaining $(\ket{00} + \ket{10})/\sqrt{2}$. Finally, by applying a \texttt{CNOT} gate (where the first qubit is the control one, as shown in Figure~\ref{Fig:05}), the resulting state is exactly the Bell state given in \eqref{Eq:2}, namely, $(\ket{00} + \ket{11})/\sqrt{2}$.

Gates \texttt{H}, \texttt{S} and \texttt{CNOT} constitute the \textit{Clifford group}~\cite{Gottesman1998}, which can be simulated efficiently on a classical computer according to the Gottesman-Knill theorem~\cite{NieChu-11}. The Clifford group is not universal, i.e., it cannot be used to describe any arbitrary quantum algorithm. However, it is sufficient to add the \texttt{T} gate to the Clifford group, and the resulting set is universal, yet it is not the only possible one. Indeed, each family of quantum computers -- e.g., IBM Q one~\cite{Abhijith2022} -- has its own specific universal gate set, which usually depends on the particulars of the underlying quantum hardware technology.

It is worthwhile to note that, regardless the particulars of the adopted gate set, deterministic cloning of quantum states is impossible. Specifically, there exists no quantum gate (or circuit) able to make a perfect copy of an arbitrary unknown quantum state. Conversely, if the state is known in advance -- specifically, if we know that the state belongs to some orthonormal basis such as $\{\ket{0},\ket{1}\}$ or $\{\ket{+},\ket{-}\}$ -- we can design a specific quantum gate to clone that state. This fundamental property is known as \textit{no-cloning theorem}, and it has deep impact on distributed quantum computing as we will discuss in Section~\ref{Sec:5}.

Another unconventional quantum phenomenon arises with the important operation constituted by \textit{measurement}, through which information from a quantum state is extracted~\cite{KouCacCal-21}, as illustrated in Figure~\ref{Fig:06}. In fact, according to the quantum measurement postulate, although a qubit may reside in a superposition of two orthogonal states as in \eqref{Eq:1}, when we want to observe or measure its value, it collapses into one of the two orthogonal states $\ket{0}$ -- with probability $|\alpha|^2$ -- and $\ket{1}$ -- with the probability $|\beta|^2$\footnote{The measurement of a qubit state may also be carried out in a basis different from that in which the qubit was prepared in~\cite{RiePol-00,NieChu-11,RiePol-11}. In the above description, for the sake of clarity, we assumed the standard basis also for the measurement.}. After its measurement/observation, the original quantum state collapses to the measured state. Hence, the measurement irreversibly alters the original qubit state~\cite{CacCalVan-20}. 

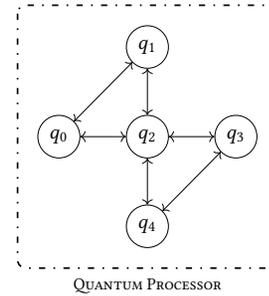
\begin{wrapfigure}[20]{R}{.45\textwidth}
	\centering
    \begin{adjustbox}{width=.25\textwidth}
        \begin{tikzpicture}[->]
            \definecolor{darkblue}{HTML}{1f4e79}
            \node [circle, draw] (0) at (0,0) {$q_{0}$};
            \node [circle, draw] (1) at (1.5,1.5) {$q_{1}$};
            \node [circle, draw] (2) at (1.5,0) {$q_{2}$};
            \node [circle, draw] (3) at (3,0) {$q_{3}$};
            \node [circle, draw] (4) at (1.5,-1.5) {$q_{4}$};
            \draw[<->] (1) -- (0);
            \draw[<->] (2) -- (0);
            \draw[<->] (2) -- (1);
            \draw[<->] (3) -- (2);
            \draw[<->] (4) -- (2);
            \draw[<->] (4) -- (3);
            \draw[thick,loosely dashdotted, rounded corners] (-0.7,2.2) rectangle (3.7,-2.2);
            \node at (1.5,-2.5) {\textsc{\footnotesize Quantum Processor}};
        \end{tikzpicture}
    \end{adjustbox}
    \caption{Coupling maps of a IBM Yorktown quantum processor~\cite{LinMasRoe-17,KanTemCor-19}. The five physical qubits stored within the processor are represented by circles. The arrows denote the possibility to realize a two-qubit \texttt{CNOT} gate between the five qubits. As an example, a \texttt{CNOT} between qubits $q_0$  and $q_1$ can be directly executed by the quantum processor, whereas a \texttt{CNOT} between qubits $q_0$ and $q_2$ cannot.}
    \label{Fig:07}
    \hrulefill
\end{wrapfigure}

Finally, it is worthwhile to mention that a quantum circuit exhibits three important quantitative features: the \textit{width}, i.e., the number of qubits, the \textit{gate count} and the \textit{depth}, i.e., the longest path in the circuit. Each and all of them affect the overall \textit{quality} of the computation result. To get an easy flavor about the aforementioned statement, it is enough to think that due to imperfect hardware the expected error propagation is upper bounded by $2(1 - (1-r)^m)$~\cite{YuLi-22} where $0 \leq r < 1$ is a constant independent of the qubit number and $m$ is the number of gates in the circuit.

\subsection{Monolithic Execution: Gate Synthesis and Circuit Compilation}
\label{Sec:3.3}

As mentioned in the previous subsection, even if there exists an uncountable number of quantum logic gates, the set of gates that can be executed on a certain quantum processor is limited, as a consequence of the constraints imposed by the underlying qubit technology~\cite{VanDev-16}. In this case, any gate outside this \textit{reduced set} must be obtained with a proper combination of the allowed gates through a process known as \textit{gate synthesis}. As an example, IBM quantum processors are realized exploiting the superconducting technology, and any logical gate that can be run on current IBM quantum processors is built from a gate set composed by the controlled-not (\texttt{CNOT}) gate and four single-qubit gates (namely, \texttt{I}, $\text{\texttt{R}}_Z$, \texttt{SX}, \texttt{X} gates).

\begin{figure*}[t]
	\centering
	\begin{adjustbox}{width=1\textwidth}
        \begin{tikzcd}
            \lstick{$q_0$} & \ctrl{2} & \qw &&&
            \ctrl{1}\gategroup[2,steps=3, style={dashed, rounded corners, inner xsep=2pt}, label style={label position=above, yshift=+0.1cm}, background]{\sc SWAP $q_0$ and $q_2$} & \targ{} & \ctrl{1} & \qw & \ctrl{1}\gategroup[2,steps=3, style={dashed, rounded corners, inner xsep=2pt}, label style={label position=above, yshift=+0.1cm}, background]{\sc SWAP back} & \targ{} & \ctrl{1} & \qw &&&
            \ctrl{1} & \qw & \ctrl{1} & \qw & \qw\\
            \lstick{$q_2$} & \qw & \qw & \equiv & &
                \targ{} & \ctrl{-1} & \targ{} & \ctrl{1} & \targ{} & \ctrl{-1} & \targ{} & \qw & \equiv & &
                \targ{} & \ctrl{1} & \targ{} & \ctrl{1} & \qw\\
            \lstick{$q_4$} & \targ{} & \qw &&&
                \qw & \qw & \qw & \targ{} & \qw & \qw & \qw & \qw &&&
                \qw & \targ{} & \qw & \targ{} & \qw
        \end{tikzcd}
    \end{adjustbox}
    \caption{Example of equivalent quantum circuits generated during quantum compiling for mapping an arbitrary \texttt{CNOT} into a sequence of \texttt{CNOT}s that can be directly executed by a given quantum processor. A \texttt{CNOT} between qubits $q_0$ (control) and $q_4$ (target) with the coupling map given in Figure~\ref{Fig:07} can be obtained through either: i) \textit{quantum state transfer}, by first swapping qubits $q_0$ and $q_2$, then by performing a \texttt{CNOT} between $q_2$ and $q_4$, and finally by swapping again qubits $q_0$ and $q_2$ so that they recover their initial position, or ii) \textit{ancilla qubit}, by performing four \texttt{CNOT} operations between neighbour qubits with the help of the intermediate qubit $q_2$. Figure reproduced from~\cite{FerCacAmo-21}.}
    \label{Fig:08}
    \hrulefill
\end{figure*}
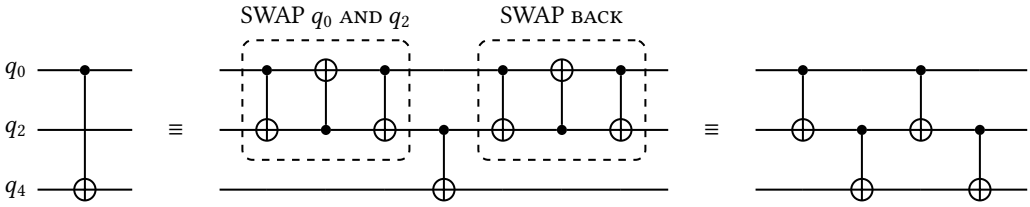

Furthermore, regardless of the underlying qubit technology, the abstract qubits subjected to quantum gates as specified by the quantum circuit, known as \textit{logical qubits}, should not be confused with the physical qubits embedded within a quantum processor~\cite{FerCacAmo-21}. With reference to the physical qubits, any quantum processor exhibits hardware constraints affecting the allowed interactions between them. As an example, \texttt{CNOT} gates cannot be applied to any physical qubit pair of an IBM quantum processor, but they are instead restricted\footnote{These limitations arise as a consequence of both the: i) noise effects induced by qubit-interactions, and ii) physical-space constraints within a single processor~\cite{FerCacAmo-21}.} to certain pairs. The allowed pairs are usually represented with the \textit{coupling map} -- namely, with a graph where vertices denote qubits and arrows denote the possibility of realizing a two-qubit \texttt{CNOT} gate between the connected qubits -- as illustrated in Figure~\ref{Fig:07}. From the above, it becomes clear that the monolithic execution of a quantum algorithm on a single quantum processor requires a circuit pre-processing known as \textit{quantum compiling}~\cite{Botea2018,KusSaeUya-21,SivDilCow-21,FerCacAmo-21}. In a nutshell, compiling a quantum circuit is a two-step\footnote{With the two steps being inter-dependent, affecting each others.} process where:
\begin{itemize}
    \item[i)] each logical qubit of the quantum circuit must be mapped onto one (or more, when adopting fault-tolerant techniques~\cite{CorKanJav-20}) physical qubit of the quantum processor, and
    \item[ii)] each two-qubit gate -- as instance, a \texttt{CNOT} -- between physical qubits non-adjacent within the coupling map must be mapped into a computational-equivalent sequence of gates between adjacent physical qubits, as exemplified in Figure~\ref{Fig:08}.
\end{itemize}
Clearly, the overall process must be optimized to account for the key performance metrics affecting quantum computation~\cite{FerAmo-18,CinSubSor-18,ZulPalWil-19}. Typically, this consists in minimizing the depth of the \textit{compiled circuit}, namely, the equivalent quantum circuit satisfying all the constrains imposed by the quantum processor coupling map.

\subsection{Distributed Execution}
\label{Sec:3.4}

\begin{wrapfigure}[19]{r}{.45\textwidth}
    \centering
        \begin{tikzcd}
            \lstick{$\ket{q_0}$} \gategroup[wires=2,steps=5,style={dashed,rounded corners,inner xsep=25pt,inner ysep=5pt}, label style={label position=above, yshift=+0.1cm}, background]{\sc Quantum Processor \#1} &
                \gate{H} & \ctrl{1} & \qw & \qw & \qw \\
            \lstick{$\ket{q_1}$} & \qw & \targ{} & \ctrl{1} & \qw & \qw \\
            \lstick{$\ket{q_2}$} \gategroup[wires=2,steps=5,style={dashed,rounded corners,inner xsep=25pt,inner ysep=5pt},label style={label position=below,yshift=-0.44cm}, background]{\sc Quantum Processor \#2} & \qw & \qw & \targ{} & \ctrl{1} & \qw \\
            \lstick{$\ket{q_3}$} & \qw & \qw & \qw & \targ{} & \qw
        \end{tikzcd}
    \caption{Toy model for distributed quantum computation. The quantum circuit is composed by three two-qubit gates, i.e, \texttt{CNOTs}. First and last gates operate \textit{locally}, namely, between qubits stored within the same QPU, whereas the intermediate gate operates \textit{remotely}, namely, between qubits stored within different QPUs.}
    \hrulefill
    \label{Fig:09}
\end{wrapfigure}
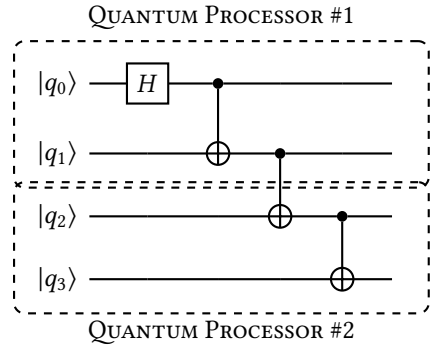

So far, we focused on quantum circuits by abstracting from the particulars of the underlying computing technology. Indeed, we mentioned that the natively-available gates depends on the underlying hardware. Yet, this is not an issue since -- as long as the available gate set is universal -- any arbitrary gate can be implemented with a finite sequence of the available gates up to arbitrary accuracy~\cite{RiePol-11}. From the above, we can safely assume that any quantum circuit can be directly executed on a given quantum processor, either in the original form or by properly replacing unavailable gates with sequences of available ones.
But the task becomes significantly harder when we move from monolithic quantum computing to distributed quantum computing. Specifically, any universal gate set must include one multi-qubit gate -- typically a 2-qubit gate such as the \texttt{CNOT} -- and any quantum circuit of some interest includes multi-qubit gates as well. As a consequence, as illustrated with the toy-model in Figure~\ref{Fig:09}, a distributed quantum computation involves operations between qubit pairs across different end-nodes (i.e., non-local gates). At a first sight, this might seem not a big deal. Also (classical) distributed computing involves operations between classical information located at different processors. And these operations are performed by simply moving the information from one processor to another. So one might be tempted to believe that the same strategy can be adopted when it comes to distributed quantum computing. Unfortunately, this is not true: quantum information requires a paradigm shift for dealing with inter-processor communications, and the rationale for this would be deeply discussed in the next sections.\\

\section{Quantum Algorithms}
\label{Sec:4}

There exists several quantum algorithms known or expected to outperform classical algorithms for problems spanning different areas, including cryptography, search and optimization, simulation of quantum systems and learning~\cite{Mon-16}. Remarkably, most known quantum algorithms use a combination of algorithmic paradigms -- namely, sub-routines -- specific to quantum computing~\cite{Abhijith2022}. These paradigms include the Quantum Fourier Transform (QFT)~\cite{Shor1994}, the Grover Operator (GO)~\cite{Gro-96}, the Harrow/Hassidim/Lloyd (HHL) method for linear systems~\cite{HarHasLlo-09}, Variational Quantum Algorithms (VQA)~\cite{CerArrBab-21}, and direct Hamiltonian simulation (SIM). A prominent example is Shor's algorithm for integer factorization~\cite{Shor1994}, which is based on QFT, illustrated by the quantum circuit in Figure~\ref{Fig:10}.

For most practical applications, quantum algorithms require large quantum computing resources -- in terms of qubit number -- much larger than those available with current noisy intermediate-scale quantum (NISQ) processors. 
For example, the recently announced IBM Quantum Osprey device has 433 qubits, which is an impressive progress with respect to state-of-the-art quantum processors, but not yet sufficient, as an example, for running practical implementations of Shor's algorithm\footnote{Factoring $L=2048$ bit primes -- for breaking current RSA implementations -- requires about $3L = 6144$ noise-free qubits~\cite{NieChu-11}. It is worth noting that merely increasing the number of physical qubits is not sufficient, as some sort of quantum error correction~\cite{Ter-15} is also required to guarantee high-quality -- namely, noise-free -- computations.}. 

Distributed quantum computing is envisioned as a scalable approach for increasing the number of qubits available for computational tasks. However, moving from monolithic to distributed quantum computing implies crucial and specific challenges.

\subsection{Partitioning of Quantum Algorithms}
\label{Sec:4.1}

A first issue that arises with quantum algorithms is whether a given algorithm -- equivalently, a given quantum circuit -- is natively suitable for distributed execution. More specifically, a \textit{perfectly distributable} quantum algorithm is a quantum algorithm that can be split into autonomous parts that do not interact -- or, at least, weakly interact -- with each others. If this is the case, each part can be assigned to some quantum processor, and each processor can contribute autonomously to the overall computation without introducing communication overhead for interacting with other processors. 

\begin{figure*}[t]
	\centering
	\begin{adjustbox}{width=1\linewidth}
        \begin{tikzcd}
            \lstick{$\ket{\varphi_1}$} & \gate{H} & \gate{R_2} & \gate{R_3} & \ \ldots\ \qw & \gate{R_{n-1}} & \gate{R_n} & \qw & \qw & \ \ldots\ \qw & \qw & \qw & \ \ldots\ \qw & \qw & \qw & \qw & \qw \\
            \lstick{$\ket{\varphi_2}$} & \qw & \ctrl{-1} & \qw & \ \ldots\ \qw & \qw & \qw & \gate{H} & \gate{R_2} & \ \ldots\ \qw & \gate{R_{n-2}} & \gate{R_{n-1}} & \ \ldots\ \qw & \qw & \qw & \qw & \qw \\
            \lstick{$\ket{\varphi_3}$} & \qw & \qw & \ctrl{-2} & \ \ldots\ \qw & \qw & \qw & \qw & \ctrl{-1} & \ \ldots\ \qw & \qw & \qw & \ \ldots\ \qw & \qw & \qw & \qw & \qw \\
            & & & & \ \ldots\ & & & & & \ \ldots\ & & & \ \ldots\ & \\
            \lstick{$\ket{\varphi_{n-1}}$} & \qw & \qw & \qw & \ \ldots\ \qw & \ctrl{-4} & \qw & \qw & \qw & \ \ldots\ \qw & \ctrl{-3} & \qw & \ \ldots\ \qw & \gate{H} & \gate{R_2} & \qw & \qw \\
            \lstick{$\ket{\varphi_n}$} & \qw & \qw & \qw & \ \ldots\ \qw & \qw & \ctrl{-5} & \qw & \qw & \ \ldots\ \qw & \qw & \ctrl{-4} & \ \ldots\ \qw & \qw & \ctrl{-1} & \gate{H} & \qw \\
        \end{tikzcd}
    \end{adjustbox}
    \caption{Quantum Fourier Transform (QFT) circuit. The $i$-th qubit is obtained through an Hadamard gate followed by $n-i$ controlled $\text{\texttt{R}}_n$ operations -- with $\text{\texttt{R}}_i = \text{\texttt{P}}_{2 \pi / i}$ denoting the phase gate given in Table~\ref{Tab:01} -- with the controlled operations controlled  by the $n-i$ higher-order qubits.}
    \label{Fig:10}
    \hrulefill
\end{figure*}

Unfortunately, this is not the usual case. As an example, let us consider the QFT algorithm, whose circuit is given in Figure~\ref{Fig:10}, notably used as sub-routine in many quantum algorithms -- e.g., Shor's algorithm and the quantum phase estimation algorithm -- as mentioned above. From Figure~\ref{Fig:10}, it is easy to assess that QFT requires each qubit to strongly interact with all the other qubits through controlled $\text{\texttt{R}}_n$ gates. Hence, QFT can be considered as the archetype of monolithic quantum algorithms, namely, of an algorithm not natively-suitable for distributed execution.

As we anticipated in Section~\ref{Sec:2}, to distribute a monolithic quantum algorithm, a quantum compiler must be used to find the best breakdown, i.e., the one that minimizes the number of gates that are applied to qubits stored at different devices. Quantum compilation is reviewed in Section~\ref{Sec:6}. Here we discuss some literature that addresses the partitioning of relevant quantum algorithms, using techniques that are tailored to the specific considered algorithms rather than general-purpose. These works may represent a good reference for a comparative evaluation of quantum compilers. 

In~\cite{Neumann2020}, Neumann et al. present two distribution schemes for the \textit{quantum phase estimation} algorithm, give the resource requirements for both and show that using less noisy shared entangled states results in a higher overall fidelity. Introduced by Kitaev~\cite{Kitaev1997}, the quantum phase estimation algorithm returns an approximation of an eigenvalue of a given unitary $U$ and a corresponding eigenvector. It has numerous applications, including Shor's algorithm~\cite{Shor1994}. The solution proposed by Neumann et al. is based on the distributed version of the QFT circuit, obtained by means of non-local controlled $U$-gates\footnote{Non-local controlled $U$-gate generalizes the telegate operation discussed in Section~\ref{Sec:5.3} to arbitrary unitary $U$~\cite{EisJAcPap-00}.}.

Another example of distributable quantum algorithm is the \textit{Variational Quantum Eigensolver} (VQE), a VQA that can be used to estimate ground state energies of molecular chemical Hamiltonians. In~\cite{DiAdamo2021}, DiAdamo et al. provide a \textit{Local to Distributed Circuit} algorithm that, given a circuit representation as a series of layers and a mapping of qubits, searches for any control gates where the control and target are physically separated between two QPUs. When found, the algorithm inserts, between the current layer and next layer in the circuit, the necessary steps to perform the control gate in a nonlocal way\footnote{\label{Footnote:1}By using the \textit{cat-entangling} method by Yimsiriwattana et al.~\cite{Yimsiriwattana2005}, which is substantially equivalent to telegate introduced in Section~\ref{Sec:5.3}.}. The size (maximum number of qubits) of the achievable Ansatz state for the VQE algorithm grows linearly with the number of QPUs, with slope linearly increasing with the number of qubits per QPU. The depth of the resulting quantum circuit is $\Omega(n)$, meaning it has a tight upper and lower bound proportional to the number $n$ of qubits.

In~\cite{NeuWez-22}, the authors present a distributed adder and a distributed distance-based classification algorithm. Both applications are framed in a way where a quantum server and $K$ other quantum nodes interact, with specific behaviors. In particular, the server is responsible for orchestrating the computation by means of non-local \texttt{CNOT} gates, while the $K$ parties provide inputs. It is possible to reframe these applications, such that the proposed quantum circuits are considered as monolithic and subsequently split in $K+1$ parts to be submitted for execution to a quantum network.
\begin{wrapfigure}[19]{r}{.55\textwidth}

    \centering
    \includegraphics[width=8cm]{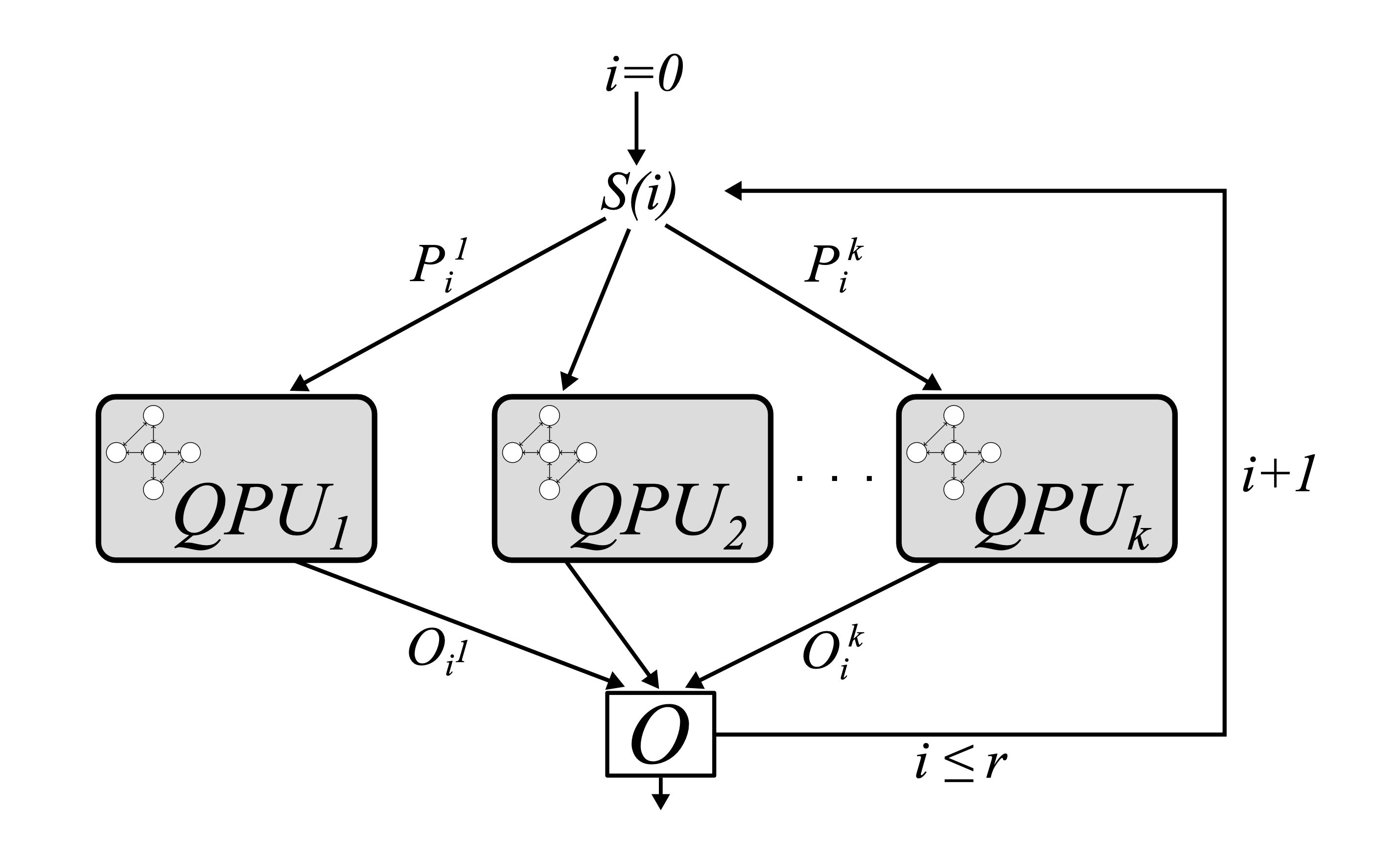}
    \caption{Execution of multiple quantum circuit instances with $k$ QPUs. For each execution round $i$, a schedule $S(i)$ maps some quantum circuit instances to the quantum network -- each QPU receiving a quantum circuit $P_i^j$ that is either a monolithic one or a sub-circuit of a monolithic one. The classical outputs are accumulated into an output vector $O$.}
    \label{Fig:11}
    \hrulefill

\end{wrapfigure}

\subsection{Execution Management}
\label{Sec:4.2}

Another challenge is related to the execution management of distributed quantum computations. In general, given a collection $\mathcal{P}$ of quantum circuit instances to be executed, this collection should be partitioned into non-overlapping subsets $\mathcal{P}_i$, such that $\mathcal{P} = \cup_i \mathcal{P}_i$. One after the other, each subset will be assigned to the available QPUs. In other words, for each execution round $i$, there exists a schedule $S(i)$ that maps some quantum circuit instances to the quantum network. If DQC is supported, some quantum circuit instances may be split into sub-circuit instances, each one to be assigned to a different QPU, as illustrated in Figure~\ref{Fig:11}). A QPU scheduling algorithm that partially address this service was proposed by Parekh et al.~\cite{Parekh2021}. 
Such an algorithm is based on a greedy approach, trying to fill all available QPUs while minimizing the number of distributed quantum circuit instances. Here the partitioning of quantum circuit instances is arbitrary, not taking into account the features of the programs.

Recalling Section~\ref{Sec:4.1}, we stress that partitioning should be an orthogonal service with respect to QPU scheduling.
It is worth noting that the QPU scheduling plane must be clearly separated from the networking plane. We demand that any subset of the available QPUs can be the target of any quantum computation, provided that the total number of physical qubits fits the circuit width. This means that the underlying network should allow to create entangled quantum states across any two QPUs. Technical details on entanglement distribution are presented in Section~\ref{Sec:5}. Here we recall a recent work by Cicconetti et al.~\cite{CicConPas-22}, which investigates the requirements and objectives of DQC from the perspective of quantum network provisioning. In particular, the authors elaborate on two different classes of traffic, namely constant-rate flows and DQC applications.

\subsection{Open Issues and Research Directions}
\label{Sec:4.3}

Future directions are both theoretical and practical. Despite a considerable amount of work on the fundamentals of distributed quantum computing~\cite{CirEkeHue-99,BeaBriGra-13,Brierley2017}, an ultimate theory of distributable quantum algorithms is still missing. It is known that the quantum circuit model and the DQC model are equivalent up to polylogarithmic depth overhead~\cite{BeaBriGra-13}, but a general framework for ranking quantum algorithms in terms of distributability has not been defined. To this purpose, it is necessary to provide a quantitative definition of quantum circuit distributability. 
Regarding execution management, the broad literature on job scheduling for high performance computing may be a starting point, but it is clear that the peculiarities of quantum computing -- quantum parallelism, no-cloning, entanglement, etc. -- demand for novel and specific strategies for the efficient execution of concurrent distributed quantum computations. A trade-off between the complexity of the distributed quantum circuit and the physical distance between quantum processors is also envisaged.

 To compare different deployments and schedules, DQC-specific key performance indicators must be defined.
Recently, two frameworks with similar names have been proposed almost at the same time, namely Quantum Network Utility Maximization (QNUM)~\cite{VARWEH-22} and Quantum Network Utility ($U_{QN}$)~\cite{LeeDaiTow-22}. While QNUM is specifically tailored to the evaluation of entanglement routing schemes in quantum networks (see Section~\ref{Sec:5} for details about entanglement), $U_{QN}$ is more abstract, aiming to capture the social and economic value of quantum networks, for a variety of applications (from secure communications to distributed sensing). Incidentally, in~\cite{LeeDaiTow-22} the example of DQC is studied in detail, through the lens of $U_{QN}$. More specifically, a quantum network utility metric is presented, which applies the Quantum Volume\footnote{Quantum Volume (QV) is a single-number metric that can be measured using a concrete protocol on near-term quantum computers of modest size. The QV method quantifies the largest random circuit of equal width and depth that the quantum processor successfully executes.} proposed in~\cite{CroBisShe-19} to the $U_{QN}$ framework. Such a metric quantifies the value derived from performing QC tasks, and it is viewed as a ``quantum volume throughput''. 
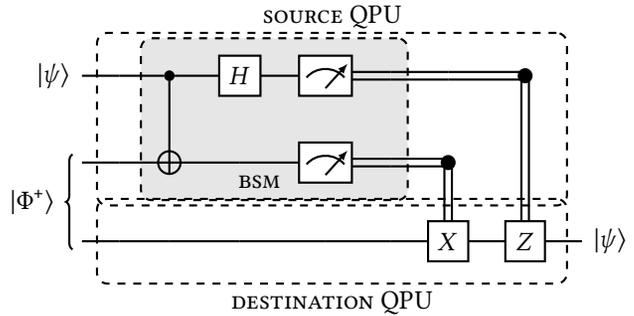
\begin{wrapfigure}[26]{r}{0.6\linewidth}
        \centering
            \begin{tikzcd}
                & \lstick{$\ket{\psi}$} &\qw\gategroup[wires=2,steps=7,style={dashed,rounded corners,inner xsep=5pt,inner ysep=5pt}, background, label style={label position=above, yshift=-0.0cm}, background]{\sc source QPU} &\ctrl{1} \gategroup[wires=2,steps=4,style={dashed,rounded corners,fill=gray!20,inner xsep=3pt,inner ysep=3pt}, background, label style={label position=below, xshift =-0.2cm, yshift=+0.05cm}, background]{\sc bsm} & \gate{H} & \meter{}& \cw & \cw & \cwbend{2}\\
            & \lstick[wires=2]{$\ket{\Phi^\texttt{+}}$} & \qw & \targ{} & \qw &  \meter{}& \cw & \cwbend{1} & & & \\
            &  & \qw\gategroup[wires=1,steps=7,style={dashed,rounded corners,inner xsep=5pt,inner ysep=5pt},background,label style={label position=below,yshift=-0.44cm}, background]{\sc destination QPU} & \qw & \qw & \qw &\qw &  \gate{X} & \gate{Z}  & \qw \rstick{\sc $\ket{\psi}$}
            \end{tikzcd}
        \caption{Pictorial representation of the quantum teleportation circuit. The first two wires belong to the source node, whereas the bottom wire belongs to the destination node. A generic state $\ket{\psi}$ is initially stored at the source, and a Bell state such as $\ket{\Phi^{\texttt{+}}}$ given in \eqref{Eq:2} must be distributed through a quantum link so that one entangled member is stored at the source and the other at the destination. Once the Bell state is available, the teleportation is obtained with some processing of $\ket{\psi}$ and the entangled member at the source, followed by two conditional gates on the entangled pair at the destination, depending on the measurement of the two qubits at the source. Each double line denotes the transmission of one classical bit -- i.e., the measurement output -- between the remote processors. The two classical bits are thus used as detailed in Table~\ref{Tab:02} for determining whether the two conditional gates \texttt{X} and \texttt{Z} must be applied to recover the original state $\ket{\psi}$ from the entangled member available at the destination.}
            \hrulefill
    \label{Fig:12}
\end{wrapfigure}
It differs from the quantum volume in two ways: i) it explicitly considers the rate at which non-local operations can be performed, and ii) it accounts for the utility derived simultaneously from tasks executed on different parts of the network.

\section{Quantum Networking}
\label{Sec:5}

As mentioned in the previous sections, when it comes to distributed quantum computing, qubits are distributed among multiple smaller quantum processors, interconnected by some sort of quantum network.

Accordingly, whenever a quantum gate must operate on \textit{remote qubits} -- namely, qubits located in different quantum processors -- some sort of \textit{communication primitive} must be available for performing inter-processor operations. Unfortunately, this communication primitive cannot be easily accomplished through classical protocols. Indeed, the different physical phenomena underlying quantum communications impose a paradigm shift.

To better understand the above statement, in the following we shed the light on the challenges arising with networking different quantum processors. To this aim, in the following we first substantiate in Section~\ref{Sec:5.1} the fundamental differences arising with interconnecting remote \textit{classical} processors versus interconnecting remote \textit{quantum} processors. Then, in Section~\ref{Sec:5.2} we introduce the marvels of quantum teleportation, which represents the underlying communication functionality enabling remote quantum operations. Stemming from this, in Section~\ref{Sec:5.3} we discuss the two possible strategies -- namely, \textit{telegate} and \textit{teledata} -- for implementing quantum gates between remote qubits. Then, in Section~\ref{Sec:5.4} we present a key strategy -- referred to as \textit{entanglement swapping} -- for virtually-augmenting the connectivity among different quantum processors. And finally, in Section~\ref{Sec:5.5}, we discuss the open problems arising with interconnecting remote quantum processors.

\subsection{Quantum Internet}
\label{Sec:5.1}

According to the on-going IETF RFC draft on Quantum Internet architectures~\cite{KozWehVan-22}, the Quantum Internet can be defined as an interconnection of heterogeneous\footnote{With heterogeneity arising, as instance, since different networks may be owned by different organizations and/or based on different quantum hardware technologies.}  quantum networks, able to exchange qubits and to generate and share entangled states among themselves. 
Hence, the Quantum Internet services ground on the manipulation and transmission of qubits as well as on the distribution of entangled states. This, in turn, imposes several challenges with no-counterpart in classical network and that cannot be solved through existing classical protocols.

As an example, Internet (and classical networks in general) extensively relies on the possibility of freely duplicating information. But this basic assumption does not hold when it comes to the Quantum Internet~\cite{IllCalMan-22,CacIllKou-22} accordingly to the \textit{no-cloning} theorem\footnote{Distributed quantum computing represents the perfect archetype for scenarios where an unknown qubit must be transmitted. Indeed, let us consider distributed computing between two remote quantum processors, where one processor executes a quantum algorithm -- more precisely, some portion or task of a quantum algorithm -- and the other processor must further process the result of the task. The partial result of the processing at the first node is indeed an unknown quantum state -- otherwise, if known in advance, the whole processing would have been pointless -- and any attempt to measure it would irreversibly alter the partial result, hence wasting the computation performed so far.}. Furthermore, according to the \textit{measurement postulate}, even the simple action of measuring a qubit -- i.e., reading the quantum information stored within -- irreversibly alters its quantum properties, such as superposition and entanglement. 

The above peculiarities of quantum mechanics have deep implications on the design of quantum communication techniques for the Quantum Internet~\cite{CacCalTaf-20}.
To further elaborate on the above statement, let us clarify that it is possible to map a qubit into a photon degree of freedom by directly transmitting this qubit to a remote node via a fiber link or free space. However, if the traveling photon is lost due to attenuation or it is corrupted by \textit{decoherence}\footnote{Any quantum system inevitably interacts with the environment and it is afflicted by decoherence, a phenomenon that irreversibly scrambles the quantum state and therefore its inner information~\cite{CacCal-19}. This kind of quantum noise affects every quantum operation, from qubit processing through qubit storing to qubit transmission, and it causes an irreversible loss of the quantum information as time passes.}, the associated quantum information cannot be recovered via a measuring process or by re-transmitting a copy of the original information. As a consequence, the techniques mitigating the imperfections imposed on the qubits cannot be directly borrowed from classical communications~\cite{CacCalVan-20}.

Thankfully, quantum entanglement~\cite{HorHorHor-09} can be exploited as a communication resource to face with the aforementioned challenges. Indeed, entanglement enables a communication technique, known as \textit{quantum teleportation}, for transmitting an unknown qubit without the physical transfer of the particle storing the qubit, as described in the following.

\subsection{Quantum Teleportation}
\label{Sec:5.2}

As introduced in Section~\ref{Sec:3}, whenever two qubits are entangled, they exist in a shared state, such that any action on a qubit affects instantaneously the other qubit as well, regardless of the distance~\cite{IllCalMan-22}.

This unconventional correlation is exploited by the so-called \textit{quantum teleportation process}~\cite{CacCalVan-20}, which enables the possibility of ``transmitting'' -- namely, \textit{teleporting} -- an unknown qubit without the physical transfer of the particle storing the qubit.

\begin{wraptable}[14]{r}{0.55\linewidth}
    \centering
    \begin{tabular}{c|c}
        \toprule
        \textbf{Measurement Output} & \textbf{Decoding operation} \\
        \toprule
        00 & \texttt{I}\\
        \midrule
        01 &\texttt{X}\\
        \midrule
        10 &\texttt{Z}\\
        \midrule
        11 &\texttt{X} followed by \texttt{Z}\\
        \bottomrule
    \end{tabular}
    \caption{Quantum teleportation: post processing operations to be performed at the destination for recovering the original quantum state. Measurement output aligned with right-most digit representing the outcome of the entangled qubit measure in Figure~\ref{Fig:12}, and gates \texttt{X} and \texttt{Z} -- corresponding to a bit- and a phase-flip, respectively -- detailed in Table~\ref{Tab:01}.}
    \label{Tab:02}
	\hrulefill
\end{wraptable}

More into details, quantum teleportation requires:
\begin{itemize}
    \item[i)] an EPR pair, namely a pair of maximally entangled qubits such as the Bell state in \eqref{Eq:2}, with one qubit of the pair distributed at the source node and the other qubit distributed at the destination;
    \item[ii)] local quantum operations both at the source and at the destination;
    \item[iii)] the transmission of two classical bits from the source to the destination.
\end{itemize}

The circuital representation of the quantum teleportation process is illustrated in Figure~\ref{Fig:12}. More into details, the source performs a pre-processing, namely, a \textit{Bell State Measurement} (BSM) on both the unknown qubit encoding the information, say $\ket{\psi}$, to be transmitted and the entangled qubit. As represented in the gray box in the figure, the BSM consists of a \texttt{CNOT} gate -- with the information qubit acting as control and the entangled qubit acting as target -- followed by an Hadamard gate on the information qubit and, finally, a measurement of both the qubits. Then, the source transmits -- though classical communications -- two classical bits encoding the measurement outcomes of the BSM. Remarkably, after the BSM, the source quantum state has been already teleported at the destination. Nevertheless, the teleported state may have been undergone a phase and/or a bit-flip, with each flip event occurring individually with a probability equal to $0.25$. Luckily, the measurement of the two qubits at the source allows the destination  -- once the measurement outcomes have been received through a classical communication channel -- to determine whether these flip events occurred. Hence, the destination performs a post-processing to reconstruct the original state $\ket{\psi}$, as detailed in Table~\ref{Tab:02}.

In conclusion, by pre-sharing a maximally-entangled pair of qubits\footnote{We may observe that direct transmission of qubits is still needed to distribute entangled states among the network nodes. However and as deeply clarified in~\cite{IllCalMan-22}, differently from unknown qubits, entangled states can be repeatedly prepared for facing with losses and/or noise corruptions.}, two nodes can reliably exchange quantum information through the teleportation process~\cite{UnnMar-20}, which represents the underlying communication functionality enabling remote quantum operations, as further elaborated in the following subsection. 

\subsection{Teledata vs Telegate}
\label{Sec:5.3}

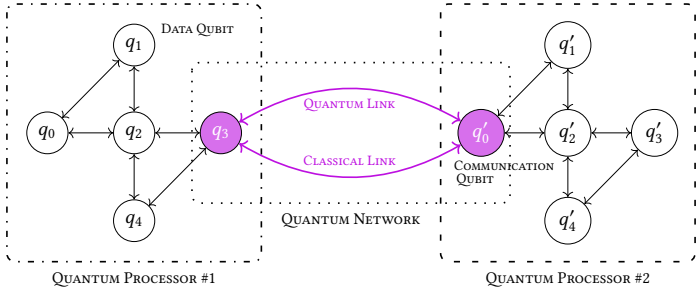
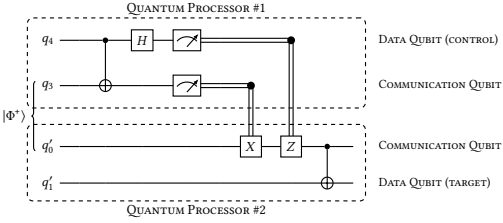
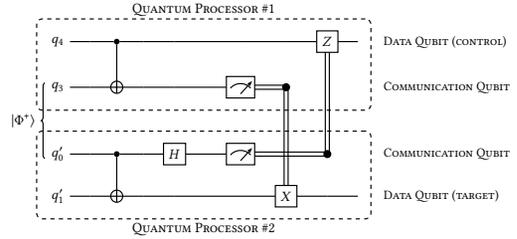
\begin{figure*}
	\centering
        \begin{minipage}[c]{1\linewidth}
		\centering
        \begin{adjustbox}{width=.66\columnwidth}
            \begin{tikzpicture}[->]
                \definecolor{darkblue}{HTML}{1f4e79}
                \definecolor{violet}{RGB}{189,16,224}
                \node [circle, draw] (0) at (0,0) {$q_{0}$};
                \node [circle, draw] (1) at (1.5,1.5) {$q_{1}$};
                \node at (2.6,1.75) {\textsc{\scriptsize Data Qubit}};
                \node [circle, draw] (2) at (1.5,0) {$q_{2}$};
                \node [circle, draw, fill=violet, fill opacity=0.6, text=white, text opacity=1] (3) at (3,0) {$q_{3}$};
                \node [circle, draw] (4) at (1.5,-1.5) {$q_{4}$};
                \draw[<->] (1) -- (0);
                \draw[<->] (2) -- (0);
                \draw[<->] (2) -- (1);
                \draw[<->] (3) -- (2);
                \draw[<->] (4) -- (2);
                \draw[<->] (4) -- (3);
                \draw[thick,loosely dashdotted, rounded corners] (-0.7,2.2) rectangle (3.7,-2.2);
                \node at (1.5,-2.5) {\textsc{\footnotesize Quantum Processor \#1}};
                \node [circle, draw, fill=violet, fill opacity=0.6, text opacity=1, text=white] (10) at (7.5,0) {$q'_{0}$};
                \node at (7.9,-0.55) {\textsc{\scriptsize Communication}};
                \node at (7.4,-0.8) {\textsc{\scriptsize Qubit}};
                \node [circle, draw] (11) at (9,1.5) {$q'_{1}$};
                \node [circle, draw] (12) at (9,0) {$q'_{2}$};
                \node [circle, draw] (13) at (10.5,0) {$q'_{3}$};
                \node [circle, draw] (14) at (9,-1.5) {$q'_{4}$};
                \draw[<->] (11) -- (10);
                \draw[<->] (12) -- (10);
                \draw[<->] (12) -- (11);
                \draw[<->] (13) -- (12);
                \draw[<->] (14) -- (12);
                \draw[<->] (14) -- (13);
                \draw[thick,loosely dashed, rounded corners] (6.8,2.2) rectangle (11.2,-2.2);
                \node at (9,-2.5) {\textsc{\footnotesize Quantum Processor \#2}};
                \draw[thick,loosely dotted, rounded corners] (2.5,1.2) rectangle (8,-1.2);
                \node at (5.25,-1.5) {\textsc{\footnotesize Quantum Network}};
                \draw[thick,<->,bend left,violet] (3) edge node[below,yshift=-1pt] {\textsc{\scriptsize Quantum Link}} (10);
                \draw[thick,<->,bend right,violet] (3) edge 
                node[above,yshift=2pt] {\textsc{\scriptsize Classical Link}}(10);            \end{tikzpicture}
    	\end{adjustbox}
	    \subcaption{Two IBM Yorktown quantum processors given in Figure~\ref{Fig:07} interconnected through a quantum network, composed by a classical and a quantum link. The classical link is used to transmit classical information, whereas the quantum link is needed for distributing entangled states between the two remote processors to enable communication functionalities. Indeed, at least one physical qubit at each processor must be reserved for entanglement generation. This kind of qubits -- dark-blue-colored in the figure -- are the \textit{communication qubits} to distinguish them from the \textit{data qubits} -- white-colored in the figure.\\
	    }
        \label{Fig:13-a}
    \end{minipage}
    \vspace{6pt}
    \begin{minipage}[c]{.49\linewidth}
		\centering
		\begin{adjustbox}{width=1\linewidth}
		    \begin{tikzcd}
			    & \lstick{$q_4$}\gategroup[wires=2,steps=9,style={dashed,rounded corners,inner xsep=20pt,inner ysep=5pt}, background, label style={label position=above, yshift=-0.0cm,}, background]{\sc Quantum Processor \#1} & \qw & \ctrl{1} & \gate{H} & \meter{} & \cw & \cw & \cwbend{3} & & & \rstick{\sc \small Data Qubit (control)}\\
			    \lstick[wires=3]{$\ket{\Phi^\texttt{+}}$} & \lstick{$q_3$} & \qw & \targ{} & \qw & \meter{} & \cw & \cwbend{2} & & & & \rstick{\sc \small	Communication Qubit}\\
			    & \\
			    & \lstick{$q'_0$}\gategroup[wires=2,steps=9,style={dashed,rounded corners,inner xsep=20pt,inner ysep=5pt},background,label style={label position=below,yshift=-0.44cm,}, background]{\sc Quantum Processor \#2} & \qw & \qw & \qw & \qw & \qw & \gate{X} & \gate{Z} & \ctrl{1} & \qw & \rstick{\sc \small Communication Qubit}\\
			    & \lstick{$q'_1$} & \qw & \qw & \qw & \qw & \qw & \qw & \qw & \targ{} & \qw & \rstick{\sc \small Data Qubit (target)}
		    \end{tikzcd}
        \end{adjustbox}
	    \subcaption{\texttt{TeleData}. To perform a \texttt{TeleData} between remote processors -- say to move the quantum state $\ket{\varphi}$ stored by data qubit $q_4$ in Figure~\ref{Fig:13-a} to communication qubit $q_0^{'}$ -- a Bell state such as $\ket{\Phi^{\texttt{+}}}$ must be distributed through the quantum link so that each pair member is stored within the \textit{communication qubit} at each processor. Once $\ket{\varphi}$ is teleported at $q_0^{'}$ (with local quantum operations and classical transmission), the remote operation -- for instance, a \texttt{CNOT} with $\ket{\varphi}$ as control and the state stored by qubit $q_1^{'}$ as target  as shown with the last \texttt{CNOT} in the figure -- can be executed through local operations.}
		\label{Fig:13-b}
	\end{minipage}    
    \hfil
    \begin{minipage}[c]{.49\linewidth}
		\centering
		\begin{adjustbox}{width=1\linewidth}
		    \begin{tikzcd}
			    & \lstick{$q_4$}\gategroup[wires=2,steps=9,style={dashed,rounded corners,inner xsep=20pt,inner ysep=5pt}, background, label style={label position=above, yshift=-0.0cm,}, background]{\sc Quantum Processor \#1} & \qw & \ctrl{1} & \qw & \qw & \qw & \qw & \qw & \gate{Z} & \qw & \rstick{\sc \small Data Qubit (control)}\\
			    \lstick[wires=3]{$\ket{\Phi^\texttt{+}}$} & \lstick{$q_3$} & \qw & \targ{} & \qw & \qw & \qw & \meter{} & \cwbend{3} & & &  \rstick{\sc \small	Communication Qubit}\\
			    & \\
			    & \lstick{$q'_0$}\gategroup[wires=2,steps=9,style={dashed,rounded corners,inner xsep=20pt,inner ysep=5pt},background,label style={label position=below,yshift=-0.44cm,}, background]{\sc Quantum Processor \#2} & \qw & \ctrl{1} & \qw & \gate{H} & \qw & \meter{} & \cw & \cwbend{-3} & & \rstick{\sc \small Communication Qubit}\\
			    & \lstick{$q'_1$} & \qw & \targ{} & \qw & \qw & \qw & \qw & \gate{X} & \qw & \qw & \rstick{\sc \small Data Qubit (target)}
		    \end{tikzcd}
        \end{adjustbox}
	    \subcaption{\texttt{TeleGate}. A \texttt{TeleGate} enables a direct gate between remote physical qubits stored at different processors without the need of quantum state teleportation, as long as a Bell state such as $\ket{\Phi^{\texttt{+}}}$ is distributed through the quantum link. As instance, a remote \texttt{CNOT} between $q_4$ and $q'_1$ in Figure~\ref{Fig:13-a} can be implemented with two local \texttt{CNOT}s between the \textit{data} and the \textit{communication qubit} at each processor, followed by a conditional gate on the data qubit depending on the measurement of the remote communication qubit.}
		\label{Fig:13-c}
	\end{minipage}
    \caption{Remote operations through either \texttt{TeleData} or \texttt{TeleGate}. Figure~\ref{Fig:13-a} shows the network topology along with the processors coupling maps, whereas Figures~\ref{Fig:13-b} and \ref{Fig:13-c} illustrate the quantum circuit detailing the classical (2 bits) and the quantum (the Bell state) resources needed to execute a \texttt{TeleData} and a \texttt{TeleGate}, respectively.  Figure reproduced from~\cite{FerCacAmo-21}.}
    \label{Fig:13}
    \hrulefill
\end{figure*}

In distributed quantum computing, quantum teleportation constitutes the fundamental communication primitive underlying the communication paradigms known as \texttt{TeleData} and \texttt{TeleGate}~\cite{VanNemMun-06}, which generalize the concept of moving quantum states among remote devices.

To provide concrete examples of the \texttt{TeleData} and \texttt{TeleGate} concepts, we must classify qubits within a QPU either as \textit{communication qubits} or as \textit{data qubits}~\cite{CuoCalCac-20}. Specifically, within each quantum processor, a subset of qubits is reserved for inter-processor communications and we refer to these qubits as communication qubits~\cite{KozWehVan-22}, to distinguish them from the remaining qubits within the device devoted to processing/storage, which we refer as data qubits. 

More into detail, entanglement distribution among network nodes requires that at least one qubit at each processor, referred to as communication qubit, must be reserved for the generation of the entangled state~\cite{KozWehVan-22}. Clearly, the more communication qubits are available within a network node, the more entanglement resource is available at that node, with an obvious positive effect on entanglement rate achievable by that node~\cite{IllCalMan-22}. But the more communication qubits are available, the less data qubits are available for quantum computing. 

As an example, consider two quantum processors interconnected via a quantum network as depicted in Figure~\ref{Fig:13}. Qubits $q_3$ and $q_0^{'}$ are communication qubits and any interaction between the two remote processors is carried out by exploiting them via either a \texttt{TeleData} or a \texttt{TeleGate} process.

With a \texttt{TeleData}, quantum information stored within a data qubit at the first processor, say $\ket{\varphi}$ in $q_4$ in Figure~\ref{Fig:13-a}, is teleported into a communication qubit of the second processor, say $q_0^{'}$ in the same figure. Once the quantum state $\ket{\varphi}$ is teleported in $q_0^{'}$, any remote operation -- originally involving $q_4$ and some data qubits at the second processor -- can be now implemented through local operations as shown with the last \texttt{CNOT} in Figure~\ref{Fig:13-b}. It must be noted, though, that whether the teleported quantum state should subsequently interact with data qubits at the first processor, a new teleportation process must be performed for teleporting the quantum state back to the first processor.
\texttt{TeleData} is not the only available option for implementing remote operations. In fact, a \texttt{TeleGate} enables to execute a direct gate between qubits belonging to remote processors by exploiting again entanglement. As instance, a remote \texttt{CNOT} with data qubit $q_4$ and $q_0^1$ in Figure~\ref{Fig:13-a} acting as control and target, respectively, can be implemented with local \texttt{CNOT}s at each quantum processor, as shown with the quantum circuit in Figure~\ref{Fig:13-c}.

\subsection{Physical vs Virtual Quantum Links}
\label{Sec:5.4}

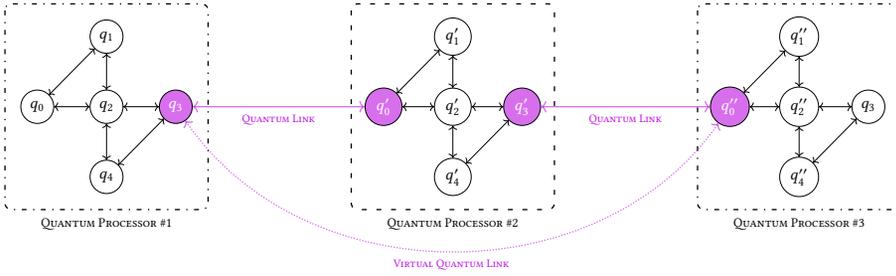
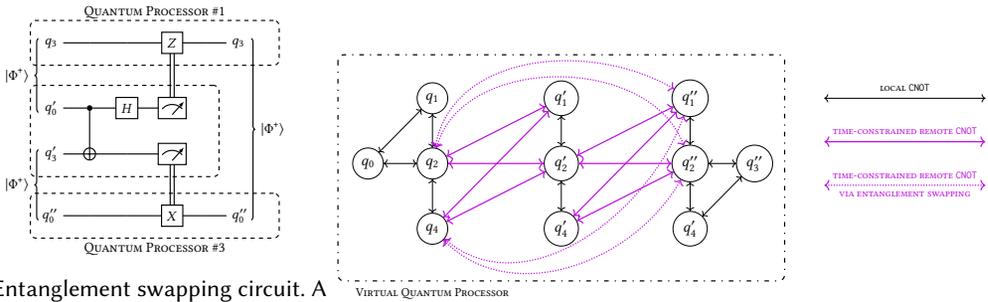
\begin{figure*}[t]
    \begin{minipage}[c]{1\linewidth}
		\centering
        \begin{adjustbox}{width=1\linewidth}
            \begin{tikzpicture}[->]
                \node (sinistra) at (-1.85,0) {};
                \node (destra) at (20.55,0) {};
                \definecolor{darkblue}{HTML}{1f4e79}
                \definecolor{violet}{RGB}{189,16,224}
                \node [circle, draw] (0) at (0,0) {$q_{0}$};
                \node [circle, draw] (1) at (1.5,1.5) {$q_{1}$};
                \node [circle, draw] (2) at (1.5,0) {$q_{2}$};
                \node [circle, draw, fill=violet, text=white, text opacity=1, fill opacity=0.6] (3) at (3,0) {$q_{3}$};
                \node [circle, draw] (4) at (1.5,-1.5) {$q_{4}$};
                \draw[<->] (1) -- (0);
                \draw[<->] (2) -- (0);
                \draw[<->] (2) -- (1);
                \draw[<->] (3) -- (2);
                \draw[<->] (4) -- (2);
                \draw[<->] (4) -- (3);
                \draw[thick,loosely dashdotted, rounded corners] (-0.7,2.2) rectangle (3.7,-2.2);
                \node at (1.5,-2.5) {\textsc{\footnotesize Quantum Processor \#1}};
                \node [circle, draw, fill=violet, text=white, text opacity=1, fill opacity=0.6] (10) at (7.5,0) {$q'_{0}$};
                \node [circle, draw] (11) at (9,1.5) {$q'_{1}$};
                \node [circle, draw] (12) at (9,0) {$q'_{2}$};
                \node [circle, draw, fill=violet, text=white, text opacity=1, fill opacity=0.6] (13) at (10.5,0) {$q'_{3}$};
                \node [circle, draw] (14) at (9,-1.5) {$q'_{4}$};
                \draw[<->] (11) -- (10);
                \draw[<->] (12) -- (10);
                \draw[<->] (12) -- (11);
                \draw[<->] (13) -- (12);
                \draw[<->] (14) -- (12);
                \draw[<->] (14) -- (13);
                \draw[thick,loosely dashed, rounded corners] (6.8,2.2) rectangle (11.2,-2.2);
                \node at (9,-2.5) {\textsc{\footnotesize Quantum Processor \#2}};
                \node [circle, draw, fill=violet, text=white, text opacity=1, fill opacity=0.6] (20) at (15,0) {$q''_{0}$};
                \node [circle, draw] (21) at (16.5,1.5) {$q''_{1}$};
                \node [circle, draw] (22) at (16.5,0) {$q''_{2}$};
                \node [circle, draw] (23) at (18,0) {$q_{3}$};
                \node [circle, draw] (24) at (16.5,-1.5) {$q''_{4}$};
                \draw[<->] (21) -- (20);
                \draw[<->] (22) -- (20);
                \draw[<->] (22) -- (21);
                \draw[<->] (23) -- (22);
                \draw[<->] (24) -- (22);
                \draw[<->] (24) -- (23);
                \draw[thick,loosely dashdotted, rounded corners] (14.3,2.2) rectangle (18.7,-2.2);
                \node at (16.5,-2.5) {\textsc{\footnotesize Quantum Processor \#3}};
                \draw[thick,<->,violet, draw opacity=0.6] (3) edge node[below,yshift=-1pt] {\textsc{\scriptsize Quantum Link}} (10);
                %
                \draw[thick,<->,violet, draw opacity=0.6] (13) edge node[below,yshift=-1pt] {\textsc{\scriptsize Quantum Link}} (20);
                %
                \draw[thick,densely dotted,<->,bend right=55,violet, draw opacity=0.6] (3) edge node[below,yshift=-1pt] {\textsc{\scriptsize Virtual Quantum Link}} (20);
                \end{tikzpicture}
    	\end{adjustbox}
        \subcaption{By swapping the entanglement at an intermediate node -- namely, quantum processor \#2 -- it is possible to distribute a Bell state between remote processors -- namely, processors \#1 and \#3 -- even if they are not directly connected through a quantum link. Hence, entanglement swapping enhances the quantum processors connectivity through \textit{virtual quantum links}.\\
        }
        \label{Fig:14-a}
    \end{minipage}
    \begin{minipage}[c]{.35\linewidth}
		\centering
		\begin{adjustbox}{width=.8\linewidth}
		    \begin{tikzcd}
			    \lstick[wires=3]{$\ket{\Phi^\texttt{+}}$} & \lstick{$q_3$}\gategroup[wires=1,steps=7,style={dashed,rounded corners,inner xsep=20pt,inner ysep=5pt}, background, label style={label position=above, yshift=-0.0cm,}, background]{\sc Quantum Processor \#1} & \qw & \qw & \gate{Z} & \qw & \qw \rstick{$q_3$} & \rstick[wires=6]{$\ket{\Phi^\texttt{+}}$} \\
			    & \\
			    & \lstick{$q'_0$}\gategroup[wires=2,steps=4,style={dashed,rounded corners,inner xsep=20pt,inner ysep=5pt}, background, label style={label position=above, yshift=-0.0cm,}, background]{} & \ctrl{1} & \gate{H} & \meter{} \vcw{-2}\\
			    \lstick[wires=3]{$\ket{\Phi^\texttt{+}}$} & \lstick{$q'_3$} & \targ{} & \qw & \meter{} \vcw{2} \\
			    & \\
			    & \lstick{$q''_0$}\gategroup[wires=1,steps=7,style={dashed,rounded corners,inner xsep=20pt,inner ysep=5pt},background,label style={label position=below,yshift=-0.44cm,}, background]{\sc Quantum Processor \#3} & \qw & \qw & \gate{X} & \qw & \qw \rstick{$q''_0$} &
		    \end{tikzcd}
        \end{adjustbox}
        \subcaption{Entanglement swapping circuit. A Bell state can be distributed between remote processors by swapping the entanglement at an intermediate node -- as instance, processors \#2 in Figure~\ref{Fig:14-a} -- through local processing and classical communication. }
        \label{Fig:14-b}
    \end{minipage}
    \hfil
    \begin{minipage}[c]{.63\linewidth}
		\centering
		\begin{adjustbox}{width=1\linewidth}
            \begin{tikzpicture}[->]
                \definecolor{darkblue}{HTML}{1f4e79}
                \definecolor{violet}{RGB}{189,16,224}
                \node [circle, draw] (0) at (0,0) {$q_{0}$};
                \node [circle, draw] (1) at (1.5,1.5) {$q_{1}$};
                \node [circle, draw] (2) at (1.5,0) {$q_{2}$};
                \node [circle, draw] (4) at (1.5,-1.5) {$q_{4}$};
                \draw[<->] (1) -- (0);
                \draw[<->] (2) -- (0);
                \draw[<->] (2) -- (1);
                \draw[<->] (4) -- (2);
                \draw[thick,loosely dashdotted, rounded corners] (-0.7,2.5) rectangle (9.7,-2.7);
                \node at (1.5,-3) {\textsc{\footnotesize Virtual Quantum Processor}};
                \node [circle, draw] (11) at (4.5,1.5) {$q'_{1}$};
                \node [circle, draw] (12) at (4.5,0) {$q'_{2}$};
                \node [circle, draw] (14) at (4.5,-1.5) {$q'_{4}$};
                \draw[<->] (12) -- (11);
                \draw[<->] (14) -- (12);
                \draw[thick,<->,violet] (2) edge (11);
                \draw[thick,<->,violet] (2) edge (12);
                \draw[thick,<->,violet] (4) edge (11);
                \draw[thick,<->,violet] (4) edge (12);
                \node [circle, draw] (21) at (7.5,1.5) {$q''_{1}$};
                \node [circle, draw] (22) at (7.5,0) {$q''_{2}$};
                \node [circle, draw] (23) at (9,0) {$q''_{3}$};
                \node [circle, draw] (24) at (7.5,-1.5) {$q'_{4}$};
                \draw[<->] (22) -- (21);
                \draw[<->] (23) -- (22);
                \draw[<->] (24) -- (22);
                \draw[<->] (24) -- (23);
                \draw[thick,<->,violet] (12) edge (21);
                \draw[thick,<->,violet] (12) edge (22);
                \draw[thick,<->,violet] (14) edge (21);
                \draw[thick,<->,violet] (14) edge (22);
                \draw[thick,densely dotted,<->,violet] (2) edge [out=75,in=110] (22);
                \draw[thick,densely dotted,<->,violet] (2) edge [out=80,in=150] (21);
                \draw[thick,densely dotted,<->,violet] (4) edge [out=-40,in=-110] (21);
                \draw[thick,densely dotted,<->,violet] (4) edge [out=-45,in=-120] (22);
                %
                \node (l1) at (10.5,1.5) {};
                \node (l2) at (14.5,1.5) {};
                \draw[<->] (l1) edge node[above,yshift=1pt] {\textsc{\scriptsize local \texttt{CNOT}}}  (l2);
                \node (l3) at (10.5,0.5) {};
                \node (l4) at (14.5,0.5) {};
                \draw[thick,<->,violet] (l3) edge node[above,yshift=1pt] {\textsc{\scriptsize time-constrained remote \texttt{CNOT}}}  (l4);
                \node (l5) at (10.5,-0.5) {};
                \node (l6) at (14.5,-0.5) {};
                \draw[thick,densely dotted,<->,violet] (l5) edge node[above,yshift=1pt] {\textsc{\scriptsize time-constrained remote \texttt{CNOT}}} node[below] {\textsc{\scriptsize via entanglement swapping}} (l6);
            \end{tikzpicture}
        \end{adjustbox}
	    \subcaption{Dynamic coupling map for the network topology shown in Figure~\ref{Fig:14-a}. The solid blue lines denote remote \texttt{CNOT}s between adjacent processors, whereas the dotted blue lines denote remote \texttt{CNOT}s between distant processors achievable via entanglement swapping.} 
		\label{Fig:14-c}
	\end{minipage}
    \caption{Augmented connectivity. Entanglement swapping increases the connectivity between physical qubits, with a number of possible remote \texttt{CNOT}s that scales at least linearly with the number of processors. Figure reproduced from~\cite{FerCacAmo-21}.}
    \label{Fig:14}
    \hrulefill
\end{figure*}

From Figure~\ref{Fig:13}, one might assume that distributed quantum computing requires a fully-connected network topology -- namely, that each quantum processor must be directly inter-connected with all the other processors -- as a consequence of the unconventional characteristics of quantum information. In other words, one might assume the connectivity between quantum processors strongly dependent on the availability of a direct entanglement generation and distribution architecture. As a matter of fact, the very opposite is true. Distributed quantum computing can exploit a strategy -- called \textit{entanglement swapping}~\cite{VanDev-16} and summarized in Fig~\ref{Fig:14} -- to implement a remote \texttt{CNOT} between qubits stored at remote processors, even if the processors are not directly connected through a quantum link.

In a nutshell, to distribute a Bell state between remote processors -- say quantum processor \#1 and \#3 in Figure~\ref{Fig:14-a} -- two Bell states must be first distributed through the quantum links so that one Bell state is shared between the first processor and an intermediate node and another Bell state is shared by the same intermediate node and the second processor.  Then, by performing a BSM on the communication qubits at the intermediate node -- i.e., qubits $q'_0$ and $q'_3$ in Figure~\ref{Fig:14-b} -- a Bell state is obtained at the remote communication qubits $q_3$ and $q''_0$ in Figure~\ref{Fig:14-b} -- by applying some local processing at the remote nodes depending on the (classical) output of the Bell state measurement.

From the above, it becomes clear that entanglement swapping significantly increases the connectivity within the virtual quantum processor. And the higher is the number of available quantum processors, the higher is the number of possible interactions. Indeed, the number of additional interactions via entanglement swapping scales linearly with the number of available processors when only two communication qubits are available at each intermediate processor. If this constraint is relaxed, the number of additional interactions via entanglement swapping scales more than linearly.

\subsection{Open Issues and Research Directions}
\label{Sec:5.5}

Stemming from the discussion carried out in the previous subsections, here we summarize some fundamental open issues and research directions towards the interconnection of different quantum processors for enabling distributed quantum computing.

First, in Section~\ref{Sec:5.3} we introduced the two possible strategies -- \texttt{TeleGate} and \texttt{TeleData} -- for implementing quantum gates between remote qubits. From a communication resource perspective, \texttt{TeleData} and \texttt{TeleGate} consume the same amount of quantum and classical resources, namely one EPR pair and the transmission of two classical bits. Yet the overall performance of the two strategies depends on a range of factors, including i) the \textit{pattern} of remote operations exhibited by the quantum circuit to be executed, ii) the characteristics of the network interconnecting the remote quantum processors, and iii) the ratio between data and communication qubits~\cite{VanNemMun-06,IllCalMan-22}.

With reference to the latter factor, a fundamental trade-off arises~\cite{FerCacAmo-21}. Specifically, each remote operation -- regardless whether it is implemented with a \texttt{TeleData} or a \texttt{TeleGate} -- consumes the entangled resource. Consequently, a new Bell state must be distributed between the remote processors before another remote operation could be executed. Hence, the more communication qubits are available within each processor, the more remote operation can be executed in parallel, reducing the communication overhead induced by the distributed computation. But the more communication qubits, the less data qubits are available for computing in each processor.

Accordingly to the above reasoning, the selection of the set of communication qubits is a crucial task for distributed quantum computing, with profound effects on the overall performance of the distributed computation.  As a matter of fact, the fundamental role played by communication qubits is further stressed by the augmented connectivity enabled by entanglement swapping, discussed in Section~\ref{Sec:5.4}. Indeed, it must be acknowledged that such an augmented connectivity does not come for free. Entanglement swapping consumes the Bell states at each intermediate processor. And the longer is the path between the two processors involved in the remote operation, the higher is the number of consumed Bell states. Clearly, the more Bell states are devoted to entanglement swapping, the less Bell states are available for implementing remote operations between neighbor quantum processors. Hence, a trade-off between ``augmented connectivity'' and ``EPR cost'' arises with entanglement swapping~\cite{FerCacAmo-21}, and the impact of this trade-off on the overall performance of distributed quantum computing must be carefully accounted for.

Another fundamental issue arising with networking remote quantum processors is represented by noise and imperfections affecting the \textit{quality} of the distributed Bell states. Clearly the noisier is the distributed Bell state, the noisier is the overall distributed quantum computation. Luckily, a well-known technique for counteracting the noise impairments affecting the entanglement generation/distribution process is constituted by \textit{entanglement distillation} (also known as \textit{entanglement purification})~\cite{BenBraPop-96,BenDivSmo-96, CirEkeHue-99,DurBri-07, RuaDaiWin-18,SchElkDoh-18,RuaKirBro-21}. Accordingly, as long as the ``quality'' of the noisy entanglement exceeds a certain threshold, it is possible to purify multiple imperfect Bell states into a single “almost-maximally entangled” pair, albeit at the price of consuming multiple noisy entangled states within the process. From the above, it follows that one of two orthogonal resources must be exploited for implementing the distillation process, namely, time or space. More into details, time-expensive distillation requires multiple rounds of entanglement generation and distribution, with each round involving few\footnote{At least two communication qubits at each processor are required.} communication qubits. Conversely, space-expensive distillation can be completed with few rounds, but with each round involving several communication qubits. Hence, there exists a fundamental trade-off between i) quality of the overall computation, ii) delay induced by entanglement distillation, and iii) communication qubits reserved for distilling a high-quality Bell state.

\section{Quantum Compiling}
\label{Sec:6}

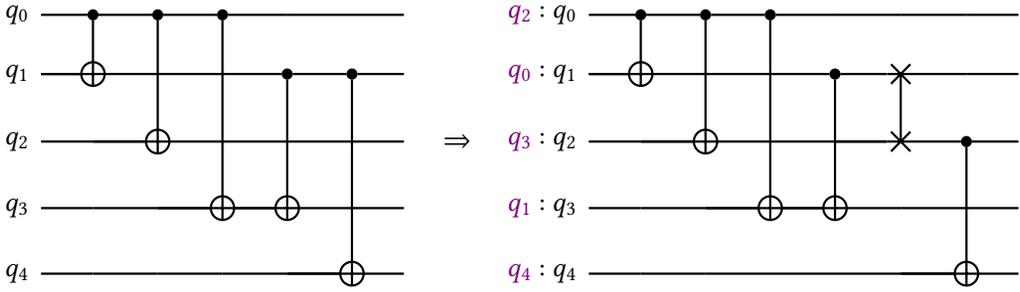
\begin{figure}[t]
    \centering
    \begin{adjustbox}{width=1\textwidth}
		\definecolor{darkblue}{HTML}{1f4e79}
		\begin{tikzcd}
		\lstick{$q_0$} & \ctrl{1} & \ctrl{2} & \ctrl{3} & \qw & \qw & \qw &&&& \lstick{\textcolor{violet}{$q_2$} $: q_0$} & \ctrl{1} & \ctrl{2} & \ctrl{3} & \qw & \qw & \qw & \qw\\
		\lstick{$q_1$} & \targ & \qw & \qw & \qw & \ctrl{2} & \ctrl{3} & \qw &&&& \lstick{\textcolor{violet}{$q_0$} $: q_1$} & \targ & \qw & \qw & \qw & \ctrl{2} & \swap{1} & \qw & \qw\\
		\lstick{$q_2$} & \qw & \targ & \qw  & \qw & \qw & \qw & \qw & \Rightarrow &&& \lstick{\textcolor{violet}{$q_3$} $: q_2$} & \qw & \targ & \qw  & \qw & \qw & \targX & \qw & \ctrl{2} & \qw\\
		\lstick{$q_3$} & \qw & \qw & \targ & \qw & \targ & \qw & \qw & \qw &&&& \lstick{\textcolor{violet}{$q_1$} $: q_3$} & \qw & \qw & \targ & \qw & \targ & \qw & \qw & \qw & \qw\\
		\lstick{$q_4$} & \qw & \qw & \qw & \qw & \targ & \qw & \qw &&&& \lstick{\textcolor{violet}{$q_4$} $: q_4$} & \qw & \qw & \qw & \qw & \qw & \targ & \qw & \qw
		\end{tikzcd}
	\end{adjustbox}
    \caption{Pictorial representation of quantum compiling. The circuit on the left is translated into the circuit on the right, in order to cope with the coupling map provided in Figure~\ref{Fig:07}. Within the rightest figure, the $q_i$ with purple font denotes the physical qubits assigned to the logical qubits $q_j$ with black font. The SWAP gate-- represented by two $\times$ symbols interconnected by a vertical line -- introduced between logical qubits $q_1$ and $q_2$ swaps their quantum states, so that the last \texttt{CNOT} gate can be applied between two neighbor physical qubits.}
    \label{Fig:15}
    \hrulefill
\end{figure}

As mentioned in Section~\ref{Sec:3.3}, quantum compilation means translating an input quantum circuit into the most efficient equivalent of itself, considering the characteristics of the device(s) that will execute the computation and minimizing the number of required multi-qubit gates. An example of quantum compilation is provided with Figure~\ref{Fig:15}, where the original quantum circuit is translated into the compiled one to account for the coupling characteristics of IBM Yorktown quantum processors, shown in Figure~\ref{Fig:07}. Clearly, as long as the hardware provides a universal set of operations, there exists a feasible transformation.

Compilers are well-established in NISQ architectures, because of their role as intermediary between the user and the hardware. Specifically, in designing a quantum algorithm using the quantum circuit formalism introduced in Section~\ref{Sec:3.2}, the designer is generally focused on expressing the computation required by the algorithm with a circuit that minimizes the number of utilized qubits and gates, regardless from the particulars of the quantum hardware that will execute the circuit. This \textit{abstract} circuit is then mapped to a circuit to be executed on a specific quantum hardware by means of a suitable compiler. Clearly, introducing such an abstract circuit has two main advantages: i) the user can focus on the logic of the circuit, namely, on the essence of the quantum algorithm, without caring too much about the hardware constraints, and ii) the designed quantum circuit is portable, in theory, to any quantum back-end. 

Intuitively, a circuit transformation may introduce some overhead, in terms of number of operations and noise. In DQC architectures, there is also a non-negligible communication cost, as discussed in Section~\ref{Sec:5}. Therefore, the compiler faces an optimization problem, i.e., finding a feasible transformation while minimizing the overhead. In general, this problem is known to be \texttt{NP}-hard~\cite{Botea2018,Soeken2019}, even for the case of a single processor.

A fundamental issue in quantum compiling is related to qubit connectivity. From the perspective of the quantum algorithm designer, any qubit is assumed to be directly connected with any other qubit. i.e., any two-qubit gate can be placed across any qubit pair. However, even on a single quantum processor as introduced in Section~\ref{Sec:3.3}, the actual connectivity degree is usually low, to mitigate the noise caused by cross-talking phenomena~\cite{ChaZhuYod-20}. \textit{Qubit routing} refers to the task of modifying quantum circuits so that they satisfy the connectivity constraints of a target quantum computer. This involves inserting SWAP gates into the circuit so that the logical gates only ever occur between adjacent physical qubits. Of course, the number of SWAP gates should be minimized, in order keep the circuit depth reasonably small. The problem gets harder when considering distributed quantum processors, where the connectivity degree of the physical qubits can be even lower. 

\begin{figure}[t]
    \centering
    \tikzset{every picture/.style={line width=0.75pt}} 
    \begin{tikzpicture}[x=0.75pt,y=0.75pt,yscale=-2,xscale=2]
        \draw [color={rgb, 255:red, 0; green, 0; blue, 0 }  ,draw opacity=0.7 ][line width=0.75]    (340.1,110) -- (400.48,110) ;
        \draw [color={rgb, 255:red, 0; green, 0; blue, 0 }  ,draw opacity=0.7 ][line width=0.75]    (340,120) -- (366.38,120) ;
        \draw [color={rgb, 255:red, 0; green, 0; blue, 0 }  ,draw opacity=0.7 ][line width=0.75]    (373.63,120) -- (400.38,120) ;
        \draw [color={rgb, 255:red, 0; green, 0; blue, 0 }  ,draw opacity=0.7 ][line width=0.75]    (340,100) -- (400,100) ;
        \draw [color={rgb, 255:red, 0; green, 0; blue, 0 }  ,draw opacity=0.7 ][line width=0.75]    (340,130) -- (400.5,130) ;
        \draw [color={rgb, 255:red, 0; green, 0; blue, 0 }  ,draw opacity=0.7 ]   (346.38,110) -- (353.63,110) ;
        \draw [color={rgb, 255:red, 0; green, 0; blue, 0 }  ,draw opacity=0.7 ]   (350,120) -- (350,133.63) ;
        \draw [shift={(350,120)}, rotate = 90] [color={rgb, 255:red, 0; green, 0; blue, 0 }  ,draw opacity=0.7 ][fill={rgb, 255:red, 0; green, 0; blue, 0 }  ,fill opacity=0.7 ][line width=0.75]      (0, 0) circle [x radius= 1.34, y radius= 1.34]   ;
        \draw  [color={rgb, 255:red, 0; green, 0; blue, 0 }  ,draw opacity=0.7 ] (346.38,130) .. controls (346.38,128) and (348,126.38) .. (350,126.38) .. controls (352,126.38) and (353.63,128) .. (353.63,130) .. controls (353.63,132) and (352,133.63) .. (350,133.63) .. controls (348,133.63) and (346.38,132) .. (346.38,130) -- cycle ;
        \draw [color={rgb, 255:red, 0; green, 0; blue, 0 }  ,draw opacity=0.7 ]   (346.38,130) -- (353.63,130) ;
        \draw [color={rgb, 255:red, 189; green, 16; blue, 224 }  ,draw opacity=0.7 ]   (370,100) -- (370,123.63) ;
        \draw [shift={(370,100)}, rotate = 90] [color={rgb, 255:red, 189; green, 16; blue, 224 }  ,draw opacity=0.7 ][fill={rgb, 255:red, 189; green, 16; blue, 224 }  ,fill opacity=0.7 ][line width=0.75]      (0, 0) circle [x radius= 1.34, y radius= 1.34]   ;
        \draw  [color={rgb, 255:red, 189; green, 16; blue, 224 }  ,draw opacity=0.7 ] (366.38,120) .. controls (366.38,118) and (368,116.38) .. (370,116.38) .. controls (372,116.38) and (373.63,118) .. (373.63,120) .. controls (373.63,122) and (372,123.63) .. (370,123.63) .. controls (368,123.63) and (366.38,122) .. (366.38,120) -- cycle ;
        \draw [color={rgb, 255:red, 189; green, 16; blue, 224 }  ,draw opacity=0.7 ]   (366.38,120) -- (373.63,120) ;
        \draw [color={rgb, 255:red, 0; green, 0; blue, 0 }  ,draw opacity=0.7 ]   (390,100) -- (390,113.63) ;
        \draw [shift={(390,100)}, rotate = 90] [color={rgb, 255:red, 0; green, 0; blue, 0 }  ,draw opacity=0.7 ][fill={rgb, 255:red, 0; green, 0; blue, 0 }  ,fill opacity=0.7 ][line width=0.75]      (0, 0) circle [x radius= 1.34, y radius= 1.34]   ;
        \draw  [color={rgb, 255:red, 0; green, 0; blue, 0 }  ,draw opacity=0.7 ] (386.38,110) .. controls (386.38,108) and (388,106.38) .. (390,106.38) .. controls (392,106.38) and (393.63,108) .. (393.63,110) .. controls (393.63,112) and (392,113.63) .. (390,113.63) .. controls (388,113.63) and (386.38,112) .. (386.38,110) -- cycle ;
        \draw [color={rgb, 255:red, 0; green, 0; blue, 0 }  ,draw opacity=0.7 ]   (386.38,110) -- (393.63,110) ;
        \draw [color={rgb, 255:red, 0; green, 0; blue, 0 }  ,draw opacity=0.7 ]   (349.91,109.96) -- (350.09,96.33) ;
        \draw [shift={(349.91,109.96)}, rotate = 270.74] [color={rgb, 255:red, 0; green, 0; blue, 0 }  ,draw opacity=0.7 ][fill={rgb, 255:red, 0; green, 0; blue, 0 }  ,fill opacity=0.7 ][line width=0.75]      (0, 0) circle [x radius= 1.34, y radius= 1.34]   ;
        \draw  [color={rgb, 255:red, 0; green, 0; blue, 0 }  ,draw opacity=0.7 ] (353.67,100.01) .. controls (353.64,102.01) and (352,103.61) .. (349.99,103.58) .. controls (347.99,103.56) and (346.39,101.91) .. (346.42,99.91) .. controls (346.44,97.91) and (348.09,96.31) .. (350.09,96.33) .. controls (352.09,96.36) and (353.69,98) .. (353.67,100.01) -- cycle ;
        \draw [color={rgb, 255:red, 0; green, 0; blue, 0 }  ,draw opacity=0.7 ]   (353.67,100.01) -- (346.42,99.91) ;
        \draw [color={rgb, 255:red, 0; green, 0; blue, 0 }  ,draw opacity=0.7 ][line width=0.75]    (281,110) -- (319,110) ;
        \draw [color={rgb, 255:red, 0; green, 0; blue, 0 }  ,draw opacity=0.7 ][line width=0.75]    (250.44,110) -- (278.81,110) ;
        \draw [line width=0.75]    (321.67,110) -- (330.04,110) ;
        \draw [line width=0.75]    (323.63,120) -- (330,120) ;
        \draw [color={rgb, 255:red, 0; green, 0; blue, 0 }  ,draw opacity=0.7 ][line width=0.75]    (250,120) -- (276.38,120) ;
        \draw [color={rgb, 255:red, 0; green, 0; blue, 0 }  ,draw opacity=0.7 ][line width=0.75]    (283.63,120) -- (316.38,120) ;
        \draw [color={rgb, 255:red, 0; green, 0; blue, 0 }  ,draw opacity=0.7 ][line width=0.75]    (250,100) -- (330,100) ;
        \draw [color={rgb, 255:red, 0; green, 0; blue, 0 }  ,draw opacity=0.7 ][line width=0.75]    (250,130) -- (296.5,130) -- (330,130) ;
        \draw [color={rgb, 255:red, 0; green, 0; blue, 0 }  ,draw opacity=0.7 ]   (256.38,110) -- (263.63,110) ;
        \draw [color={rgb, 255:red, 0; green, 0; blue, 0 }  ,draw opacity=0.7 ]   (260,120) -- (260,133.63) ;
        \draw [shift={(260,120)}, rotate = 90] [color={rgb, 255:red, 0; green, 0; blue, 0 }  ,draw opacity=0.7 ][fill={rgb, 255:red, 0; green, 0; blue, 0 }  ,fill opacity=0.7 ][line width=0.75]      (0, 0) circle [x radius= 1.34, y radius= 1.34]   ;
        \draw  [color={rgb, 255:red, 0; green, 0; blue, 0 }  ,draw opacity=0.7 ] (256.38,130) .. controls (256.38,128) and (258,126.38) .. (260,126.38) .. controls (262,126.38) and (263.63,128) .. (263.63,130) .. controls (263.63,132) and (262,133.63) .. (260,133.63) .. controls (258,133.63) and (256.38,132) .. (256.38,130) -- cycle ;
        \draw [color={rgb, 255:red, 0; green, 0; blue, 0 }  ,draw opacity=0.7 ]   (256.38,130) -- (263.63,130) ;
        \draw [color={rgb, 255:red, 189; green, 16; blue, 224 }  ,draw opacity=0.7 ]   (280,110) -- (280,123.63) ;
        \draw [shift={(280,110)}, rotate = 90] [color={rgb, 255:red, 189; green, 16; blue, 224 }  ,draw opacity=0.7 ][fill={rgb, 255:red, 189; green, 16; blue, 224 }  ,fill opacity=0.7 ][line width=0.75]      (0, 0) circle [x radius= 1.34, y radius= 1.34]   ;
        \draw  [color={rgb, 255:red, 189; green, 16; blue, 224 }  ,draw opacity=0.7 ] (276.38,120) .. controls (276.38,118) and (278,116.38) .. (280,116.38) .. controls (282,116.38) and (283.63,118) .. (283.63,120) .. controls (283.63,122) and (282,123.63) .. (280,123.63) .. controls (278,123.63) and (276.38,122) .. (276.38,120) -- cycle ;
        \draw [color={rgb, 255:red, 189; green, 16; blue, 224 }  ,draw opacity=0.7 ]   (276.38,120) -- (283.63,120) ;
        \draw [color={rgb, 255:red, 0; green, 0; blue, 0 }  ,draw opacity=0.7 ]   (300,100) -- (300,113.63) ;
        \draw [shift={(300,100)}, rotate = 90] [color={rgb, 255:red, 0; green, 0; blue, 0 }  ,draw opacity=0.7 ][fill={rgb, 255:red, 0; green, 0; blue, 0 }  ,fill opacity=0.7 ][line width=0.75]      (0, 0) circle [x radius= 1.34, y radius= 1.34]   ;
        \draw  [color={rgb, 255:red, 0; green, 0; blue, 0 }  ,draw opacity=0.7 ] (296.38,110) .. controls (296.38,108) and (298,106.38) .. (300,106.38) .. controls (302,106.38) and (303.63,108) .. (303.63,110) .. controls (303.63,112) and (302,113.63) .. (300,113.63) .. controls (298,113.63) and (296.38,112) .. (296.38,110) -- cycle ;
        \draw [color={rgb, 255:red, 0; green, 0; blue, 0 }  ,draw opacity=0.7 ]   (296.38,110) -- (303.63,110) ;
        \draw [color={rgb, 255:red, 189; green, 16; blue, 224 }  ,draw opacity=0.7 ]   (320,110) -- (320,123.63) ;
        \draw [shift={(320,110)}, rotate = 90] [color={rgb, 255:red, 189; green, 16; blue, 224 }  ,draw opacity=0.7 ][fill={rgb, 255:red, 189; green, 16; blue, 224 }  ,fill opacity=0.7 ][line width=0.75]      (0, 0) circle [x radius= 1.34, y radius= 1.34]   ;
        \draw  [color={rgb, 255:red, 189; green, 16; blue, 224 }  ,draw opacity=0.7 ] (316.38,120) .. controls (316.38,118) and (318,116.38) .. (320,116.38) .. controls (322,116.38) and (323.63,118) .. (323.63,120) .. controls (323.63,122) and (322,123.63) .. (320,123.63) .. controls (318,123.63) and (316.38,122) .. (316.38,120) -- cycle ;
        \draw [color={rgb, 255:red, 189; green, 16; blue, 224 }  ,draw opacity=0.7 ]   (316.38,120) -- (323.63,120) ;
        \draw [color={rgb, 255:red, 0; green, 0; blue, 0 }  ,draw opacity=0.7 ]   (259.91,109.96) -- (260.09,96.33) ;
        \draw [shift={(259.91,109.96)}, rotate = 270.74] [color={rgb, 255:red, 0; green, 0; blue, 0 }  ,draw opacity=0.7 ][fill={rgb, 255:red, 0; green, 0; blue, 0 }  ,fill opacity=0.7 ][line width=0.75]      (0, 0) circle [x radius= 1.34, y radius= 1.34]   ;
        \draw  [color={rgb, 255:red, 0; green, 0; blue, 0 }  ,draw opacity=0.7 ] (263.67,100.01) .. controls (263.64,102.01) and (262,103.61) .. (259.99,103.58) .. controls (257.99,103.56) and (256.39,101.91) .. (256.42,99.91) .. controls (256.44,97.91) and (258.09,96.31) .. (260.09,96.33) .. controls (262.09,96.36) and (263.69,98) .. (263.67,100.01) -- cycle ;
        \draw [color={rgb, 255:red, 0; green, 0; blue, 0 }  ,draw opacity=0.7 ]   (263.67,100.01) -- (256.42,99.91) ;
        \draw (335.19,114.2) node  [font=\normalsize]  {$\equiv $};
    \end{tikzpicture}
    \caption{Example of ebit optimization for the circuit of Figure~\ref{Fig:16}: the left part of the equivalence can be optimized to the right one, which reduces the number of non-local gates.}
    \label{Fig:17}
    \hrulefill
\end{figure}
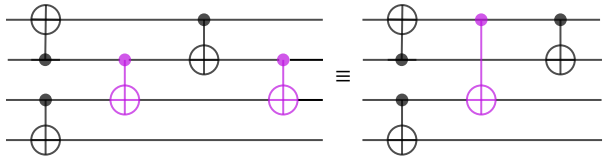

For DQC to be effective and efficient, the quantum compiler must perform some preliminary ebit optimization (such as the one illustrated in Figure~\ref{Fig:17}), then find the best split for the abstract circuit, i.e., the split that minimizes the overall communication cost required to execute the distributed circuit. At the same time, the quantum compiler must find the best local transformation for each piece of computation.

From the above, it should be clear that designing an efficient compiler is a tough task. Because of this, a plethora of proposals to tackle the problem emerges from the literature. In future work, some of them may be combined to more sophisticated compilers. This already happened for local computing. For example, the quantum compiler from the IBM Q framework~\cite{IBMQtranspiler} has several layers of optimization, each tackling the problem from different perspectives.

Most quantum compilers for DQC are characterized by two fundamental steps, namely \textit{qubit assignment} and \textit{non-local gate handling}. In the following, we present these two compilation steps, with reference to the most relevant literature. In Table~\ref{Tab:03}, we compare some prominent DQC-oriented quantum compiling strategies. To this purpose, we consider the programming language, the supported network topologies, the qubit assignment strategy, the non-local gate handling strategy, and the availability of an open source release of the software.

In the remainder of the section, we first present some of the most representative strategies for qubit assignment and non-local gate handling. Then, we discuss some open issues.   

\begin{table*}[t]
    \scriptsize
    \centering
    \begin{tabular}{c|cccccc}
		\toprule
        \textbf{Compiler} & \textbf{Language} & \textbf{Network Topologies} & \textbf{Qubit Assignment} & \textbf{Non-local Gate Handling} & \textbf{Open Source}\\
		\toprule
       ~\cite{AndresMartinez2019} & Haskell & hypergraph & minimum k-cut & telegate and teledata & YES\\
		\midrule
       ~\cite{Sundaram2021} & unknown & hypergraph & minimum k-cut & telegate & NO\\
        \midrule  
       ~\cite{SunGupRam-22} & unknown & any & Tabu search & telegate and teledata & NO\\
        \midrule
       ~\cite{Daei2020} & MATLAB & / & heuristic & teledata & NO\\
        \midrule
       ~\cite{Dava2020} & MATLAB & / & dynamic programming & teledata & NO\\
        \midrule
       ~\cite{Nikahd2021} & C++ and CPLEX & n.a. & minimum k-cut & telegate and teledata & NO\\
        \midrule
       ~\cite{FerCacAmo-21} & Python & LLN & sorting & telegate and teledata & NO\\
        \midrule
       ~\cite{CuoCalKrs-21} & pseudo-code & any & integer linear programming & telegate & /\\
        \midrule
       ~\cite{Dadkhah2021} & MATLAB & n.a. & genetic alg. & teledata & NO\\
        \midrule
       ~\cite{BeaBriGra-13} & / & any & sorting & teledata & /\\
        \midrule
       ~\cite{Brierley2017} & / & hypercube & sorting & teledata & /\\
		\bottomrule
    \end{tabular}
    \caption{Comparison of DQC-oriented quantum compiling strategies. Some strategies find the best partition of the input monolithic quantum circuit in a completely network-agnostic fashion. Some strategies are purely theoretical, not supported by a software implementation.}
    \label{Tab:03}
    \hrulefill
\end{table*}

\subsection{Qubit Assignment}
\label{Sec:6.1}

An abstract circuit is composed by \textit{logical qubits}, while a quantum processor is equipped with a register of \textit{physical qubits}. An assignment, in its most basic form, is a one-to-one mapping between logical and physical qubits.\footnote{One can also consider fault-tolerant mappings, where more than one physical qubit encode a single logical qubit. However we consider this as side work, out from the scope of this survey.}  Whether it is better to tackle it dynamically -- changing the assignment while computing -- or statically -- defining the assignment at the beginning and keeping it for the whole execution of the computation -- is an open problem, which also depends on whether the partition between communication qubits and computing qubits is static or dynamic. 

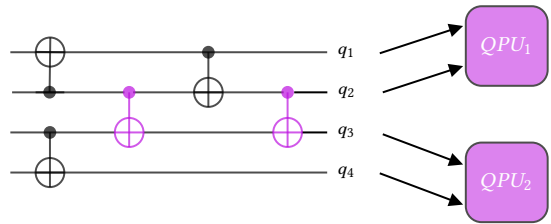
\begin{figure}[t]
    \centering
    \tikzset{every picture/.style={line width=0.75pt}}
    \begin{tikzpicture}[x=0.75pt,y=0.75pt,yscale=-2,xscale=2]
        \draw  [color={rgb, 255:red, 0; green, 0; blue, 0 }  ,draw opacity=0.7 ][fill={rgb, 255:red, 189; green, 16; blue, 224 }  ,fill opacity=0.52 ] (265,192.5) .. controls (265,190.31) and (266.78,188.53) .. (268.97,188.53) -- (281.7,188.53) .. controls (283.89,188.53) and (285.67,190.31) .. (285.67,192.5) -- (285.67,204.4) .. controls (285.67,206.59) and (283.89,208.37) .. (281.7,208.37) -- (268.97,208.37) .. controls (266.78,208.37) and (265,206.59) .. (265,204.4) -- cycle ;
        \draw [color={rgb, 255:red, 0; green, 0; blue, 0 }  ,draw opacity=0.7 ][line width=0.75]    (181,210) -- (219,210) ;
        \draw [color={rgb, 255:red, 0; green, 0; blue, 0 }  ,draw opacity=0.7 ][line width=0.75]    (150.44,210) -- (178.81,210) ;
        \draw [line width=0.75]    (221.67,210) -- (230.04,210) ;
        \draw [line width=0.75]    (223.63,220) -- (230,220) ;
        \draw [color={rgb, 255:red, 0; green, 0; blue, 0 }  ,draw opacity=0.7 ][line width=0.75]    (150,220) -- (176.38,220) ;
        \draw [color={rgb, 255:red, 0; green, 0; blue, 0 }  ,draw opacity=0.7 ][line width=0.75]    (183.63,220) -- (216.38,220) ;
        \draw [color={rgb, 255:red, 0; green, 0; blue, 0 }  ,draw opacity=0.7 ][line width=0.75]    (150,200) -- (230,200) ;
        \draw [color={rgb, 255:red, 0; green, 0; blue, 0 }  ,draw opacity=0.7 ][line width=0.75]    (150,230) -- (196.5,230) -- (230,230) ;
        \draw [color={rgb, 255:red, 0; green, 0; blue, 0 }  ,draw opacity=0.7 ]   (156.38,210) -- (163.63,210) ;
        \draw [color={rgb, 255:red, 0; green, 0; blue, 0 }  ,draw opacity=0.7 ]   (160,220) -- (160,233.63) ;
        \draw [shift={(160,220)}, rotate = 90] [color={rgb, 255:red, 0; green, 0; blue, 0 }  ,draw opacity=0.7 ][fill={rgb, 255:red, 0; green, 0; blue, 0 }  ,fill opacity=0.7 ][line width=0.75]      (0, 0) circle [x radius= 1.34, y radius= 1.34];
        \draw  [color={rgb, 255:red, 0; green, 0; blue, 0 }  ,draw opacity=0.7 ] (156.38,230) .. controls (156.38,228) and (158,226.38) .. (160,226.38) .. controls (162,226.38) and (163.63,228) .. (163.63,230) .. controls (163.63,232) and (162,233.63) .. (160,233.63) .. controls (158,233.63) and (156.38,232) .. (156.38,230) -- cycle ;
        \draw [color={rgb, 255:red, 0; green, 0; blue, 0 }  ,draw opacity=0.7 ]   (156.38,230) -- (163.63,230) ;
        \draw [color={rgb, 255:red, 189; green, 16; blue, 224 }  ,draw opacity=0.7 ]   (180,210) -- (180,223.63) ;
        \draw [shift={(180,210)}, rotate = 90] [color={rgb, 255:red, 189; green, 16; blue, 224 }  ,draw opacity=0.7 ][fill={rgb, 255:red, 189; green, 16; blue, 224 }  ,fill opacity=0.7 ][line width=0.75]      (0, 0) circle [x radius= 1.34, y radius= 1.34];
        \draw  [color={rgb, 255:red, 189; green, 16; blue, 224 }  ,draw opacity=0.7 ] (176.38,220) .. controls (176.38,218) and (178,216.38) .. (180,216.38) .. controls (182,216.38) and (183.63,218) .. (183.63,220) .. controls (183.63,222) and (182,223.63) .. (180,223.63) .. controls (178,223.63) and (176.38,222) .. (176.38,220) -- cycle ;
        \draw [color={rgb, 255:red, 189; green, 16; blue, 224 }  ,draw opacity=0.7 ]   (176.38,220) -- (183.63,220) ;
        \draw [color={rgb, 255:red, 0; green, 0; blue, 0 }  ,draw opacity=0.7 ]   (200,200) -- (200,213.63) ;
        \draw [shift={(200,200)}, rotate = 90] [color={rgb, 255:red, 0; green, 0; blue, 0 }  ,draw opacity=0.7 ][fill={rgb, 255:red, 0; green, 0; blue, 0 }  ,fill opacity=0.7 ][line width=0.75]      (0, 0) circle [x radius= 1.34, y radius= 1.34]   ;
        \draw  [color={rgb, 255:red, 0; green, 0; blue, 0 }  ,draw opacity=0.7 ] (196.38,210) .. controls (196.38,208) and (198,206.38) .. (200,206.38) .. controls (202,206.38) and (203.63,208) .. (203.63,210) .. controls (203.63,212) and (202,213.63) .. (200,213.63) .. controls (198,213.63) and (196.38,212) .. (196.38,210) -- cycle ;
        \draw [color={rgb, 255:red, 0; green, 0; blue, 0 }  ,draw opacity=0.7 ]   (196.38,210) -- (203.63,210) ;
        \draw [color={rgb, 255:red, 189; green, 16; blue, 224 }  ,draw opacity=0.7 ]   (220,210) -- (220,223.63) ;
        \draw [shift={(220,210)}, rotate = 90] [color={rgb, 255:red, 189; green, 16; blue, 224 }  ,draw opacity=0.7 ][fill={rgb, 255:red, 189; green, 16; blue, 224 }  ,fill opacity=0.7 ][line width=0.75]      (0, 0) circle [x radius= 1.34, y radius= 1.34]   ;
        \draw  [color={rgb, 255:red, 189; green, 16; blue, 224 }  ,draw opacity=0.7 ] (216.38,220) .. controls (216.38,218) and (218,216.38) .. (220,216.38) .. controls (222,216.38) and (223.63,218) .. (223.63,220) .. controls (223.63,222) and (222,223.63) .. (220,223.63) .. controls (218,223.63) and (216.38,222) .. (216.38,220) -- cycle ;
        \draw [color={rgb, 255:red, 189; green, 16; blue, 224 }  ,draw opacity=0.7 ]   (216.38,220) -- (223.63,220) ;
        \draw [color={rgb, 255:red, 0; green, 0; blue, 0 }  ,draw opacity=0.7 ]   (159.91,209.96) -- (160.09,196.33) ;
        \draw [shift={(159.91,209.96)}, rotate = 270.74] [color={rgb, 255:red, 0; green, 0; blue, 0 }  ,draw opacity=0.7 ][fill={rgb, 255:red, 0; green, 0; blue, 0 }  ,fill opacity=0.7 ][line width=0.75]      (0, 0) circle [x radius= 1.34, y radius= 1.34]   ;
        \draw  [color={rgb, 255:red, 0; green, 0; blue, 0 }  ,draw opacity=0.7 ] (163.67,200.01) .. controls (163.64,202.01) and (162,203.61) .. (159.99,203.58) .. controls (157.99,203.56) and (156.39,201.91) .. (156.42,199.91) .. controls (156.44,197.91) and (158.09,196.31) .. (160.09,196.33) .. controls (162.09,196.36) and (163.69,198) .. (163.67,200.01) -- cycle ;
        \draw [color={rgb, 255:red, 0; green, 0; blue, 0 }  ,draw opacity=0.7 ]   (163.67,200.01) -- (156.42,199.91) ;
        \draw  [color={rgb, 255:red, 0; green, 0; blue, 0 }  ,draw opacity=0.7 ][fill={rgb, 255:red, 189; green, 16; blue, 224 }  ,fill opacity=0.52 ] (265,226.5) .. controls (265,224.31) and (266.78,222.53) .. (268.97,222.53) -- (281.7,222.53) .. controls (283.89,222.53) and (285.67,224.31) .. (285.67,226.5) -- (285.67,238.4) .. controls (285.67,240.59) and (283.89,242.37) .. (281.7,242.37) -- (268.97,242.37) .. controls (266.78,242.37) and (265,240.59) .. (265,238.4) -- cycle ;
        \draw    (243.28,200.26) -- (261.48,194.37) ;
        \draw [shift={(264.33,193.45)}, rotate = 162.08] [fill={rgb, 255:red, 0; green, 0; blue, 0 }  ][line width=0.08]  [draw opacity=0] (3.57,-1.72) -- (0,0) -- (3.57,1.72) -- cycle    ;
        \draw    (244.04,210) -- (261.48,204.37) ;
        \draw [shift={(264.33,203.45)}, rotate = 162.11] [fill={rgb, 255:red, 0; green, 0; blue, 0 }  ][line width=0.08]  [draw opacity=0] (3.57,-1.72) -- (0,0) -- (3.57,1.72) -- cycle    ;
        \draw    (243.28,230.19) -- (261.52,236.96) ;
        \draw [shift={(264.33,238)}, rotate = 200.35] [fill={rgb, 255:red, 0; green, 0; blue, 0 }  ][line width=0.08]  [draw opacity=0] (3.57,-1.72) -- (0,0) -- (3.57,1.72) -- cycle    ;
        \draw    (244.04,220.45) -- (261.52,226.95) ;
        \draw [shift={(264.33,228)}, rotate = 200.41] [fill={rgb, 255:red, 0; green, 0; blue, 0 }  ][line width=0.08]  [draw opacity=0] (3.57,-1.72) -- (0,0) -- (3.57,1.72) -- cycle;
        \draw (230,200) node [anchor=west] [inner sep=0.75pt]  [font=\tiny]  {${\displaystyle\ \ q_{1}}$};
        \draw (230.04,210) node [anchor=west] [inner sep=0.75pt]  [font=\tiny]  {$\ \ q_{2}$};
        \draw (230,220) node [anchor=west] [inner sep=0.75pt]  [font=\tiny]  {$\ \ q_{3}$};
        \draw (230,230) node [anchor=west] [inner sep=0.75pt]  [font=\tiny]  {$\ \ q_{4}$};
        \draw (275.33,198.45) node  [font=\footnotesize,color={rgb, 255:red, 255; green, 255; blue, 255 }  ,opacity=1 ]  {$QPU_{1}$};
        \draw (275.33,232.45) node  [font=\footnotesize,color={rgb, 255:red, 255; green, 255; blue, 255 }  ,opacity=1 ]  {$QPU_{2}$};
    \end{tikzpicture}
    \caption{Toy example of qubit assignment. Once the logical qubits composing the quantum circuit have been assigned to the different QPUs, the \texttt{CNOT}s between remote qubits -- highlighted in violet -- becomes non-local.}
    \label{Fig:16}
    \hrulefill
\end{figure}

In DQC, qubit assignment is a general-purpose approach to the partitioning problem, introduced in Section~\ref{Sec:4.2}. Specifically, for a given set of logical qubits, we need choose a partition that maps sub-sets of logical qubits to processors, while minimizing the number of required interactions among different sub-sets, as shown in Figure~\ref{Fig:16}.

Several authors investigate this research direction~\cite{AndresMartinez2019,FerCacAmo-21,Sundaram2021,SunGupRam-22}. The reader will find in these works different proposals to address the qubit assignment problem. Not all the papers match in the minimum assumptions for the technology. Specifically, as described in Section~\ref{Sec:5}, we are at a stage where one need to make predictions on the most likely DQC architecture that will run in the next future. If one assumes any connectivity, the resulting model is general-purpose, but it is also hard to tackle. Restricting the connectivity to one that satisfies some properties makes the model less general, but a good set of assumptions in this direction may shape future implementations as well. Currently, the preferred line is to keep connectivity general~\cite{SunGupRam-22}.

Andr\'es-Mart\'inez and Heunen~\cite{AndresMartinez2019} propose to encode a logical circuit as an hypergraph. An hyperedge represents one \textit{ebit} -- i.e., one EPR shared between QPUs --  which allows for a telegate to be performed. Qubit assignment works by minimizing the number of cuts, as each cut corresponds to an ebit. Sundaram et al.~\cite{Sundaram2021} present a two-step solution, where the first step is quantum assignment. Circuits are represented as edge-weighted graphs with qubits as vertices. The edge weights correspond to an estimation for the number of \textit{cat-entanglements}\footref{Footnote:1}. The problem is then solved as a minimum k-cut, where partitions have roughly the same size. In~\cite{SunGupRam-22}, the same authors extend their approach to the case of an arbitrary-topology network of heterogeneous quantum computers by means of a Tabu search algorithm. In~\cite{Daei2020}, by Daei et al., the circuit becomes an undirected graph with qubits as vertices, while edge weights correspond to the number of two-qubit gates between them. In~\cite{Dava2020}, the authors represent circuits as bipartite graphs with two sets of vertices -- one set for the qubits and one for the gates -- and edges to encode dependencies of qubits and gates. Then for the qubit assignment problem, they propose a partitioning algorithm via dynamic programming to minimize the number of teledata operations.

When qubit assignment is dynamic, new challenges -- as well as new possibilities -- arise.  Nikahd et al.~\cite{Nikahd2021} propose a minimum k-cut partitioning algorithm formulated as an ILP optimization problem, to minimize the number of remote interactions. They use a moving window and apply the partitioning algorithm to small sections of the circuit, thus the partition may change with the moving window by means of teledata operations. In~\cite{FerCacAmo-21}, Ferrari et al. consider the worst-case scenario of QPUs interconnected through an LNN topology\footnote{The Linear Nearest Neighbor (LNN) topology~\cite{Fowler2004} consists of processors arranged in a single line -- namely, in a 1-dimensional lattice -- where each processor is interconnected with two neighbors. In the worst-case scenario -- namely, the most challenging one -- each QPU is equipped with a single computational qubit, and only neighboring qubits can interact each others.}. Rather than focusing on the number of remote interactions, they design a sorting algorithm to reduce the depth overhead induced by such time consuming operations. The authors show that the overhead is upper-bounded by a factor that grows linearly with the number of qubits. Cuomo et al. in~\cite{CuoCalKrs-21} model the compilation problem with an Integer Linear Programming formulation. The formulation is inspired to the vast theory on dynamic network problems. Authors managed to define the problem as a special case of \textit{quickest multi-commodity flow}. Such a result allows to perform optimization by means of techniques coming from the literature, such as a \textit{time-expanded} representation of the distributed architecture.

\subsection{Non-local Gate Handling}
\label{Sec:6.2}

As described in Section~\ref{Sec:5}, assumptions on the architectures not only concern connectivity. Predicting the best kind of remote interactions is of critical importance as well. In this sense, the general agreement is that the generation and distribution of entangled states is a fundamental resource to be used sparingly. Indeed, a common goal in the literature is to minimize the number of consumed ebits, as it is the main bottleneck to distributed quantum computation. To this aim, qubit assignment discussed above represents a starting point for further optimization steps, which now concern circuit manipulation.

As described in Section~\ref{Sec:5.3}, there are two main approaches for implementing non-local gates, namely teledata and telegate.

The teledata approach is considered, for example, in~\cite{BeaBriGra-13, Brierley2017, Daei2020, Dava2020, Dadkhah2021}.  Beals et al.~\cite{BeaBriGra-13} prove that the quantum circuit model, the quantum parallel RAM model, and the DQC model are equivalent up to polylogarithmic depth overhead. Other than this major result, they provide an algorithm for emulating circuits on any network graph.
Brierley~\cite{Brierley2017} focuses on $n$-qubit cyclic butterfly networks (a special case of hypercubic network) and proves that there is a sequence of local gates with depth $6\log n$ such that the qubit at node $a$ is sent to node $\pi(a)$ for all $a = 1,...,n$ and any permutation $\pi : [1,n] \rightarrow [1,n]$. In other words, the butterfly network can implement any quantum algorithm with an overhead of $6\log n$. Such a network topology is suitable for multi-chip quantum devices or small controlled networks. In medium-scale or global networks, it is hard to implement such a constrained architecture. Daei et al.~\cite{Daei2020} propose a method to minimize the number of quantum teleportations between DQC partitions. The main idea is to turn the monolithic quantum circuit into an undirected weighted graph, where the weight of each edge represents the number of gates involving a specific pair of qubits for execution. Then, the graph is partitioned using the Kernighan-Lin (K-L) algorithm for VLSI design~\cite{KerLin-1970}, so that the number of edges between partitions is minimized. Finally, each graph partition is converted to a quantum circuit. Davarzani et al.~\cite{Dava2020} propose an algorithm for minimizing teleportations consisting of two steps: first, the quantum circuit is converted into a bipartite graph model, and then a dynamic programming approach (DP) is used to partition the model into low-capacity quantum circuits. Finally, Dadkhah et al.~\cite{Dadkhah2021} propose a heuristic approach to replace the equivalent circuits in the initial quantum circuit. Then, they use a genetic algorithm to partition the placement of qubits so that the number of teleportations could be optimized for the communications of a DQC.

The telegate direction is pursued, for example, in~\cite{AndresMartinez2019,Sundaram2021,CuoCalKrs-21}. Andr\'es-Martinez et al.~\cite{AndresMartinez2019} use cat-entanglement\footref{Footnote:1} to implement non-local quantum gates. The chosen gate set contains every one-qubit gate and a single two-qubit gate, namely the CZ gate (i.e., the controlled version of the Z gate). The authors consider no restriction on the ebit connectivity between QPUs. Then, they reduce the problem of distributing a circuit across multiple QPUs to hypergraph partitioning. The proposed approach is evaluated against five quantum circuits, including QFT. The proposed solution has some drawbacks, in particular that there is no way to customize the number of communication qubits of each QPU.
As previously mentioned, in Sundaram's et al. paper~\cite{Sundaram2021}, a two-step quantum compiling approach is introduced. The first step is qubit assignment, while the second step is finding the smallest set of cat-entanglement operations that will enable the execution of all telegates. The authors state that, in a special setting, this problem can be reduced to a vertex-cover problem, allowing for a polynomial-time optimal solution based on integer linear programming. They also provide a $O(\log n)$-approximate solution, where $n$ is the total number of global gates, for a generalized setting by means of greedy search algorithm. Also the aforementioned work by Cuomo et al.~\cite{CuoCalKrs-21} adopts the telegate approach.

\subsection{Open Issues and Research Directions}
\label{Sec:6.3}

The most advanced quantum compilers for execution on single quantum processors are noise-aware, i.e., they take the noise statistics of the device into account, for some or all steps~\cite{MurBakJav-19,NisPanSat-19,NiuSuaSta-20,SivDilCow-21,FerAmo-22}. A noise-aware quantum compiler for DQC is still missing. Indeed, it is still an open question what kind of noise-awareness such a compiler should have. The different options range from a compiler that has complete knowledge of the target execution platform (quantum processors, quantum links, etc.) to a compiler that only knows generic features of the target quantum processors and network -- as the execution manager will decide the actual execution platform assigned to the computation.  

Further work could be done regarding the integration of quantum compilers with simulation tools -- in line with the preliminary attempt that was made by Ferrari et al.~\cite{FerNasAmo-21} -- allowing for automated workflows that would allow for faster comparative evaluation of compiling strategies. 

So far, testing the quality of compiled circuits on real execution platforms has not been possible for the majority of researchers. Once a quantum network will be available to the public -- much like current IBM Q, Rigetti, etc. single quantum devices -- it will be possible to evaluate DQC compilers more effectively , with key performance indicators including the resulting computation quality, state fidelity, and other performance metrics~\cite{WanGuoSha-22}.

\section{Simulation Tools}
\label{Sec:7}

To support the research community in the design and evaluation of quantum computing and quantum network technologies, including hardware, protocols and applications, many simulation tools have been developed recently.

Simulations are very important for several reasons. First of all, they allow for defining hardware requirements using a top-down approach, i.e., starting from applications and protocols. In this way, hardware design is driven by high-level KPIs (key performance indicators), rather than proceeding by trial and error. Another advantage of simulations is related to network sizing. Given the number of potential users and the number of available quantum processors, simulation allows for devising and evaluating different network topologies and entanglement routing schemes, which results in saving time and money. Regarding DQC, simulation plays a crucial role for establishing the correctness of the compiled distributed quantum programs, and evaluating the quality of their execution against different hardware platforms, network configurations and scheduling algorithms.

In Table~\ref{Tab:04}, we compare some prominent simulation tools that, in our view, can be used for designing and evaluating DQC systems. We propose to classify each tool as belonging to one of three possible classes: i) hardware-oriented (HW), ii) protocol-oriented (PR), and iii) application-oriented (AP). In the remainder of the section, we first present each class with some of the most representative simulation tools. Then, we discuss some open issues.   

\subsection{Hardware-oriented}
\label{Sec:7.1}

We denote as HW simulation tools those that allow the user to model the physical entities with the desired degree of detail, including noise models. Prominent examples are SQUANCH~\cite{Bartlett2018} and NetSquid~\cite{Coopmans2021}, discussed in the following. Regarding DQC, we note that HW simulation tools are useful for evaluating the impact of different hardware technologies (including noise models) on the quality of the distributed program execution.

\begin{table*}[t]
    \scriptsize
    \centering
    \begin{tabular}{c|cccccc}
		\toprule
        \textbf{Simulation Tool} & \textbf{Language} & \textbf{Multiprocessing} & \textbf{Multithreading} & \textbf{Noise Models} & \textbf{Open Source} & \textbf{Class}\\
		\toprule
        SQUANCH~\cite{Bartlett2018} & Python & NO & NO & YES & YES & HW\\
		\midrule
        NetSquid~\cite{Coopmans2021} & Python & NO & NO & YES & NO & HW\\
		\midrule
        SimulaQron~\cite{Dahlberg2018} & Python & YES & NO & NO & YES & PR\\
		\midrule
        SeQUeNCe~\cite{Wu2021} & C++/Python & YES & NO & YES & YES & PR\\
		\midrule
        QuiSP~\cite{Matsuo2021} & C++ & NO & NO & YES & YES & PR\\ 
		\midrule
        QuNetSim~\cite{DiadamoNotzel2021} & Python & NO & YES & NO & YES & PR\\
		\midrule
        NetQASM SDK~\cite{DahVecDon-22} & C++/Python & NO & YES & YES & YES & AP\\
		\midrule
        QNE-ADK~\cite{QNEADK-22} & C++/Python & NO & NO & YES & NO & AP\\
		\bottomrule
    \end{tabular}
    \caption{Comparison of simulation tools that can be used for designing and evaluating DQC systems.}
    \label{Tab:04}
    \hrulefill
\end{table*}

The \textit{Simulator for Quantum Networks and Channels} (SQUANCH)~\cite{Bartlett2018} is an open-source Python framework for creating parallelized simulations of distributed quantum information processing. Despite the framework includs many features of a general-purpose quantum computing simulator, it is optimized specifically for simulating quantum networks. It includes functionality to allow users to design complex multi-party quantum networks, extensible classes for modeling noisy quantum channels, and a multiprocessed NumPy backend for performant simulations. The core modules are \textsf{QSystem}, representing a multi-body quantum system as a density matrix in the computational basis, and \textsf{QStream}, which is an iterable ensemble of separable $N$-qubit \textsf{QSystem}s optimized for cache locality. By default \textsf{QStream} state is stored in a shared memory as a C-type array of doubles, which is type-casted as a 3D array of \textsf{np.complex64} values. During simulations, \textsf{Agent}s run in parallel from separate processes, synchronizing clocks and passing information between each other through \textsf{Channel}s. There is no explicit concurrency safety when a \textsf{QSystem} is modified by multiple agents, as sending and receiving \textsf{Qubit}s are blocking operations that allow for naturally safe parallelism. However, the scalability of this simulation tool is hindered by the lack of support for distributed multiprocessing, as all the processes must run on the same machine. The source code is not maintained since 2018.

NetSquid~\cite{Coopmans2021} is one of the most advanced platforms for simulating quantum networking and modular computing systems subject to physical non-idealities. It ranges from the physical layer and its control plane up to the application level. This is achieved by integrating several key technologies: a discrete-event simulation engine, a specialized quantum computing library, a modular framework for modeling quantum hardware devices, and an asynchronous programming framework for describing quantum protocols. NetSquid has been used for different purposes, such as the evaluation of a benchmarking procedure for quantum protocols~\cite{LiaBahSil-22}, the evaluation of end-to-end entanglement generation strategies in terms of capacity bounds and impact on Quantum Key Distribution (QKD)~\cite{NehNieRas-20,ManAmo-22}, and the performance evaluation of request scheduling algorithms for quantum networks~\cite{CicConPas-21}. 

\subsection{Protocol-oriented}
\label{Sec:7.2}

In the proposed classification, PR simulation tools are mostly devoted to the design and evaluation of general-purpose quantum protocols, -- such as quantum state teleportation, quantum leader election, etc.~\cite{ProtocolZoo-22} -- with the possibility to model hardware-agnostic networked quantum processors, with very limited (if not missing) support for noise modeling. Relevant examples are SimulaQron~\cite{Dahlberg2018}, SeQUeNCe~\cite{Wu2021}, QuiSP~\cite{Matsuo2021} and QuNetSim~\cite{DiadamoNotzel2021}. Regarding DQC, PR simulation tools are useful for evaluating the impact of different compiling and execution management strategies on the quality of the distributed program execution, in (almost) ideal conditions.

SimulaQron~\cite{Dahlberg2018} is a tool for developing distributed software that runs on real or simulated classical and quantum end-nodes, connected by classical and quantum links. SimulaQron spawns three stacked processes per network node: the lowest one for wrapping a simulated quantum registry, based on an hardware-specific third-party simulator; the intermediate process exposing simulated qubits that map 1-to-1 to those of the quantum registry; the upper process providing virtual qubits that are manipulated within a platform-independent application. For example, if two virtual qubits belonging to different processes, running on physically-separated servers, are manipulated in order to share an entangled state (let say, a Bell state), the corresponding simulated qubits (and quantum register ones) are both stored in the memory of one server, in order to make it possible to simulate measurements in a consistent fashion. This process-oriented approach makes SimulaQron quite scalable and able to leverage multicore server architecture in order to speed up the execution of the simulations. However, SimulaQron does not come with noise model support, thus preventing the simulation of quantum protocols over non-ideal networks.

SeQUeNCe~\cite{Wu2021} is an open-source discrete-event quantum network simulator, whose latest release fully supports parallel simulation. The authors designed and developed a quantum state manager (QSM) that maintains shared quantum information distributed across multiple processes, and also optimized their parallel code by minimizing the overhead of the QSM and by decreasing the amount of synchronization among processes.  

QuiSP~\cite{Matsuo2021} is an event-driven Quantum Internet simulation package. QuiSP is built on top of the OMNeT++ discrete event simulation framework. Compared to the simulators discussed so far, many of which focus on physically realistic simulation of a single small network, QuiSP is oriented to protocol design for complex, heterogeneous networks at large scale while keeping the physical layer as realistic as possible. Emphasis has been placed on realistic noise models. The declared long-term goal for the simulator is to be able to handle an internetwork with 100 networks of 100 nodes each. To simulate quantum networks at the cost of only a few classical bits per qubit, QuiSP works \textit{in the error basis}, i.e., tracking only errors, not states. The premise is that the desired quantum state is known and only deviations from this ideal state must be tracked. This is a novel approach for simulating quantum networks, adapted from quantum error correction~\cite{Devitt2013}. 
The performance of QuISP was investigated in terms of events processed per second and the duration of CPU time taken to generate one end-to-end Bell pair, using the Docker environment that QuISP provides. It was shown in~\cite{Matsuo2021} that the average CPU time (in seconds) per end-to-end Bell pair generated grows no worse than polynomially in the number of quantum repeaters. Increasing the number of repeaters results in longer simulation time in the scaling, as expected. It also emerged that that QuISP might have some kind of unintended overhead which scales linearly on the number of buffer qubits, which the authors expect to fix in a near-term release~\cite{Matsuo2021}.

QuNetSim~\cite{DiadamoNotzel2021} implements a layered model of network component objects inspired by the OSI model. In particular, application, transport, and network layers are considered. QuNetSim does not explicitly incorporate features of the link and physical layers. Indeed, QuNetSim relies on open-source qubit simulators that are used to simulate the physical qubits in the network, namely SimulaQron~\cite{Dahlberg2018}, ProjectQ~\cite{Steiger2018} and EQSN~\cite{EQSN2020} (the latter one being the default backend, as it was developed by the QuNetSim team).
In QuNetSim, network nodes can run both classical and quantum applications. The transport layer component prepares classical packets, encodes qubits for superdense message transmission, handles the generation of the two correction bits for quantum state teleportation, etc. The network layer component can route classical and quantum information using two internal network graphs and two different routing algorithms. 
The network component objects are implemented using threading and observing queues. Extensive use of threading allows each task to wait without blocking the main program thread, which simulates the behavior of sending information and waiting for an acknowledgment, or expecting information to arrive for some period of time from another host. 
QuNetSim works well for small scale simulations using five to ten hosts that are separated by a small number of hops, while it tends to reach its limits when many entangled qubits are being generated across the network with many parallel operations.

\subsection{Application-oriented}
\label{Sec:7.3}

The third class is devoted to AP simulation tools, which are tailored to the design and implementation of quantum network applications. Usually, these tools rely on simulated backends offered by other packages that are not directly accessible to the user -- for example, NetQASM SDK~\cite{DahVecDon-22} relying on NetSquid~\cite{Coopmans2021}.
Regarding DQC, AP simulation tools are useful for quickly assessing the quality of quantum circuit splits produced by quantum compilers. The execution management scheme (i.e., job scheduling, entanglement routing, etc.) is hidden to the user, which is at most allowed to specify the network topology (from a short list of preconfigured networks) and the values of a few parameters characterizing the hardware of the quantum processors.

The process of setting up a simulation requires strong expertise in the simulator itself, thus being inconvenient for those who are only interested in quantum protocol evaluation or in the design of supporting tools such as quantum compilers. Recently, Ferrari \textit{et al.}~\cite{FerNasAmo-21} presented a software tool, denoted as DQC Executor, that accepts as input the description of the network and the code of the algorithm, and then executes the simulation by automatically constructing the network topology and mapping the computation onto it, in a framework-agnostic way and transparently to the user. The tool is in its early stages and currently supports automatic deployment of distributed quantum algorithms to the NetSquid~\cite{Coopmans2021} simulator. The description of the network is provided by the user in a specific YAML format. The distributed algorithm, instead, is defined with the OpenQASM~\cite{CroBisSmo-17} language.

NetQASM SDK~\cite{DahVecDon-22} is a high-level software development kit, in Python, whose purpose is to make easier to write quantum network applications, to simulate them through NetSquid~\cite{Coopmans2021} or SimulaQron~\cite{Dahlberg2018}, and (expected in the near future) to execute them on real hardware. Indeed, the quantum programs developed with NetQASM SDK are translated into low-level programs based on the NetQASM language, similar in nature to classical assembly languages. With respect to other QASM languages, NetQASM provides elements for remote entanglement generation. On the other hand, NetQASM contains no provision for classical communication with remote nodes. Synchronization between the NetQASM programs (through classical send/recv primitives) of multiple nodes is the responsibility of the application programmer.

The Quantum Network Explorer Application Development Kit (QNE-ADK)~\cite{QNEADK-22} allows the user to create applications and experiments and run them on a simulator. When configuring an application, the user specifies the different roles and what types of inputs the application uses. In addition, the user writes the functionality of the application using the NetQASM SDK~\cite{DahVecDon-22}. When configuring an experiment, the user can give values to the inputs that were specified when creating the application. The user also chooses which channels and nodes are used in the network and which role is linked to which node. Once configured, the experiment is parsed and sent to the NetSquid simulator~\cite{Coopmans2021}. QNE-ADK is particularly useful when the application code developed with NetQASM SDK is provided to the user, whose only duty is to configure and perform experiments. Indeed, using the execution environment is straightforward. There is also a visual interface that further simplifies the experiment configuration.

Both NetQASM SDK~\cite{DahVecDon-22} and QNE-ADK~\cite{QNEADK-22} are very useful tools. Without them, configuring DQC simulations is quite a complex task.

\subsection{Open Issues and Research Directions}
\label{Sec:7.4}

There is a sufficiently variegated choice of simulation tools for quantum networks and backends to support DQC research, with specialization on hardware, protocols, or applications. On the other hand, a simulation tool allowing for full-stack simulation of large networks is still missing. Such a tool should be support multiprocessing and multithreading, and simple deployment of DQC simulations on high performance computing facilities.

Another possible direction is the development of tools for orchestrating DQC simulations, with automated instantiation of simulation objects representing QPUs and quantum network components. Having quantum compilers for DQC in the loop would be also very useful. Last but not least, it would be great to have the possibility to seamlessly replace simulated hardware with real devices.

\section{Conclusions and Future Perspectives}
\label{Sec:8}

\begin{figure}[t]
    \centering
    \includegraphics[width=0.95\linewidth]{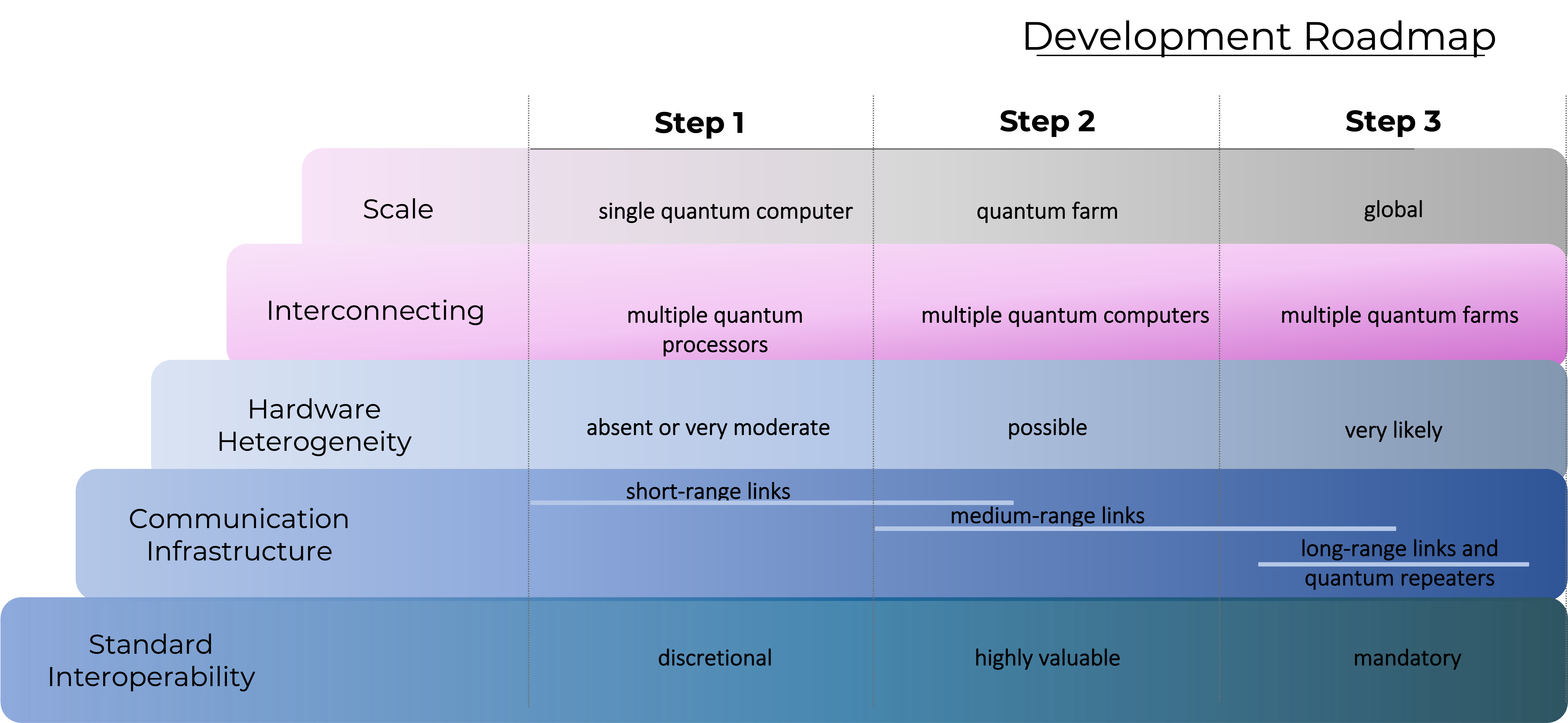}
    \caption{Possible stages for the development of distributed quantum computing. It is reasonable to assume that the underlying hardware complexity scales proportional with three dimensions: i) the extension of the communication infrastructure, ii) the number of interconnected quantum devices, iii) and the hardware heterogeneity among the quantum devices.}
    \label{Fig:18}
    \hrulefill
\end{figure}

Here we conclude the survey by first providing an industrial perspective on distributed quantum computing, and then by discussing the possible stages of distributing quantum computing development.

\subsection{Industrial and Standardization Perspective}
\label{Sec:8.1}
A first quantum revolution has already exploited quantum technologies in our everyday life, creating a deep techno-economic and social impact. Today, a second revolution is underway, and it is safe to predict it will have a major impact in many markets, ranging from Telecom and ICT, through Medicine, to Finance and Transportation, and so on. 

Clearly, significant work is still needed to develop enabling components and systems for DQC. Yet, considering the foreseen industrial opportunities, significant investments are being made worldwide across public and private organizations.

One major obstacle on the way of industrial exploitation of distributed quantum computing is that, nowadays, the industry has not yet consolidated around one type of quantum hardware technology. In this scenario, a \textit{quantum hardware abstraction layer} (Quantum-HAL) -- embracing the two killer domains of quantum technologies for ICT, namely, quantum computing and quantum networking -- would allow applications and services developers to start using the abstractions of the underneath quantum hardware, even if still under consolidation. This would definitely simplify and speed-up the development of quantum platforms, services, and applications. Indeed, a Quantum-HAL for distributed quantum computing would provide unified northbound quantum application programming interfaces (APIs) for the higher layers, decoupling from the different types of quantum hardware technologies (e.g., trapped ions, superconducting qubits, silicon photonic qubits).

Another key aspect for increasing the TRL (Technology Readiness Level) of distributed quantum computing concerns its integration with current Telecom and ICT infrastructures. This implies the definition and standardization of a management and control approach (architectures and APIs) able of interworking with current solutions. All these activities require coordinated and joint efforts including -- where appropriate -- existing projects, industry bodies and standard (ITU-T, ETSI, IETF and IEEE just to mention a few) active in the area of quantum technologies.

Overall, the final goal is to bridge the gap between DQC and the established cloud and edge computing platforms, tools and methods, and to focus in on the inter-related constraints between the different aspects of the architectural design, so to enable the development of practical DQC solutions. To achieve this goal, research and innovation activities are required in diverse and complementary fields, ranging from computational complexity and networked systems through quantum information and optics to communications and computer science engineering.

\subsection{The Journey Ahead}
\label{Sec:8.2}

Paving a journey towards distributed quantum computing is a challenging task, as hard as any other prediction about technological developments. Yet, we can sketch roughly three stages, as discussed in the following and summarized with Figure~\ref{Fig:18}.

The first step involves distributed quantum computing exploiting multiple quantum processors within a single quantum computer. The quantum hardware underlying the qubits is likely to be homogeneous among the different processors. Yet, some sort of hardware heterogeneity may arise within each processor due to the differences in terms of requirements\footnote{Quantum memory requires coherence times several order of magnitude larger than computational qubits.} between memory qubits and computational qubits. The physical distance between remote qubits is clearly very short. Hence, it is reasonable to assume, as communication infrastructure, short-range microwave links. The network topology is likely static, so that only simple quantum network functionalities are required. Clearly, quantum decoherence must be carefully accounted for, so that the decoherence time can be used as overall key metric. Local operations between qubits within a single processor must be complemented by remote operations between qubits placed at different processors. The trade-off between qubits devoted to computation and entangled qubits devoted to communication represents a fundamental issue with no counterpart in classical distributed computing. The very challenging task of designing distributed quantum algorithms must explicitly take such trade-off -- as well as the delay induced by remote operations -- into consideration.

The second step involves inter-rack distributed quantum computing, where the computation is performed collectively by multiple quantum computers located within the same farm. At this stage, some sort of hardware heterogeneity might arise, given that different quantum computers are involved in the computation. Clearly, such heterogeneity must be taken into consideration at each layer of a distributed quantum computing ecosystem. Yet, entanglement distribution still benefits from a tightly controlled environment -- reasonable to assume available within a single quantum farm -- and the relatively short distances. As a matter of fact, the communication infrastructure can still be composed by cold microwave links~\cite{MagStoKur-20} although optical links would greatly simplify the hardware requirements albeit at the price of significant technological advances in the microwave-optical conversion. Delay imposed by classical and quantum communication times is slightly longer -- when compared to stage one -- hence more sophisticated timing and synchronization functionalities are required. The network topology becomes more complex, and it may present some sort of temporal dynamics as the number of interconnected quantum computers might change in time. This, in turn, induces network functionalities dynamics that must be carefully taken into account. The problem of remote operations compiling -- and, hence, the trade-off between computational and communication qubits -- becomes even more intricate. Finally, at this stage, the execution management problem (previously discussed in Section~\ref{Sec:4.2}) will arise, with multiple users performing concurrent access to the resources. 

The third step involves interconnecting multiple geographically-distributed quantum farms. Two are the key challenges here. First, there exists a likely spread heterogeneity -- given that the different quantum farms will be likely operated by different companies --, which requires significant efforts in terms of standardization and interoperability. Furthermore, the heterogeneity among quantum links, e.g., optical, free-space or satellite, will arise. The delays induced by the distances will introduce severe challenges on the entanglement generation and distribution. The increasing number of quantum devices to be wired and the heterogeneity of the environments hosting the quantum computers must be taken into account as well. At this stage, the compiling and execution management problems would be even more complex, demanding for specific network services to be integrated with those of the classical Internet (such as DNS, DHCP, etc.).

We underline that, although each successive stage is distinguished by an increasing amount of interconnected quantum resources, the actual deployment evolution will strongly depend on the technological advances and the experimental implementations of the different entities composing a distributed quantum computing ecosystem~\cite{CuoCalCac-20}.

One of the judicious questions raised from this discussion is: when will we see the distributed quantum computing? There is no definite answer to this question. However, we firmly believe this is a goal that requires a collaborative effort and a multi-disciplinary approach between academics and industries. The required competences and skills are many and diverse and each is interconnected with and vital to the others.


\bibliographystyle{unsrtnat}
\bibliography{main.bib}

\begin{thebibliography}{118}
\providecommand{\natexlab}[1]{#1}
\providecommand{\url}[1]{\texttt{#1}}
\expandafter\ifx\csname urlstyle\endcsname\relax
  \providecommand{\doi}[1]{doi: #1}\else
  \providecommand{\doi}{doi: \begingroup \urlstyle{rm}\Url}\fi

\bibitem[Cacciapuoti et~al.(2020{\natexlab{a}})Cacciapuoti, Caleffi, Tafuri,
  Cataliotti, Gherardini, and Bianchi]{CacCalTaf-20}
Angela~Sara Cacciapuoti, Marcello Caleffi, Francesco Tafuri, Francesco~Saverio
  Cataliotti, Stefano Gherardini, and Giuseppe Bianchi.
\newblock Quantum internet: Networking challenges in distributed quantum
  computing.
\newblock \emph{IEEE Network}, 34\penalty0 (1):\penalty0 137--143,
  2020{\natexlab{a}}.
\newblock \doi{10.1109/MNET.001.1900092}.

\bibitem[Rohde(2021)]{Rohde2021}
P.~P. Rohde.
\newblock \emph{{The Quantum Internet -- The Second Quantum Revolution}}.
\newblock {Cambridge University Press}, 2021.

\bibitem[{EU Quantum Flagship}()]{EUQF}
{EU Quantum Flagship}.
\newblock {The Future is Quantum}.
\newblock URL \url{https://qt.eu/}.

\bibitem[{QIA Team}()]{QIA}
{QIA Team}.
\newblock {Quantum Internet Alliance}.
\newblock URL \url{https://quantum-internet.team/}.

\bibitem[{European Commission and European Space Agency}()]{EuroQCI}
{European Commission and European Space Agency}.
\newblock {The European Quantum Communication Infrastructure (EuroQCI)
  Initiative}.
\newblock URL
  \url{https://digital-strategy.ec.europa.eu/en/policies/european-quantum-communication-infrastructure-euroqci}.

\bibitem[Finigan()]{Finigan2022}
Will Finigan.
\newblock {Quantum network research centers in the US}.
\newblock URL
  \url{https://www.aliroquantum.com/blog/quantum-network-research-centers-in-the-us}.

\bibitem[{Q-NEXT}()]{QNEXT}
{Q-NEXT}.
\newblock {Q-NEXT website}.
\newblock URL \url{https://q-next.org/}.

\bibitem[{HQAN}()]{HQAN}
{HQAN}.
\newblock {HQAN website}.
\newblock URL \url{https://hqan.illinois.edu/}.

\bibitem[{CQN}()]{CQN}
{CQN}.
\newblock {CQN website}.
\newblock URL \url{http://cqn-erc.org/}.

\bibitem[Yin et~al.(2017)Yin, Cao, Li, et~al.]{Yin2017}
Juan Yin, Yuan Cao, Yu-Huai Li, et~al.
\newblock {Satellite-based entanglement distribution over 1200 kilometers}.
\newblock \emph{Science}, 356\penalty0 (6343):\penalty0 1140--1144, 2017.

\bibitem[Graps()]{Graps2022}
Amara Graps.
\newblock {How Much Money Has China Already Invested into Quantum Technology? -
  Part 2}.
\newblock URL
  \url{https://quantumcomputingreport.com/how-much-money-has-china-already-invested-into-quantum-technology/}.

\bibitem[Cuomo et~al.(2020)Cuomo, Caleffi, and Cacciapuoti]{CuoCalCac-20}
Daniele Cuomo, Marcello Caleffi, and Angela~Sara Cacciapuoti.
\newblock Towards a distributed quantum computing ecosystem.
\newblock \emph{IET Quantum Communication}, 1:\penalty0 3--8(5), July 2020.
\newblock \doi{10.1049/iet-qtc.2020.0002}.

\bibitem[{QCommHub}()]{QCommHub}
{QCommHub}.
\newblock {QCommHub website}.
\newblock URL \url{https://www.quantumcommshub.net/}.

\bibitem[{Amazon}()]{AWSQN}
{Amazon}.
\newblock {Announcing the AWS Center for Quantum Networking}.
\newblock URL
  \url{https://aws.amazon.com/blogs/quantum-computing/announcing-the-aws-center-for-quantum-networking/}.

\bibitem[Wang et~al.(2022{\natexlab{a}})Wang, Rahman, Li, Aelmans, and
  Chakraborty]{WanRahLi-22}
Chonggang Wang, Akbar Rahman, Ruidong Li, Melchior Aelmans, and Kaushik
  Chakraborty.
\newblock Application scenarios for the quantum internet.
\newblock Internet-Draft draft-irtf-qirg-quantum-internet-use-cases-12,
  Internet Engineering Task Force, 2022{\natexlab{a}}.
\newblock Work in Progress.

\bibitem[Van~Meter and Devitt(2016)]{VanMeter2016}
Rodney Van~Meter and Simon~J. Devitt.
\newblock The {{Path}} to {{Scalable Distributed Quantum Computing}}.
\newblock \emph{Computer}, 49\penalty0 (9):\penalty0 31--42, September 2016.
\newblock ISSN 0018-9162.
\newblock \doi{10.1109/MC.2016.291}.

\bibitem[Caleffi et~al.(2018)Caleffi, Cacciapuoti, and Bianchi]{CalCacBia-18}
Marcello Caleffi, Angela~Sara Cacciapuoti, and Giuseppe Bianchi.
\newblock Quantum internet: From communication to distributed computing!
\newblock In \emph{Proc. of ACM NANOCOM '18}, pages 1--4. Association for
  Computing Machinery, 2018.
\newblock ISBN 9781450357111.
\newblock \doi{10.1145/3233188.3233224}.

\bibitem[Ferrari et~al.(2021{\natexlab{a}})Ferrari, Cacciapuoti, Amoretti, and
  Caleffi]{FerCacAmo-21}
Davide Ferrari, Angela~Sara Cacciapuoti, Michele Amoretti, and Marcello
  Caleffi.
\newblock Compiler design for distributed quantum computing.
\newblock \emph{IEEE Transactions on Quantum Engineering}, 2:\penalty0 1--20,
  2021{\natexlab{a}}.
\newblock \doi{10.1109/TQE.2021.3053921}.

\bibitem[Avron et~al.(2021)Avron, Casper, and Rozen]{Avron2021}
J.~Avron, Ofer Casper, and Ilan Rozen.
\newblock Quantum advantage and noise reduction in distributed quantum
  computing.
\newblock \emph{Phys. Rev. A}, 104:\penalty0 052404, Nov 2021.
\newblock \doi{10.1103/PhysRevA.104.052404}.

\bibitem[Wehner et~al.(2018)Wehner, Elkouss, and Hanson]{WehElkHan-18}
Stephanie Wehner, David Elkouss, and Ronald Hanson.
\newblock {Quantum Internet: a Vision for the Road Ahead}.
\newblock \emph{Science}, 362\penalty0 (6412), 2018.

\bibitem[Caleffi et~al.(2020)Caleffi, Chandra, Cuomo, Hasaanpour, and
  Cacciapuoti]{CalChaCuo-20}
Marcello Caleffi, Daryus Chandra, Daniele Cuomo, Shima Hasaanpour, and
  Angela~Sara Cacciapuoti.
\newblock {The Rise of the Quantum Internet.}
\newblock \emph{IEEE Computer}, 2020.

\bibitem[{IBM}()]{IBM2025}
{IBM}.
\newblock {Expanding the IBM Quantum roadmap to anticipate the future of
  quantum-centric supercomputing}.
\newblock URL \url{https://research.ibm.com/blog/ibm-quantum-roadmap-2025}.

\bibitem[Parekh et~al.(2021)Parekh, Ricciardi, Darwish, and
  DiAdamo]{Parekh2021}
R.~Parekh, A.~Ricciardi, A.~Darwish, and S.~DiAdamo.
\newblock Quantum algorithms and simulation for parallel and distributed
  quantum computing.
\newblock In \emph{2021 IEEE/ACM Second International Workshop on Quantum
  Computing Software (QCS)}, pages 9--19, Los Alamitos, CA, USA, nov 2021. IEEE
  Computer Society.
\newblock \doi{10.1109/QCS54837.2021.00005}.
\newblock URL
  \url{https://doi.ieeecomputersociety.org/10.1109/QCS54837.2021.00005}.

\bibitem[Zhong et~al.(2021)Zhong, Chang, Bienfait, et~al.]{ZhoChaBie-21}
Youpeng Zhong, Hung-Shen Chang, Audrey Bienfait, et~al.
\newblock Deterministic multi-qubit entanglement in a quantum network.
\newblock \emph{Nature}, 590\penalty0 (7847):\penalty0 571--575, 2021.

\bibitem[Pompili et~al.(2021)Pompili, Hermans, Baier, et~al.]{PomHerBai-21}
M.~Pompili, S.~L.~N. Hermans, S.~Baier, et~al.
\newblock Realization of a multinode quantum network of remote solid-state
  qubits.
\newblock \emph{Science}, 372\penalty0 (6539):\penalty0 259--264, 2021.

\bibitem[Hermans et~al.(2022)Hermans, Pompili, Beukers, et~al.]{HerPomBeu-22}
S.~L.~N. Hermans, M.~Pompili, H.~K.~C. Beukers, et~al.
\newblock Qubit teleportation between non-neighbouring nodes in a quantum
  network.
\newblock \emph{Nature}, 605\penalty0 (7911):\penalty0 663--668, 2022.

\bibitem[Cacciapuoti et~al.(2020{\natexlab{b}})Cacciapuoti, Caleffi, Van~Meter,
  and Hanzo]{CacCalVan-20}
Angela~Sara Cacciapuoti, Marcello Caleffi, Rodney Van~Meter, and Lajos Hanzo.
\newblock When entanglement meets classical communications: Quantum
  teleportation for the quantum internet.
\newblock \emph{IEEE Transactions on Communications}, 68\penalty0 (6):\penalty0
  3808--3833, 2020{\natexlab{b}}.
\newblock invited paper.

\bibitem[Rieffel and Polak(2000)]{RiePol-00}
Eleanor Rieffel and Wolfgang Polak.
\newblock An introduction to quantum computing for non-physicists.
\newblock \emph{ACM Computing Survey}, 32\penalty0 (3):\penalty0 300–335, sep
  2000.

\bibitem[Rieffel and Polak(2011)]{RiePol-11}
Eleanor Rieffel and Wolfgang Polak.
\newblock \emph{{Quantum Computing: A Gentle Introduction}}.
\newblock The MIT Press, 2011.

\bibitem[Nielsen and Chuang(2011)]{NieChu-11}
Michael~A. Nielsen and Isaac~L. Chuang.
\newblock \emph{Quantum Computation and Quantum Information}.
\newblock Cambridge University Press, 2011.

\bibitem[Dirac(1939)]{Dirac1939}
P.~A.~M. Dirac.
\newblock {A new notation for quantum mechanics}.
\newblock \emph{Mathematical Proceedings of the Cambridge Philosophical
  Society}, 35\penalty0 (3):\penalty0 416--418, 1939.

\bibitem[Jozsa and Linden(2003)]{JozLin-03}
Richard Jozsa and Noah Linden.
\newblock On the role of entanglement in quantum-computational speed-up.
\newblock \emph{Proceedings of the Royal Society of London. Series A:
  Mathematical, Physical and Engineering Sciences}, 459\penalty0
  (2036):\penalty0 2011--2032, 2003.
\newblock \doi{10.1098/rspa.2002.1097}.

\bibitem[Markham and Sanders(2008)]{MarSan-11}
Damian Markham and Barry~C. Sanders.
\newblock Graph states for quantum secret sharing.
\newblock \emph{Phys. Rev. A}, 78:\penalty0 042309, 2008.

\bibitem[Bell(1964)]{Bel-64}
J.~S. Bell.
\newblock {On the Einstein Podolsky Rosen paradox}.
\newblock \emph{Physics Physique Fizika}, 1:\penalty0 195--200, Nov 1964.

\bibitem[Einstein et~al.(1935)Einstein, Podolsky, and Rosen]{EinPodRos-35}
Albert Einstein, Boris Podolsky, and Nathan Rosen.
\newblock Can quantum-mechanical description of physical reality be considered
  complete?
\newblock \emph{Physical review}, 47\penalty0 (10):\penalty0 777, 1935.

\bibitem[Gottesman(1998)]{Gottesman1998}
Daniel Gottesman.
\newblock Theory of fault-tolerant quantum computation.
\newblock \emph{Phys. Rev. A}, 57:\penalty0 127--137, Jan 1998.

\bibitem[J. et~al.(2022)J., Adedoyin, Ambrosiano, et~al.]{Abhijith2022}
Abhijith J., Adetokunbo Adedoyin, John Ambrosiano, et~al.
\newblock Quantum algorithm implementations for beginners.
\newblock \emph{ACM Transactions on Quantum Computing}, 3\penalty0 (4), jul
  2022.
\newblock ISSN 2643-6809.

\bibitem[Koudia et~al.(2022)Koudia, Cacciapuoti, Simonov, and
  Caleffi]{KouCacCal-21}
Seid Koudia, Angela~Sara Cacciapuoti, Kyrylo Simonov, and Marcello Caleffi.
\newblock How deep the theory of quantum communications goes: Superadditivity,
  superactivation and causal activation.
\newblock \emph{IEEE Communications Surveys \& Tutorials}, 2022.
\newblock In press.

\bibitem[Linke et~al.(2017)Linke, Maslov, Roetteler, et~al.]{LinMasRoe-17}
Norbert~M. Linke, Dmitri Maslov, Martin Roetteler, et~al.
\newblock Experimental comparison of two quantum computing architectures.
\newblock \emph{Proceedings of the National Academy of Sciences}, 114\penalty0
  (13):\penalty0 3305--3310, 2017.

\bibitem[Kandala et~al.(2019)Kandala, Temme, C{\'o}rcoles,
  et~al.]{KanTemCor-19}
Abhinav Kandala, Kristan Temme, Antonio~D. C{\'o}rcoles, et~al.
\newblock Error mitigation extends the computational reach of a noisy quantum
  processor.
\newblock \emph{Nature}, 567\penalty0 (7749):\penalty0 491--495, 2019.

\bibitem[Yu and Li(2022)]{YuLi-22}
Ziang Yu and Yingzhou Li.
\newblock Analysis of error propagation in quantum computers.
\newblock \emph{arXiv e-prints}, 2022.
\newblock arXiv:2209.01699.

\bibitem[Meter and Devitt(2016)]{VanDev-16}
R.~Van Meter and S.~J. Devitt.
\newblock The path to scalable distributed quantum computing.
\newblock \emph{Computer}, 49\penalty0 (9):\penalty0 31--42, Sept 2016.
\newblock ISSN 0018-9162.
\newblock \doi{10.1109/MC.2016.291}.

\bibitem[Botea et~al.(2018)Botea, Kishimoto, and Marinescu]{Botea2018}
A.~Botea, A.~Kishimoto, and R.~Marinescu.
\newblock {On the Complexity of Quantum Circuit Compilation}.
\newblock In \emph{The Eleventh International Symposium on Combinatorial Search
  (SOCS 2018)}, 2018.

\bibitem[Kusyk et~al.(2021)Kusyk, Saeed, and Uyar]{KusSaeUya-21}
Janusz Kusyk, Samah~M. Saeed, and Muharrem~Umit Uyar.
\newblock Survey on quantum circuit compilation for noisy intermediate-scale
  quantum computers: Artificial intelligence to heuristics.
\newblock \emph{IEEE Transactions on Quantum Engineering}, 2:\penalty0 1--16,
  2021.
\newblock \doi{10.1109/TQE.2021.3068355}.

\bibitem[Sivarajah et~al.(2020)Sivarajah, Dilkes, Cowtan, et~al.]{SivDilCow-21}
Seyon Sivarajah, Silas Dilkes, Alexander Cowtan, et~al.
\newblock t|ket⟩: a retargetable compiler for nisq devices.
\newblock \emph{Quantum Science and Technology}, 6\penalty0 (1):\penalty0
  014003, nov 2020.

\bibitem[{Carcoles} et~al.(2020){Carcoles}, {Kandala}, {Javadi-Abhari},
  et~al.]{CorKanJav-20}
A.~D. {Carcoles}, A.~{Kandala}, A.~{Javadi-Abhari}, et~al.
\newblock Challenges and opportunities of near-term quantum computing systems.
\newblock \emph{Proc. of the IEEE}, pages 1--15, 2020.
\newblock in press.

\bibitem[Ferrari and Amoretti(2018)]{FerAmo-18}
Davide Ferrari and Michele Amoretti.
\newblock Efficient and effective quantum compiling for entanglement-based
  machine learning on ibm q devices.
\newblock \emph{International Journal of Quantum Information}, 16\penalty0
  (08):\penalty0 1840006, 2018.

\bibitem[Cincio et~al.(2018)Cincio, Suba{\c{s}}{\i}, Sornborger, and
  Coles]{CinSubSor-18}
Lukasz Cincio, Yi{\u{g}}it Suba{\c{s}}{\i}, Andrew~T Sornborger, and Patrick~J
  Coles.
\newblock Learning the quantum algorithm for state overlap.
\newblock \emph{New Journal of Physics}, 20\penalty0 (11):\penalty0 113022,
  Nov. 2018.

\bibitem[{Zulehner} et~al.(2019){Zulehner}, {Paler}, and {Wille}]{ZulPalWil-19}
A.~{Zulehner}, A.~{Paler}, and R.~{Wille}.
\newblock An efficient methodology for mapping quantum circuits to the ibm qx
  architectures.
\newblock \emph{IEEE Transactions on Computer-Aided Design of Integrated
  Circuits and Systems}, 38\penalty0 (7):\penalty0 1226--1236, 2019.

\bibitem[Montanaro(2016)]{Mon-16}
Ashley Montanaro.
\newblock Quantum algorithms: An overview.
\newblock \emph{npj Quantum Information}, 2\penalty0 (1):\penalty0 15023,
  January 2016.
\newblock ISSN 2056-6387.
\newblock \doi{10.1038/npjqi.2015.23}.

\bibitem[Shor(1994)]{Shor1994}
Peter~W. Shor.
\newblock Polynomial time algorithms for discrete logarithms and factoring on a
  quantum computer.
\newblock In \emph{Algorithmic Number Theory}, pages 289--289. Springer Berlin
  Heidelberg, 1994.
\newblock ISBN 978-3-540-49044-9.

\bibitem[Grover(1996)]{Gro-96}
Lov~K. Grover.
\newblock A fast quantum mechanical algorithm for database search.
\newblock In \emph{Proceedings of the Twenty-Eighth Annual ACM Symposium on
  Theory of Computing}, STOC '96, page 212–219, 1996.
\newblock ISBN 0897917855.

\bibitem[Harrow et~al.(2009)Harrow, Hassidim, and Lloyd]{HarHasLlo-09}
Aram~W. Harrow, Avinatan Hassidim, and Seth Lloyd.
\newblock Quantum algorithm for linear systems of equations.
\newblock \emph{Phys. Rev. Lett.}, 103:\penalty0 150502, Oct 2009.

\bibitem[Cerezo et~al.(2021)Cerezo, Arrasmith, Babbush, et~al.]{CerArrBab-21}
M.~Cerezo, Andrew Arrasmith, Ryan Babbush, et~al.
\newblock Variational quantum algorithms.
\newblock \emph{Nature Reviews Physics}, 3\penalty0 (9):\penalty0 625--644,
  September 2021.

\bibitem[Terhal(2015)]{Ter-15}
Barbara~M. Terhal.
\newblock Quantum error correction for quantum memories.
\newblock \emph{Rev. Mod. Phys.}, 87:\penalty0 307--346, Apr 2015.

\bibitem[Neumann et~al.(2020)Neumann, {van Houte}, and Attema]{Neumann2020}
Niels M.~P. Neumann, Roy {van Houte}, and Thomas Attema.
\newblock Imperfect {{Distributed Quantum Phase Estimation}}.
\newblock In \emph{Computational {{Science}} \textendash{} {{ICCS}} 2020},
  Lecture {{Notes}} in {{Computer Science}}, pages 605--615. {Springer
  International Publishing}, 2020.

\bibitem[Kitaev(1997)]{Kitaev1997}
{A Yu} Kitaev.
\newblock Quantum computations: algorithms and error correction.
\newblock \emph{Russian Mathematical Surveys}, 52\penalty0 (6):\penalty0
  1191--1249, dec 1997.

\bibitem[Eisert et~al.(2000)Eisert, Jacobs, Papadopoulos, and
  Plenio]{EisJAcPap-00}
J.~Eisert, K.~Jacobs, P.~Papadopoulos, and M.~B. Plenio.
\newblock Optimal local implementation of nonlocal quantum gates.
\newblock \emph{Phys. Rev. A}, 62:\penalty0 052317, Oct 2000.

\bibitem[DiAdamo et~al.(2021)DiAdamo, Ghibaudi, and Cruise]{DiAdamo2021}
Stephen DiAdamo, Marco Ghibaudi, and James Cruise.
\newblock Distributed {{Quantum Computing}} and {{Network Control}} for
  {{Accelerated VQE}}.
\newblock \emph{IEEE Transactions on Quantum Engineering}, 2:\penalty0 1--21,
  2021.
\newblock ISSN 2689-1808.
\newblock \doi{10.1109/TQE.2021.3057908}.

\bibitem[Yimsiriwattana and Lomonaco(2005)]{Yimsiriwattana2005}
Anocha Yimsiriwattana and {Samuel J. Jr.} Lomonaco.
\newblock Generalized {GHZ} states and distributed quantum computing.
\newblock \emph{Contemp. Math.}, 381, 2005.
\newblock \doi{10.1090/conm/381}.

\bibitem[Neumann and Wezeman(2022)]{NeuWez-22}
Niels M.~P. Neumann and Robert~S. Wezeman.
\newblock Distributed quantum machine learning.
\newblock In \emph{Innovations for Community Services}, pages 281--293.
  Springer International Publishing, 2022.

\bibitem[Cicconetti et~al.(2022)Cicconetti, Conti, and
  Passarella]{CicConPas-22}
Claudio Cicconetti, Marco Conti, and Andrea Passarella.
\newblock Resource allocation in quantum networks for distributed quantum
  computing.
\newblock In \emph{2022 IEEE International Conference on Smart Computing
  (SMARTCOMP)}, pages 124--132, 2022.

\bibitem[Cirac et~al.(1999)Cirac, Ekert, Huelga, and
  Macchiavello]{CirEkeHue-99}
J.~I. Cirac, A.~K. Ekert, S.~F. Huelga, and C.~Macchiavello.
\newblock Distributed quantum computation over noisy channels.
\newblock \emph{Phys. Rev. A}, 59:\penalty0 4249--4254, Jun 1999.

\bibitem[Beals et~al.(2013)Beals, Brierley, Gray, et~al.]{BeaBriGra-13}
Robert Beals, Stephen Brierley, Oliver Gray, et~al.
\newblock Efficient distributed quantum computing.
\newblock \emph{Proc. of the Royal Society A: Mathematical, Physical and
  Engineering Sciences}, 469\penalty0 (2153):\penalty0 20120686, 2013.

\bibitem[Brierley(2017)]{Brierley2017}
Stephen Brierley.
\newblock Efficient implementation of quantum circuits with limited qubit
  interactions.
\newblock \emph{Quantum Info. Comput.}, 17\penalty0 (13–14):\penalty0
  1096–1104, November 2017.
\newblock ISSN 1533-7146.

\bibitem[Vardoyan and Wehner(2022)]{VARWEH-22}
Gayane Vardoyan and Stephanie Wehner.
\newblock {Quantum Network Utility Maximization}.
\newblock \emph{arXiv e-prints}, 2022.
\newblock arXiv:2210.08135v1.

\bibitem[Lee et~al.(2022)Lee, Dai, Towsley, and Englund]{LeeDaiTow-22}
Yuan Lee, Wenhen Dai, Dan Towsley, and Dirk Englund.
\newblock {Quantum Network Utility: A Framework for Benchmarking Quantum
  Networks}.
\newblock \emph{arXiv e-prints}, 2022.
\newblock arXiv:2210.10752v1.

\bibitem[Cross et~al.(2019)Cross, Bishop, Sheldon, Nation, and
  Gambetta]{CroBisShe-19}
Andrew~W. Cross, Lev~S. Bishop, Sarah Sheldon, Paul~D. Nation, and Jay~M.
  Gambetta.
\newblock Validating quantum computers using randomized model circuits.
\newblock \emph{Phys. Rev. A}, 100:\penalty0 032328, Sep 2019.
\newblock \doi{10.1103/PhysRevA.100.032328}.

\bibitem[Kozlowski et~al.(2022)Kozlowski, Wehner, Meter, Rijsman, Cacciapuoti,
  Caleffi, and Nagayama]{KozWehVan-22}
Wojciech Kozlowski, Stephanie Wehner, Rodney~Van Meter, Bruno Rijsman,
  Angela~Sara Cacciapuoti, Marcello Caleffi, and S.~Nagayama.
\newblock Architectural principles for a quantum internet.
\newblock Internet-Draft draft-irtf-qirg-principles-10, Internet Engineering
  Task Force, 2022.
\newblock Work in Progress.

\bibitem[Illiano et~al.(2022)Illiano, Caleffi, Manzalini, and
  Cacciapuoti]{IllCalMan-22}
Jessica Illiano, Marcello Caleffi, Antonio Manzalini, and Angela~Sara
  Cacciapuoti.
\newblock Quantum internet protocol stack: a comprehensive survey.
\newblock \emph{Computer Networks}, 213, 2022.

\bibitem[Cacciapuoti et~al.(2022)Cacciapuoti, Illiano, and
  Caleffi]{CacIllKou-22}
Angela~Sara Cacciapuoti, Seid Illiano, Jessica~Koudia, and Marcello Caleffi.
\newblock The quantum internet: Enhancing classical services one qubit at a
  time.
\newblock \emph{IEEE Networks}, 2022.

\bibitem[Cacciapuoti and Caleffi(2019)]{CacCal-19}
Angela~Sara Cacciapuoti and Marcello Caleffi.
\newblock Toward the quantum internet: A directional-dependent noise model for
  quantum signal processing.
\newblock In \emph{IEEE ICASSP '19}, pages 7978--7982, 2019.
\newblock \doi{10.1109/ICASSP.2019.8683195}.

\bibitem[Horodecki et~al.(2009)Horodecki, Horodecki, Horodecki, and
  Horodecki]{HorHorHor-09}
R.~Horodecki, Pawe{\l} Horodecki, Micha{\l} Horodecki, and Karol Horodecki.
\newblock Quantum entanglement.
\newblock \emph{Reviews of modern physics}, 81\penalty0 (2):\penalty0
  865\color{black}, 2009.

\bibitem[Unnikrishnan and Markham(2020)]{UnnMar-20}
Anupama Unnikrishnan and Damian Markham.
\newblock Authenticated teleportation and verification in a noisy network.
\newblock \emph{Phys. Rev. A}, 102:\penalty0 042401, 2020.

\bibitem[Van~Meter et~al.(2006)Van~Meter, Nemoto, Munro, and
  Itoh]{VanNemMun-06}
R.~Van~Meter, K.~Nemoto, W.J. Munro, and K.M. Itoh.
\newblock Distributed arithmetic on a quantum multicomputer.
\newblock In \emph{33rd International Symposium on Computer Architecture
  (ISCA'06)}, pages 354--365, 2006.

\bibitem[Bennett et~al.(1996{\natexlab{a}})Bennett, Brassard, Popescu,
  et~al.]{BenBraPop-96}
Charles~H Bennett, Gilles Brassard, Sandu Popescu, et~al.
\newblock Purification of noisy entanglement and faithful teleportation via
  noisy channels.
\newblock \emph{Phys. Rev. Lett.}, 76\penalty0 (5):\penalty0 722,
  1996{\natexlab{a}}.

\bibitem[Bennett et~al.(1996{\natexlab{b}})Bennett, DiVincenzo, Smolin, and
  Wootters]{BenDivSmo-96}
C.~H. Bennett, David~P DiVincenzo, John~A Smolin, and William~K Wootters.
\newblock Mixed-state entanglement and quantum error correction.
\newblock \emph{Physical Review A}, 54\penalty0 (5):\penalty0
  3824\color{black}, 1996{\natexlab{b}}.

\bibitem[D{\"u}r and Briegel(2007)]{DurBri-07}
Wolfgang D{\"u}r and Hans~J Briegel.
\newblock Entanglement purification and quantum error correction.
\newblock \emph{Reports on Progress in Physics}, 70\penalty0 (8):\penalty0
  1381, 2007.

\bibitem[Ruan et~al.(2018)Ruan, Dai, and Win]{RuaDaiWin-18}
L.~Ruan, Wenhan Dai, and Moe~Z Win.
\newblock Adaptive recurrence quantum entanglement distillation for
  two-kraus-operator channels.
\newblock \emph{Physical Review A}, 97\penalty0 (5):\penalty0
  052332\color{black}, 2018.

\bibitem[Rozp{\k{e}}dek et~al.(2018)Rozp{\k{e}}dek, Schiet, Elkouss,
  et~al.]{SchElkDoh-18}
Filip Rozp{\k{e}}dek, Thomas Schiet, David Elkouss, et~al.
\newblock Optimizing practical entanglement distillation.
\newblock \emph{Physical Review A}, 97\penalty0 (6):\penalty0 062333, 2018.

\bibitem[Ruan et~al.(2021)Ruan, Kirby, Brodsky, and Win]{RuaKirBro-21}
L.~Ruan, Brian~T Kirby, Michael Brodsky, and Moe~Z Win.
\newblock Efficient entanglement distillation for quantum channels with
  polarization mode dispersion.
\newblock \emph{Physical Review A}, 103\penalty0 (3):\penalty0
  032425\color{black}, 2021.

\bibitem[Soeken et~al.(2019)Soeken, Meuli, Schmitt, et~al.]{Soeken2019}
M.~Soeken, G.~Meuli, B.~Schmitt, et~al.
\newblock Boolean satisfiability in quantum compilation.
\newblock \emph{Phil. Trans. Royal Soc. A}, 378\penalty0 (2164):\penalty0
  1--16, 2019.
\newblock \doi{10.1098/rsta.2019.0161}.

\bibitem[Chamberland et~al.(2020)Chamberland, Zhu, Yoder, et~al.]{ChaZhuYod-20}
C.~Chamberland, G.~Zhu, T.~J. Yoder, et~al.
\newblock {Topological and Subsystem Codes on Low-Degree Graphs with Flag
  Qubits}.
\newblock \emph{Physical Review X}, 10\penalty0 (011022), 2020.

\bibitem[{IBM Q}(2022)]{IBMQtranspiler}
{IBM Q}.
\newblock {Transpiler}.
\newblock \\https://qiskit.org/documentation/apidoc/transpiler.html, 2022.

\bibitem[Andr\'es-Mart\'{\i}nez and Heunen(2019)]{AndresMartinez2019}
Pablo Andr\'es-Mart\'{\i}nez and Chris Heunen.
\newblock Automated distribution of quantum circuits via hypergraph
  partitioning.
\newblock \emph{Phys. Rev. A}, 100:\penalty0 032308, Sep 2019.
\newblock \doi{10.1103/PhysRevA.100.032308}.

\bibitem[G.~Sundaram et~al.(2021)G.~Sundaram, Gupta, and
  Ramakrishnan]{Sundaram2021}
Ranjani G.~Sundaram, Himanshu Gupta, and C.~R. Ramakrishnan.
\newblock {Efficient Distribution of Quantum Circuits}.
\newblock In \emph{35th International Symposium on Distributed Computing (DISC
  2021)}, 2021.

\bibitem[Sundaram et~al.(2022)Sundaram, Gupta, and Ramakrishnan]{SunGupRam-22}
R.~G. Sundaram, H.~Gupta, and C.~R. Ramakrishnan.
\newblock {Distribution of Quantum Circuits Over General Quantum Networks}.
\newblock In \emph{2022 IEEE International Conference on Quantum Computing and
  Engineering (QCE)}, pages 415--425, 2022.

\bibitem[Daei et~al.(2020)Daei, Navi, and {Zomorodi-Moghadam}]{Daei2020}
Omid Daei, Keivan Navi, and Mariam {Zomorodi-Moghadam}.
\newblock Optimized quantum circuit partitioning.
\newblock \emph{Int J Theor Phys}, 59\penalty0 (12):\penalty0 3804--3820,
  December 2020.
\newblock ISSN 1572-9575.
\newblock \doi{10.1007/s10773-020-04633-8}.

\bibitem[Davarzani et~al.(2020)Davarzani, Zomorodi-Moghadam, Houshmand, and
  Nouri-baygi]{Dava2020}
Z.~Davarzani, M.~Zomorodi-Moghadam, M.~Houshmand, and M.~Nouri-baygi.
\newblock A dynamic programming approach for distributing quantum circuits by
  bipartite graphs.
\newblock \emph{Quantum Information Processing}, 19, 2020.
\newblock \doi{10.1007/s11128-020-02871-7}.

\bibitem[Nikahd et~al.(2021)Nikahd, Mohammadzadeh, Sedighi, and
  Zamani]{Nikahd2021}
Eesa Nikahd, Naser Mohammadzadeh, Mehdi Sedighi, and Morteza~Saheb Zamani.
\newblock Automated window-based partitioning of quantum circuits.
\newblock \emph{Phys. Scr.}, 96\penalty0 (3):\penalty0 035102, January 2021.
\newblock ISSN 1402-4896.
\newblock \doi{10.1088/1402-4896/abd57c}.

\bibitem[Cuomo et~al.(2021)Cuomo, Caleffi, Krsulich, Tramonto, Agliardi, Prati,
  and Cacciapuoti]{CuoCalKrs-21}
Daniele Cuomo, Marcello Caleffi, Kevin Krsulich, Filippo Tramonto, Gabriele
  Agliardi, Enrico Prati, and Angela~Sara Cacciapuoti.
\newblock Optimized compiler for distributed quantum computing, 2021.

\bibitem[Dadkhah et~al.(2021)Dadkhah, Zomorodi, and Hosseini]{Dadkhah2021}
Davood Dadkhah, Mariam Zomorodi, and Seyed~Ebrahim Hosseini.
\newblock A {{New Approach}} for {{Optimization}} of {{Distributed Quantum
  Circuits}}.
\newblock \emph{International Journal of Theoretical Physics}, 60\penalty0
  (9):\penalty0 3271--3285, September 2021.
\newblock ISSN 0020-7748, 1572-9575.
\newblock \doi{10.1007/s10773-021-04904-y}.

\bibitem[Fowler et~al.(2004)Fowler, Devitt, and Hollenberg]{Fowler2004}
A.~G. Fowler, S.~J. Devitt, and L.~C.~L. Hollenberg.
\newblock Implementation of {S}hor's algorithm on a linear nearest neighbor
  qubit array.
\newblock \emph{Quantum Information and Computation}, 4:\penalty0 237--251,
  2004.
\newblock \doi{10.26421/QIC4.4}.

\bibitem[Kernighan and Lin(1970)]{KerLin-1970}
B.~W. Kernighan and S.~Lin.
\newblock An efficient heuristic procedure for partitioning graphs.
\newblock \emph{The Bell System Technical Journal}, 49\penalty0 (2):\penalty0
  291--307, 1970.
\newblock \doi{10.1002/j.1538-7305.1970.tb01770.x}.

\bibitem[Murali et~al.(2019)Murali, Baker, Abhari, et~al.]{MurBakJav-19}
Prakash Murali, Jonathan~M. Baker, Ali~Javadi Abhari, et~al.
\newblock Noise-adaptive compiler mappings for noisy intermediate-scale quantum
  computers.
\newblock \emph{arXiv e-prints}, 2019.
\newblock arXiv:1901.11054.

\bibitem[Nishio et~al.(2020)Nishio, Pan, Satoh, et~al.]{NisPanSat-19}
Shin Nishio, Yulu Pan, Takahiko Satoh, et~al.
\newblock Extracting success from ibm’s 20-qubit machines using error-aware
  compilation.
\newblock \emph{J. Emerg. Technol. Comput. Syst.}, 16\penalty0 (3), may 2020.

\bibitem[Niu et~al.(2020)Niu, Suau, Staffelbach, and
  Todri-Sanial]{NiuSuaSta-20}
Siyuan Niu, Adrien Suau, Gabriel Staffelbach, and Aida Todri-Sanial.
\newblock A hardware-aware heuristic for the qubit mapping problem in the nisq
  era.
\newblock \emph{IEEE Transactions on Quantum Engineering}, 1:\penalty0 1--14,
  2020.

\bibitem[Ferrari and Amoretti(2022)]{FerAmo-22}
Davide Ferrari and Michele Amoretti.
\newblock Noise-adaptive quantum compilation strategies evaluated with
  application-motivated benchmarks.
\newblock In \emph{Proceedings of the 19th ACM International Conference on
  Computing Frontiers}, CF '22, page 237–243, 2022.
\newblock \doi{10.1145/3528416.3530250}.

\bibitem[Ferrari et~al.(2021{\natexlab{b}})Ferrari, Nasturzio, and
  Amoretti]{FerNasAmo-21}
Davide Ferrari, Saverio Nasturzio, and Michele Amoretti.
\newblock A software tool for mapping and executing distributed quantum
  computations on a network simulator, 2021{\natexlab{b}}.
\newblock URL \url{https://2021.qcrypt.net/speakers/#list-of-accepted-posters}.

\bibitem[Wang et~al.(2022{\natexlab{b}})Wang, Guo, and Shan]{WanGuoSha-22}
Junchao Wang, Guoping Guo, and Zheng Shan.
\newblock Sok: Benchmarking the performance of a quantum computer.
\newblock \emph{Entropy}, 24\penalty0 (10), 2022{\natexlab{b}}.
\newblock \doi{10.3390/e24101467}.

\bibitem[Bartlett(2018)]{Bartlett2018}
Ben Bartlett.
\newblock A distributed simulation framework for quantum networks and channels.
\newblock \emph{arXiv e-prints}, 2018.
\newblock arXiv:1808.07047.

\bibitem[Coopmans et~al.(2021)Coopmans, Knegjens, Dahlberg,
  et~al.]{Coopmans2021}
Tim Coopmans, Robert Knegjens, Axel Dahlberg, et~al.
\newblock {{NetSquid}}, a {{NETwork Simulator}} for {{QUantum Information}}
  using {{Discrete}} events.
\newblock \emph{Communications Physics}, 4\penalty0 (1):\penalty0 164, December
  2021.

\bibitem[Dahlberg and Wehner(2018)]{Dahlberg2018}
Axel Dahlberg and Stephanie Wehner.
\newblock {SimulaQron - a simulator for developing quantum internet software}.
\newblock \emph{Quantum Science and Technology}, 4\penalty0 (1):\penalty0
  015001, Sep 2018.

\bibitem[Wu et~al.(2021)Wu, Kolar, Chung, et~al.]{Wu2021}
Xiaoliang Wu, Alexander Kolar, Joaquin Chung, et~al.
\newblock Sequence: a customizable discrete-event simulator of quantum
  networks.
\newblock \emph{Quantum Science and Technology}, 6\penalty0 (4):\penalty0
  045027, 2021.

\bibitem[Matsuo(2021)]{Matsuo2021}
T.~Matsuo.
\newblock {Simulation of a Dynamic, RuleSet-based Quantum Network}.
\newblock \emph{arXiv e-prints}, 2021.
\newblock arXiv:1908.10758.

\bibitem[Diadamo et~al.(2021)Diadamo, Notzel, Zanger, and
  Bese]{DiadamoNotzel2021}
Stephen Diadamo, Janis Notzel, Benjamin Zanger, and Mehmet~Mert Bese.
\newblock {QuNetSim: A Software Framework for Quantum Networks}.
\newblock \emph{IEEE Transactions on Quantum Engineering}, 2:\penalty0 1–12,
  2021.

\bibitem[Dahlberg et~al.(2022)Dahlberg, van~der Vecht, Donne,
  et~al.]{DahVecDon-22}
Axel Dahlberg, Bart van~der Vecht, Carlo~Delle Donne, et~al.
\newblock Netqasm - a low-level instruction set architecture for hybrid
  quantum–classical programs in a quantum internet.
\newblock \emph{Quantum Science and Technology}, 7\penalty0 (3):\penalty0
  035023, jun 2022.

\bibitem[QuTech(2022)]{QNEADK-22}
QuTech.
\newblock {Quantum Network Explorer ADK}, 2022.
\newblock URL \url{https://github.com/QuTech-Delft/qne-adk}.

\bibitem[Liao et~al.(2022)Liao, Bahrani, {da Silva}, and Kashefi]{LiaBahSil-22}
Chin-Te Liao, Sima Bahrani, Francisco~Ferreira {da Silva}, and Elham Kashefi.
\newblock Benchmarking of quantum protocols.
\newblock \emph{Scientific Reports}, 12\penalty0 (1):\penalty0 5298, March
  2022.
\newblock ISSN 2045-2322.
\newblock \doi{10.1038/s41598-022-08901-x}.

\bibitem[Mehic et~al.(2020)Mehic, Niemiec, Rass, et~al.]{NehNieRas-20}
Miralem Mehic, Marcin Niemiec, Stefan Rass, et~al.
\newblock Quantum key distribution: A networking perspective.
\newblock \emph{ACM Comput. Surv.}, 53\penalty0 (5), sep 2020.

\bibitem[Manzalini and Amoretti(2022)]{ManAmo-22}
Antonio Manzalini and Michele Amoretti.
\newblock End-to-end entanglement generation strategies: Capacity bounds and
  impact on quantum key distribution.
\newblock \emph{Quantum Reports}, 4\penalty0 (3):\penalty0 251--263, 2022.

\bibitem[Cicconetti et~al.(2021)Cicconetti, Conti, and
  Passarella]{CicConPas-21}
Claudio Cicconetti, Marco Conti, and Andrea Passarella.
\newblock Request scheduling in quantum networks.
\newblock \emph{IEEE Transactions on Quantum Engineering}, 2:\penalty0 2--17,
  2021.

\bibitem[{Various Authors}(2022)]{ProtocolZoo-22}
{Various Authors}.
\newblock {Quantum Protocol Zoo}, 2022.
\newblock URL \url{https://wiki.veriqloud.fr/index.php}.

\bibitem[Devitt et~al.(2013)Devitt, Munro, and Nemoto]{Devitt2013}
S.J. Devitt, W.J. Munro, and K.~Nemoto.
\newblock Quantum error correction for beginners.
\newblock \emph{Reports on Progress in Physics}, 76\penalty0 (7), 2013.

\bibitem[Steiger et~al.(2018)Steiger, H{\"a}ner, and Troyer]{Steiger2018}
Damian~S. Steiger, Thomas H{\"a}ner, and Matthias Troyer.
\newblock {{ProjectQ}}: {{An Open Source Software Framework}} for {{Quantum
  Computing}}.
\newblock \emph{Quantum}, 2:\penalty0 49, January 2018.
\newblock ISSN 2521-327X.
\newblock \doi{10.22331/q-2018-01-31-49}.

\bibitem[Zanger(2020)]{EQSN2020}
S.~Zanger, B. andd~DiAdamo.
\newblock {EQSN: Effective Quantum Simulator for Networks}, 2020.
\newblock URL \url{https://github.com/tqsd/EQSN\_python}.

\bibitem[Cross et~al.(2017)Cross, Bishop, Smolin, and Gambetta]{CroBisSmo-17}
Andrew~W. Cross, Lev~S. Bishop, John~A. Smolin, and Jay~M. Gambetta.
\newblock Open quantum assembly language.
\newblock \emph{arXiv e-prints}, July 2017.
\newblock arXiv:1707.03429.

\bibitem[Magnard et~al.(2020)Magnard, Storz, Kurpiers, et~al.]{MagStoKur-20}
P.~Magnard, S.~Storz, P.~Kurpiers, et~al.
\newblock Microwave quantum link between superconducting circuits housed in
  spatially separated cryogenic systems.
\newblock \emph{Phys. Rev. Lett.}, 125:\penalty0 260502, Dec 2020.

\end{thebibliography}

\end{document}